\documentclass[twocolumn,pra,superscriptaddress,english,nofootinbib]{revtex4-2}
\usepackage[T1]{fontenc}
\usepackage{color}
\usepackage{babel}
\usepackage{mathtools}
\usepackage{bm}
\usepackage{graphicx}
\usepackage{esint}
\usepackage{amsfonts}
\usepackage{graphicx}
\usepackage{float}
\usepackage{amsthm}
\usepackage{amssymb}
\usepackage[ruled, longend]{algorithm2e}
\usepackage{listings}
\usepackage[usenames,divipsnames,svgnames,table]{xcolor}
\usepackage{booktabs}
\usepackage{multirow}

\newcommand{\codefont}{\fontfamily{bch}\selectfont\small\itshape}

\lstdefinestyle{mypython}{
language=Python,
basicstyle=\codefont\footnotesize\itshape,
numberstyle={\tiny\color{lightgray}},
frame=single,
framerule=0.5pt,
xleftmargin=0.5cm,
xrightmargin=0.5cm,
numberstyle=\tiny\textsf,
showstringspaces=false,
numberfirstline=true,
backgroundcolor=\color{lightgray!10},
commentstyle=\color{tropicalrainforest},    % comment style
keywordstyle=\color{ForestGreen}\bfseries,       % keyword style
classoffset = 2,
inputencoding=utf8,
frameround=fftt,
}
\usepackage[unicode=true,pdfusetitle,
bookmarks=true,bookmarksnumbered=false,bookmarksopen=false,
breaklinks=true,pdfborder={0 0 0},backref=false,colorlinks=true]{hyperref}
\hypersetup{linkcolor=NavyBlue,urlcolor=NavyBlue,citecolor=NavyBlue}
\usepackage{breakurl}

\begin{document}

\title{QuanEstimation: An open-source toolkit for quantum parameter estimation}

\author{Mao Zhang}
\affiliation{National Precise Gravity Measurement Facility, MOE Key Laboratory
of Fundamental Physical Quantities Measurement, School of Physics,
Huazhong University of Science and Technology, Wuhan 430074, China}

\author{Huai-Ming Yu}
\affiliation{National Precise Gravity Measurement Facility, MOE Key Laboratory
of Fundamental Physical Quantities Measurement, School of Physics,
Huazhong University of Science and Technology, Wuhan 430074, China}

\author{Haidong Yuan}
\affiliation{Department of Mechanical and Automation Engineering, The Chinese
University of Hong Kong, Shatin, Hong Kong}

\author{Xiaoguang Wang}
\affiliation{Zhejiang Institute of Modern Physics, Department of Physics,
Zhejiang University, Hangzhou 310027, China}

\author{Rafa\l{} Demkowicz-Dobrza\'{n}ski}
\affiliation{Faculty of Physics, University of Warsaw, 02-093 Warszawa, Poland}

\author{Jing Liu}
\email{liujingphys@hust.edu.cn}
\affiliation{National Precise Gravity Measurement Facility, MOE Key Laboratory
of Fundamental Physical Quantities Measurement, School of Physics,
Huazhong University of Science and Technology, Wuhan 430074, China}

\begin{abstract}
Quantum parameter estimation promises a high-precision measurement in theory; however, how to design
the optimal scheme in a specific scenario, especially under a practical condition, is still a serious
problem that needs to be solved case by case due to the existence of multiple mathematical bounds and
optimization methods. Depending on the scenario considered, different bounds may be more or less
suitable, both in terms of computational complexity and the tightness of the bound itself. At the same
time, the metrological schemes provided by different optimization methods need to be tested against
realization complexity, robustness, etc. Hence, a comprehensive toolkit containing various bounds and
optimization methods is essential for the scheme design in quantum metrology. To fill this vacancy,
here we present a Python-Julia-based open-source toolkit for quantum parameter estimation, which
includes many well-used mathematical bounds and optimization methods. Utilizing this toolkit, all
procedures in the scheme design, such as the optimizations of the probe state, control and measurement,
can be readily and efficiently performed.
\end{abstract}

\maketitle

\section{Introduction}

Quantum metrology is an emerging cross-disciplinary field between precision measurement and quantum technology,
and has now become one of the most promising fields in quantum technology due to the general belief that it
could step into the industrial-grade applications in a short time~\cite{Giovannetti2004,Giovannetti2011,Degen2017,
Braun2018,Pezze2018}. Meanwhile, its development not only benefits the applied technologies like the magnetometry,
thermometry, and gravimetry, but also the studies in fundamental physics such as the detection of gravitational
waves~\cite{LIGO2013} and the search of dark matters~\cite{Backes2021,Jiang2021}. As the theoretical support of
quantum metrology, quantum parameter estimation started from 1960s~\cite{Helstrom1967}, and has become an
indispensable component of quantum metrology nowadays~\cite{Paris2009,Toth2014,Szczykulska2016,Liu2020,Rafal2020,
Sidhu2020,Albarelli2020,Liu2022,Erol2014}.

One of the key challenges in quantum parameter estimation is to design optimal schemes with quantum apparatuses
and quantum resources, leading to enhanced precision when compared with their classical counterparts. A typical
scheme in quantum parameter estimation usually contains four steps: (1) preparation; (2) parameterization; (3)
measurement; and (4) classical estimation. The first step is the preparation of the probe state. The parameters
to be estimated are involved in the second step, which is also known as sensing in the field of quantum sensing.
With the parameterized state given in the second step, the third step is to perform the quantum measurement, which
results in a set of probability distributions. Estimating the unknown parameters from the obtained probability
distributions is finished in the last step. The design of an optimal scheme usually requires the optimizations of
some or all of the steps above.

In quantum parameter estimation, there exist various mathematical bounds to depict the theoretical precision
limit. Depending on the type of the bound considered, it  will be more or less informative depending on the type
of estimation scenario considered, be it: single-shot versus many-repetition scenario, single versus multiple-parameter
scenario, etc. Moreover, by choosing different objective functions when optimizing quantum estimation schemes,
one may arrive at solutions with contrastingly different robustness properties, complexity of practical implementation
and so on. Hence, the design of optimal schemes has to be performed case by case most of the time. This is the reason
why a general quantum parameter estimation toolkit is needed. In the meantime, thanks to the fast development of
quantum metrology and its promising future, many scientists working on specific physical systems, such as the
nitrogen-vacancy centers, quantum circuits, trapped ions, and atoms, are also eager to use the cutting-edge
technologies in quantum parameter estimation for the design of metrological schemes on their platforms. The existence
of a comprehensive toolkit will definitely reduce the technical difficulty for them to fulfill this mission and
greatly improve the efficiency of research. Therefore, developing such a toolkit is the major motivation of this
paper.

Currently, there exist many useful toolkits based on various coding platforms in quantum information. A famous one is
the QuTiP developed by Johansson, Nation, and Nori~\cite{Johansson2012,Johansson2013} in 2012, which can execute many
basic calculations in quantum information. In the field of quantum control, Machnes et al.~\cite{Machnes2011}
developed DYNAMO and Hogben et al. developed Spinach~\cite{Hogben2011} based on Matlab. Goerz et al. developed
Krotov~\cite{Goerz2019}, which owns three versions based on Fortran, Python, and Julia, respectively. G\"{u}nther
et al. developed Quandary~\cite{Gunther2021} based on C++. Moreover, there exist other packages like
Kwant~\cite{Groth2014} for quantum transport and ProjectQ~\cite{Steiger2018} for quantum computing. In quantum
metrology, Chabuda and Demkowicz-Dobrza\'{n}ski developed TNQMetro~\cite{Chabuda2021}, a tensor-network
based Python package to perform efficient quantum metrology computations.

Hereby we present a new numerical toolkit, QuanEstimation, based on both Python and Julia for the quantum parameter
estimation and provide some examples to demonstrate its usage and performance. QuanEstimation is designed to fill
the lack of a general toolkit in quantum parameter estimation, not a general one in quantum information. Hence,
it only focuses on the missions in quantum parameter estimation, and the main features are significantly different
from the existing toolkits in quantum information. Specifically, it contains several widely-used metrological tools,
such as the asymptotic Fisher information based quantities as well as their Bayesian counterparts (including direct
Bayesian cost minimization, Bayesian versions of the classical and quantum Cram\'{e}r-Rao bounds as well as the
quantum Ziv-Zakai bound). For the sake of scheme design, QuanEstimation can execute the optimizations of the probe
state, control, and measurement, as well as the simultaneous optimizations among them with both gradient-based and
gradient-free methods. Due to the fact that most of the time adaptive measurement schemes are the best practical way
to realize the asymptotic advantage indicated by the quantum Fisher information, QuanEstimation can also execute
online adaptive measurement schemes, such as the adaptive phase estimation, and provide the real-time values of the
tunable parameters that can be directly used in an experiment.

\section{Overview}

%====================== Figure ==============================
\begin{figure*}[bt]
\centering\includegraphics[width=17cm]{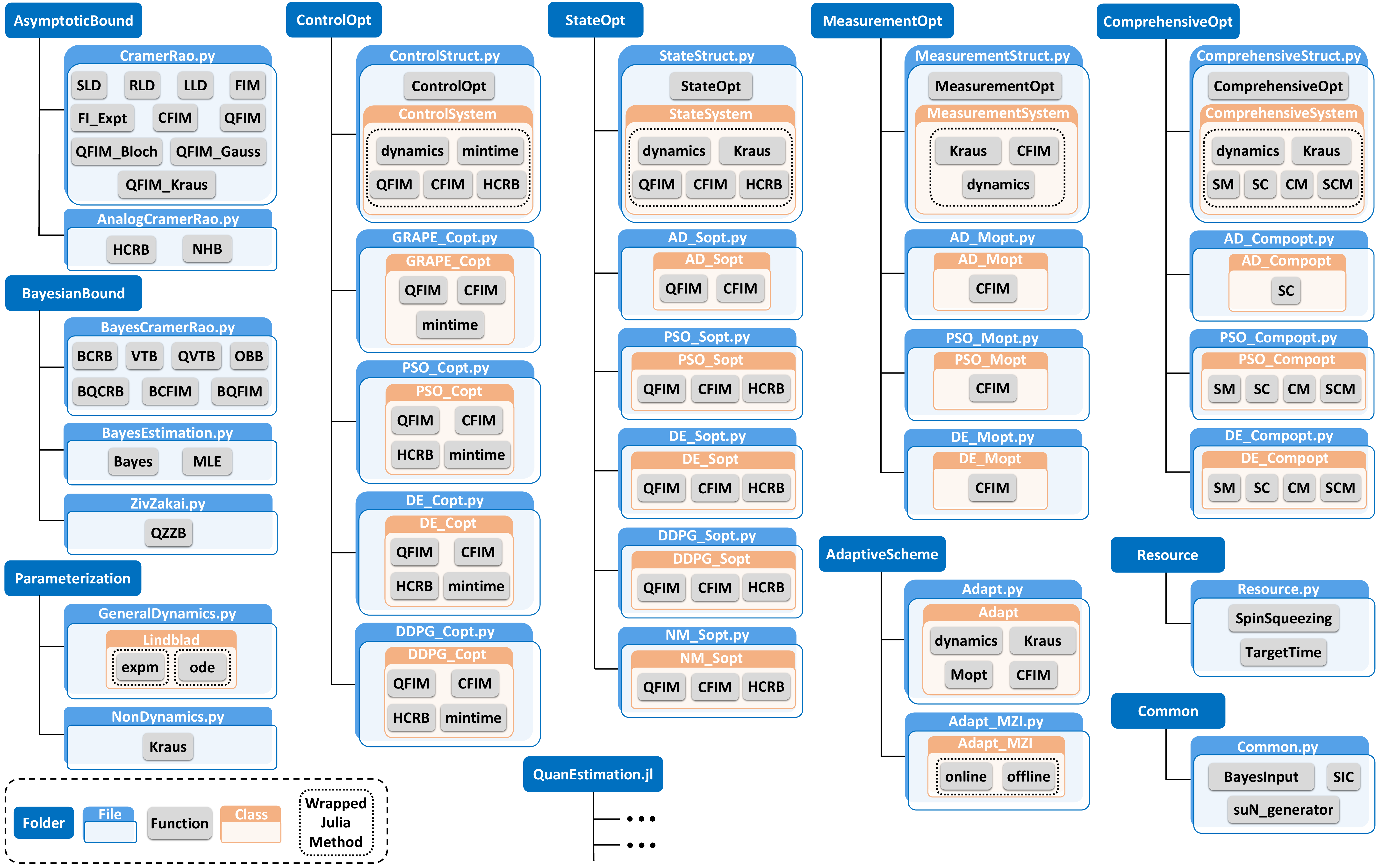}
\caption{Schematic of the package structure of QuanEstimation.
The blue boxes and the light blue boxes represent the folders and
the files. The orange boxes and the gray boxes represent the classes
and the functions/methods. The boxes circled by the dotted lines
represents the wrapped Julia methods, which are solved in Julia scripts.}
\label{fig:package_structure}
\end{figure*}
%===========================================================

QuanEstimation is a scientific computing package focusing on the calculations and optimizations in quantum
parameter estimation. It is based on both Python and Julia. The interface is written in Python due to the fact
that nowadays Python is one of the most popular platforms for scientific computing. However, QuanEstimation
contains many optimization processes which need to execute massive numbers of elementary processes such as the loops.
These elementary processes could be very time-consuming in Python, and thus strongly affect the efficiency of the
optimizations. This is why Julia is involved in this package. Julia has many wonderful features, such as optional
typing and multiple dispatch, and these features let the loop and other calculation processes cost way less time
than those in Python. Hence, the optimizations in QuanEstimation are all performed in Julia. Nevertheless, currently
the community of Julia is not comparable to that of Python, and the hybrid structure of this package would allow
the people who are not familiar with Julia use the package without any obstacle. In the meantime, QuanEstimation
has a full Julia version for the users experienced in Julia.

The package structure of QuanEstimation is illustrated in Fig.~\ref{fig:package_structure}. The blue boxes and the
light blue boxes represent the folders and the files. The orange boxes and the gray boxes represent the classes
and the functions/methods. The boxes circled by the dotted lines represents the wrapped Julia methods, which are
solved in Julia, namely, this part of calculation are sent to Julia to execute.

The functions for the calculation of the parameterization process and dynamics are in the folder named
"Parameterization". In this folder, the file "GeneralDynamics.py" contains the functions to solve the Lindblad-type
master equation. Currently, the master equation can be solved directly, i.e., solving the corresponding ordinary
differential equation, or via the matrix exponential. To improve the efficiency, the calculation of the dynamics
via the matrix exponential are executed in Julia and when the calculation is finished, the data is sent back to
Python for  further use. The file "NonDynamics.py" contains the non-dynamical methods for the parameterization,
which currently includes the description via Kraus operators. Details and the usage of these functions will be
thoroughly introduced in Sec.~\ref{sec:para}.

The functions for the calculation of the metrological tools and bounds are distributed in two folders named
"AsymptoticBound" and "BayesianBound". In the folder "AsymptoticBound", the file "CramerRao.py" contains the
functions to calculate the quantities related to the quantum Cram\'{e}r-Rao bounds, and the file "AnalogCramerRao.py"
contains those to calculate the Holevo-type quantum Cram\'{e}r-Rao bound and Nagaoka-Hayashi bound. In the folder
"BayesianBound", the file "BayesCramerRao.py" contains the functions to calculate several versions of the Bayesian
classical and quantum Cram\'{e}r-Rao bounds and "ZivZakai.py" contains the function to calculate the quantum
Ziv-Zakai bound. The file "BayesEstimation.py" contains the functions to execute the Bayesian estimation and the
maximum likelihood estimation. The aforementioned metrological tools and the corresponding rules to call them will
be given in Sec.~\ref{sec:tools}.

The functions for the calculation of metrological resources are placed in the folder named "Resource". In this folder,
the file "Resource.py" currently contains two types of resources, the spin squeezing and the target time to reach a
given value of an objective function, which will be thoroughly introduced in Sec.~\ref{sec:resource}. The resources
that can be readily calculated via QuTiP~\cite{Johansson2012,Johansson2013} are not included at this moment.

The scripts for the control optimization, state optimization, measurement optimization, and comprehensive
optimization are in the folders named "ControlOpt", "StateOpt", "MeasurementOpt", and "ComprehensiveOpt",
respectively. The structures of these folders are basically the same, and here we only take the folder of "ControlOpt"
as an demonstration to explain the basic structure. In this folder, the file "ControlStruct.py" contains a
function named {\codefont ControlOpt()} and a class named {\codefont ControlSystem()}. The function
{\codefont ControlOpt()} is used to receive the initialized parameters given by the user, and then delivers them
to one of the classes in the files "GRAPE\_Copt.py", "PSO\_Copt.py", "DE\_Copt.py", and "DDPG\_Copt.py" according to
the user's choice of the algorithm. These classes inherit the attributes in {\codefont ControlSystem()}. Then based on
the choice of the objective function, the related parts in {\codefont ControlSystem()} is called in these classes to
further run the scripts in Julia. {\codefont ControlSystem()} contains all the common parts that different algorithms
would use and the interface with the scripts in Julia. This design is to avoid the repetition code in the algorithm
files and let the extension neat and simple when more algorithms need to be included in the future. The usage of
QuanEstimation for control optimization, state optimization, measurement optimization, and comprehensive optimization,
as well as the corresponding illustrations will be thoroughly discussed in Secs.~\ref{sec:control_opt}, \ref{sec:state_opt},
\ref{sec:measurement_opt}, and~\ref{sec:comprehensive_opt}, respectively.

The scripts for the adaptive measurement are in the folder named "AdaptiveScheme". In this folder, the file "Adapt.py"
contains the class to execute the adaptive measurement scheme, and "Adapt\_MZI.py" contains the class to generate
online and offline adaptive schemes in the Mach-Zehnder interferometer. The details of the adaptive scheme and how to
perform it with QuanEstimation will be given in Sec.~\ref{sec:adapt}.

The folder "Common" contains some common functions that are regularly called in QuanEstimation. Currently it
contains three functions. {\codefont SIC()} is used to generate a set of rank-one symmetric informationally complete
positive operator-valued measure. {\codefont suN\_generator()} is used to generate a set of su($N$) generators.
{\codefont BayesInput()} is used to generate a legitimate form of Hamiltonian (or a set of Kraus operators) and its
derivative, which can be used as the input in some functions in "BayesEstimation.py" and "Adapt.py".

All the Julia scripts are wrapped up as an independent Julia package named "QuanEstimation.jl", which has already been
added in the official Julia registry and can be directly called in Julia via {\codefont using QuanEstimation}. One
design principle of QuanEstimation for the optimizations is that once the calculation goes into the parts in Julia, it
will stay in Julia until all the calculations are finished and data generated. Hence, "QuanEstimation.jl" also contains
the scripts to calculate the metrological tools and resources for the sake of internal calling in Julia. To keep a high
extendability, the optimizations are divided into four elements in Julia, including the scenario of optimization, the
algorithm, the parameterization process and the objective function, which are distributed in the files "OptScenario.jl",
"Algorithm.jl", "Parameterization.jl", and "ObjectiveFunc.jl" in the folders "OptScenario", "Algorithm", "Parameterization",
and "ObjectiveFunc", respectively. Once the information and parameter settings of all elements are input by the user,
they are sent to the file "run.jl", which is further used to execute the program.

Similar to other packages, the usage of QuanEstimation requires the existence of some other packages in the environment.
In python it requires the pre-installation of numpy, scipy, sympy, cvxpy, and more-itertools. In Julia it requires
the pre-installation of LinearAlgebra, Zygote, Convex, SCS, ReinforcementLearning, SparseArrays, DelimitedFiles,
StatsBase, BoundaryValueDiffEq, Random, Trapz, Interpolations, Printf, IntervalSets, StableRNGs, Flux, Distributions,
DifferentialEquations, and QuadGK. The calling of the package in Python can be done with the following line of code:
\begin{lstlisting}[breaklines=true,numbers=none,frame=trBL]
from quanestimation import *
\end{lstlisting}
All the scripts demonstrated in the following are based on this calling form.

\section{Parameterization process}
\label{sec:para}

The parameterization process is a key step in the quantum parameter estimation, and in physical terms this process
corresponds to a parameter dependent quantum dynamics. Hence, the ability to solve the dynamics is an indispensable
element of numerical calculations in quantum parameter estimation. In QuanEstimation, we mainly focus on the dynamics
governed by the quantum master equation
\begin{align}
\partial_t\rho &=\mathcal{L}\rho \nonumber \\
&=-i[H,\rho]+\sum_i \gamma_i\left(\Gamma_i\rho\Gamma^{\dagger}_i
-\frac{1}{2}\left\{\rho,\Gamma^{\dagger}_i \Gamma_i \right\}\right) \label{eq:mastereq},
\end{align}
where $\rho$ is the evolved density matrix, $H$ is the Hamiltonian of the system, and $\Gamma_i$ and $\gamma_i$
are the $i$th decay operator and decay rate, respectively. Here $\gamma_i$ could either be fixed or time-dependent.
The total Hamiltonian $H$ includes two terms, the free Hamiltonian $H_0(\bold{x})$, which is a function of the
parameters $\bold{x}$, and control Hamiltonian $H_{\mathrm{c}}$. In the quantum parameter estimation, most
calculations require the dynamical information of $\rho$ and its derivatives with respect to $\bold{x}$, which
is denoted by $\partial_{\bold{x}}\rho:=(\partial_{0}\rho,\partial_1\rho,\dots)$ with $\partial_a$ short for
$\partial_{x_a}$. Hence, in the package $\rho$ and $\partial_{\bold{x}}\rho$ can be solved simultaneously via
the code:
\begin{lstlisting}[breaklines=true,numbers=none,frame=trBL]
dynamics = Lindblad(tspan,rho0,H0,dH,
              decay=[],Hc=[],ctrl=[])
rho,drho = dynamics.expm()
\end{lstlisting}
Here the input {\codefont tspan} is an array representing the time length for the evolution and {\codefont rho0}
is a matrix representing the initial (probe) state. {\codefont H0} is a matrix or a list of matrices representing the free
Hamiltonian. It is a matrix when the free Hamiltonian is time-independent and a list (the length equals to
that of {\codefont tspan}) when it is time-dependent. {\codefont dH} is a list containing the derivatives of
$H_0(\bold{x})$ on $\bold{x}$, i.e., $[\partial_a H_0,\partial_b H_0,\dots]$. {\codefont decay} is a list including
both decay operators and decay rates, and its input rule is {\codefont decay=[[Gamma1,gamma1],[Gamma2,gamma2],\dots]},
where {\codefont Gamma1} ({\codefont Gamma2}) and {\codefont gamma1} ({\codefont gamma2}) represent $\Gamma_1$
($\Gamma_2$) and $\gamma_1$ ($\gamma_2$), respectively. {\codefont gamma1} ({\codefont gamma2}) could be either a
float number (representing a fixed decay rate) or an array (representing a time-dependent decay rate), and when it is
an array, its length should be the same with {\codefont tspan}. Currently all the length of the decay rates should be
the same, i.e., all be float numbers or arrays with the same length. The default value is empty, which means the
dynamics is unitary. {\codefont Hc} is a list of matrices representing the control Hamiltonians and when it is empty,
the dynamics is only governed by the free Hamiltonian. {\codefont ctrl} (default value is empty) is a list of arrays
containing the control amplitudes with respect the control Hamiltonians in {\codefont Hc}. The output {\codefont rho}
is a list representing density matrices in the dynamics. {\codefont drho} is also a list and its $i$th entry is
a list containing all derivatives $\partial_{\bold{x}}\rho$ at $i$th time interval. Moreover, {\codefont dynamics.expm()}
in this demonstrating code means the dynamics is solved by the matrix exponential, i.e., the density matrix
at $j$th time interval is calculated via $\rho_j=e^{\Delta t_j\mathcal{L}}\rho_{j-1}$ with $\Delta t_j$ a small time
interval and $\rho_{j-1}$ the density matrix at the previous time interval. $\partial_{\bold{x}}\rho_j$ is solved
by the iterative equation
\begin{align}
\partial_{\bold{x}}\rho_j &=\Delta t_j(\partial_{\bold{x}}\mathcal{L})\rho_j
+e^{\Delta t_j \mathcal{L}}(\partial_{\bold{x}}\rho_{j-1}) \nonumber \\
&=-i\Delta t_j[\partial_{\bold{x}}H_0, \rho_j]+e^{\Delta t_j \mathcal{L}}(\partial_{\bold{x}}\rho_{j-1}).
\end{align}
Here the decay operators and decay rates are assumed to be independent of $\bold{x}$. In this method $\Delta t_j$ is
automatically obtained by calculating the difference between the $j$th and $(j-1)$th entries in {\codefont tspan}.
The numerical accuracy of the equation above is limited by the set of $\{\Delta t_j\}$, indicating that a smaller
$\{\Delta t_j\}$ would always benefit the improvement of the accuracy in general. However, a smaller $\{\Delta t_j\}$
also means a larger number of calculation steps for a fixed evolution time, resulting in a greater time consumption.
Hence, in practice a reasonable values of $\{\Delta t_j\}$ should be chosen to balance the accuracy and time consumption.

Alternatively, the dynamics can also be solved by directly solving the ordinary differential equation (ODE) in
Eq.~(\ref{eq:mastereq}), which can be realized by replacing {\codefont dynamics.expm()} with {\codefont dynamics.ode()}
in the demonstrating code. In this method, $\partial_{\bold{x}}\rho$ is solved by the equation
\begin{equation}
\partial_t(\partial_{\bold{x}}\rho)=-i\left[\partial_{\bold{x}}H,\rho\right]
+\mathcal{L}\left(\partial_{\bold{x}}\rho\right).
\end{equation}

The calculation of metrological bounds, which will be discussed in the next section, does not rely on the calling
of above intrinsic dynamics in the package as they only require the input of $\rho$ and $\partial_{\bold{x}}\rho$
(and other essential parameters), not any dynamical information. Hence, the dynamics can also be solved by other
packages like QuTiP~\cite{Johansson2012,Johansson2013}.

In certain cases, the parameterization process can be described by some non-dynamical methods, such as the Kraus
operators. In this case, the parameterized density matrix can be expressed by
\begin{equation}
\rho(\bold{x})=\sum_i K_i(\bold{x})\rho_0 K_i^{\dagger}(\bold{x}),
\label{eq:kraus_opt}
\end{equation}
where $K_i(\bold{x})$ is a Kraus operator satisfying $\sum_{i}K^{\dagger}_i K_i=\openone$ with $\openone$ the
identity operator, $\rho_0$ is the probe state which is independent of the unknown parameters. In QuanEstimation,
$\rho$ and $\partial_{\bold{x}}\rho$ obtained from Kraus operators can be solved via the code:
\begin{lstlisting}[breaklines=true,numbers=none,frame=trBL]
rho,drho = Kraus(rho0,K,dK)
\end{lstlisting}
Here {\codefont rho0} is a matrix representing the probe state, {\codefont K} is a list of matrices with each
entry a Kraus operator, and {\codefont dK} is a list with $i$th entry also a list representing the derivatives
$\partial_{\bold{x}}K_i$.

The aforementioned functions only calculate $\rho$ and $\partial_{\bold{x}}\rho$ at a fixed point of $\bold{x}$.
However, in the Bayesian scenarios, the values of $\rho$ and $\partial_{\bold{x}}\rho$ with respect to a regime
of $\bold{x}$ may be in need. In this case, if the users can provide the specific functions of $H$ and
$\partial_{\bold{x}}H$, or Kraus operators $\{K_i\}$ and derivatives $\{\partial_{\bold{x}} K_i\}$, the variables
{\codefont H}, {\codefont dH} (or {\codefont K}, {\codefont dK}) can be generated by the function
\begin{lstlisting}[breaklines=true,numbers=none,frame=trBL,mathescape=true]
H0,dH = BayesInput(x,func,dfunc,
                   channel="dynamics")
\end{lstlisting}
Here {\codefont x} is a list of arrays representing the regime of $\bold{x}$. {\codefont H0} is a list of matrices
representing the free Hamiltonian with respect to the values in {\codefont x}, and it is multidimensional in the
case that {\codefont x} has more than one entry. {\codefont dH} is a (multidimensional) list with each entry also
a list representing $\partial_{\bold{x}}H$ with respect to the values in {\codefont x}. {\codefont func} and
{\codefont dfunc} are the handles of the functions {\codefont func()} and {\codefont dfunc()}, which are defined
by the users representing $H(\bold{x})$ and $\partial_{\bold{x}}H(\bold{x})$. Notice that the output of
{\codefont dfunc()} should also be a list representing $[\partial_0 H,\partial_1 H,\dots]$. The output of
{\codefont BayesInput()} can be switched between {\codefont H}, {\codefont dH} and {\codefont K}, {\codefont dK}
by setting {\codefont channel="dynamics"} or {\codefont channel="Kraus"}. After calling {\codefont BayesInput()},
$\rho$ and $\partial_{\bold{x}}\rho$ can be further obtained via the calling of {\codefont Lindblad()} and
{\codefont Kraus()}.

\section{Quantum metrological tools}
\label{sec:tools}

In this section, we will briefly introduce the metrological tools that have been involved in QuanEstimation and
demonstrate how to calculate them with our package. Both asymptotic and Bayesian tools are included, such as the
quantum Cram\'{e}r-Rao bounds, Holevo Cram\'{e}r-Rao bound, Nagaoka-Hayashi bound, Bayesian estimation, and Bayesian
type of Cram\'{e}r-Rao bounds like Van Trees bound and Tsang-Wiseman-Caves bound.

\subsection{Quantum Cram\'{e}r-Rao bounds}
\label{sec:QCRB}

Quantum Cram\'{e}r-Rao bounds~\cite{Helstrom1976,Holevo1982} are the most renown metrological tools in quantum
parameter estimation. Let $\rho=\rho(\bold{x})$ be a parameterized density matrix and
$\{\Pi_y\}$ a set of positive operator-valued measure (POVM), then the covariance matrix
$\mathrm{cov}(\hat{\bold{x}},\{\Pi_y\}):=\sum_y\mathrm{Tr}(\rho\Pi_y)(\hat{\bold{x}}-\bold{x})(\hat{\bold{x}}
-\bold{x})^{\mathrm{T}}$ for the unknown parameters $\bold{x}=(x_0,x_1,\dots)^{\mathrm{T}}$ and the corresponding
unbiased estimators $\hat{\bold{x}}=(\hat{x}_0,\hat{x}_1,\dots)^{\mathrm{T}}$ satisfies the following
inequalities~\cite{Helstrom1976,Holevo1982}
\begin{equation}
\mathrm{cov}\left(\hat{\bold{x}}, \{\Pi_y\}\right)
\geq \frac{1}{n}\mathcal{I}^{-1}\left(\{\Pi_y\}\right)
\geq \frac{1}{n} \mathcal{F}^{-1},
\end{equation}
where $n$ is the repetition of the experiment, $\mathcal{I}$ is the classical Fisher information matrix (CFIM) and
$\mathcal{F}$ is the quantum Fisher information matrix (QFIM). Note that the estimators $\hat{\bold{x}}$ are in
fact functions of the measurement outcomes $y$, and formally should always be written as $\hat{{\bold{x}}}(y)$.
Still, we drop this explicit dependence on $y$ for conciseness of formulas. A thorough derivation
of this bound can be found in a recent review~\cite{Liu2020}.

For a set of discrete probability
distribution $\{p(y|\bold{x})=\mathrm{Tr}(\rho\Pi_y)\}$, the CFIM is defined by
\begin{equation}
\mathcal{I}_{ab}=\sum_{y}\frac{1}{p(y|\bold{x})}[\partial_a p(y|\bold{x})][\partial_b p(y|\bold{x})].
\label{eq:CFIM}
\end{equation}
Here $\mathcal{I}_{ab}$ is short for $\mathcal{I}_{x_a,x_b}$, the $ab$th entry of the CFIM. For a
continuous probability density, the equation above becomes $\mathcal{I}_{ab}=\int \frac{1}{p(y|\bold{x})}
[\partial_a p(y|\bold{x})][\partial_b p(y|\bold{x})]\mathrm{d}y$. The diagonal entry $\mathcal{I}_{aa}$
is the classical Fisher information (CFI) for $x_a$.

The QFIM does not depend on the actual measurement performed, and one can encounter a few equivalent definitions
of this quantity. The one the most often used reads:
\begin{equation}
\mathcal{F}_{ab}=\frac{1}{2}\mathrm{Tr}(\rho\{L_a, L_b\})
\end{equation}
with $\mathcal{F}_{ab}$ being the $ab$th entry of $\mathcal{F}$ and $L_{a(b)}$ the symmetric logarithmic
derivative (SLD) operator for $x_{a(b)}$. $\{\cdot,\cdot\}$ represents the anti-commutator. The
SLD operator is Hermitian and determined by the equation
\begin{equation}
\partial_{a}\rho=\frac{1}{2}(\rho L_{a}+L_{a}\rho).
\end{equation}
The mathematical properties of the SLD operator and QFIM can be found in a recent review~\cite{Liu2020}.
The diagonal entry of $\mathcal{F}_{aa}$ is the quantum Fisher information (QFI) for $x_a$.
Utilizing the spectral decomposition $\rho=\sum_{i}\lambda_i |\lambda_i\rangle\langle \lambda_i|$, the
SLD operator can be calculated via the equation
\begin{equation}
\langle\lambda_i|L_{a}|\lambda_j\rangle=\frac{2\langle\lambda_i| \partial_{a}\rho |\lambda_j\rangle}
{\lambda_i+\lambda_j},  \label{eq:SLD_eigen}
\end{equation}
for $\lambda_i$ or $\lambda_j$ not equal to zero. For $\lambda_i=\lambda_j=0$, the corresponding matrix entry of
$L_a$ can be set to zero.

In QuanEstimation, the SLD operator can be calculated via the function:
\begin{lstlisting}[breaklines=true,numbers=none,frame=trBL]
SLD(rho,drho,rep="original",eps=1e-8)
\end{lstlisting}
Here the input {\codefont rho} is a matrix representing the parameterized density matrix, and {\codefont drho}
is a list of matrices representing the derivatives of the density matrix on $\bold{x}$, i.e.,
$[\partial_0\rho,\partial_1\rho,\dots]$. When {\codefont drho} only contains one entry ($[\partial_0 \rho]$),
the output of {\codefont SLD()} is a matrix ($L_0$), and it is a list ($[L_0,L_1,\dots]$) otherwise. The basis
of the output SLD can be adjusted via the variable {\codefont rep}. The default choice {\codefont rep="original"}
means the basis is the same with that of the input density matrix. The other choice is {\codefont rep="eigen"},
which means the SLD is written in the eigenspace of the density matrix. Due to the fact that the entries of SLD
in the kernel are arbitrary, in the package they are just set to be zeros for simplicity. The default machine
epsilon is {\codefont eps=1e-8}, which can be modified as required. Here the machine epsilon means that if a
eigenvalue of the density matrix is less than the given number ($10^{-8}$ by default), it will be treated as
zero in the calculation of SLD.

Apart from the SLD operator, the QFIM can also be defined via other types of logarithmic derivatives.
Some well-used ones are the right and left logarithmic derivatives (RLD, LLD)~\cite{Holevo1982,Yuen1973}. The RLD
and LLD are determined by $\partial_{a}\rho=\rho \mathcal{R}_a$ and $\partial_{a}\rho=\mathcal{R}_a^{\dagger}\rho$,
respectively. Utilizing the spectral decomposition, the entries of RLD and LLD can be calculated as
\begin{align}
\langle\lambda_i| \mathcal{R}_{a} |\lambda_j\rangle
&= \frac{1}{\lambda_i}\langle\lambda_i| \partial_{a}\rho |\lambda_j\rangle,~~\lambda_i\neq 0; \\
\langle\lambda_i| \mathcal{R}_{a}^{\dagger} |\lambda_j\rangle
&= \frac{1}{\lambda_j}\langle\lambda_i| \partial_{a}\rho |\lambda_j\rangle,~~\lambda_j\neq 0.
\end{align}
The corresponding QFIM is $\mathcal{F}_{ab}=\mathrm{Tr}(\rho \mathcal{R}_a \mathcal{R}^{\dagger}_b)$. In QuanEstimation,
the LLD and RLD can be calculated via the functions {\codefont RLD()} and {\codefont LLD()}. The inputs are the same
with {\codefont SLD()}. Notice that the RLD and LLD only exist when the support of $\rho$ contains the the support of
$\partial_a\rho$. Hence, if this condition is not satisfied, the calculation will be terminated and a line of
reminder will arise to remind that {\codefont RLD()} and {\codefont LLD()} do not exist in this case.

In QuanEstimation, the QFIM and QFI can be calculated via the function:
\begin{lstlisting}[breaklines=true,numbers=none,frame=trBL]
QFIM(rho,drho,LDtype="SLD",exportLD=False,
     eps=1e-8)
\end{lstlisting}
Here {\codefont LDtype=" "} is the type of logarithmic derivatives, including {\codefont "SLD"}, {\codefont "RLD"},
and {\codefont "LLD"}. Notice that the values of QFIM based on RLD and LLD are actually the same when
the RLD and LLD exist. If {\codefont exportLD=True}, apart from the QFIM, the corresponding values of logarithmic
derivatives in the original basis will also be exported.

In the case that the parameterization is described via the Kraus operators, the QFIM can be calculated via the function:
\begin{lstlisting}[breaklines=true,numbers=none,frame=trBL]
QFIM_Kraus(rho0,K,dK,LDtype="SLD",
           exportLD=False,eps=1e-8)
\end{lstlisting}
The input {\codefont rho0} is a matrix representing the density matrix of the initial state. {\codefont K} is a
list of matrices with each entry a Kraus operator, and {\codefont dK} is a list with $i$th entry being also a list
representing the derivatives $\partial_{\bold{x}}K_i$.

The CFIM and CFI for a fully classical scenario can be calculated by the function
\begin{lstlisting}[breaklines=true,numbers=none,frame=trBL]
FIM(p,dp,eps=1e-8)
\end{lstlisting}
The input {\codefont p} is an array representing the probability distribution and {\codefont dp} is a list
with the $i$th entry being itself also a list containing the derivatives of $p_i$ on $\bold{x}$, i.e.,
$[\partial_0 p_i,\partial_1 p_i,\dots]$. In the realistic experiments, the derivatives of the conditional probability
are difficult to obtain, and therefore the CFI cannot be measured directly. However, it is possible that a known small
drift $\delta\bold{x}$ can be experimentally invoked into the system and let the true value of $\bold{x}$ moves slightly.
In such cases, the CFI can be calculated once the distributions $p(y|\bold{x})$ and $p(y|\bold{x}+\delta\bold{x})$ are
acquired, which are usually obtained via the distribution fitting. In the case of single-parameter estimation, the CFI
can be expressed by $\mathcal{I}=8\big[1-\sum_y\sqrt{p(y|x)p(y|x+\delta x)}\big]/\delta^2 x$ due to the relation
between the CFI and the classical fidelity $\sum_y \sqrt{p(y|\bold{x})q(y|\bold{x})}$. In QuanEstimation, if the users
have two sets of data (results of $y$) with respect to the value $x$ and $x+\delta x$ in experiments, then the CFI can
be calculated via the following function:
\begin{lstlisting}[breaklines=true,numbers=none,frame=trBL]
FI_Expt(y1,y2,dx,ftype="norm")
\end{lstlisting}
Here {\codefont y1} and {\codefont y2} are two arrays containing the data of $y$ in experiments with respect to the
values $x$ and $x+\delta x$, and {\codefont dx} represents the value of $\delta x$. Currently, four types of distributions
are available for distribution fitting, including the normal (Gaussian) distribution ({\codefont ftype="norm"}), gamma
distribution ({\codefont ftype="gamma"}), rayleigh distribution ({\codefont ftype="rayleigh"}), and poisson distribution
({\codefont ftype="poisson"}).

In a quantum scenario, the CFIM can be calculated by
\begin{lstlisting}[breaklines=true,numbers=none,frame=trBL]
CFIM(rho,drho,M=[],eps=1e-8)
\end{lstlisting}
The variable {\codefont M} is a list containing a set of POVM. The default measurement is a set of rank-one symmetric
informationally complete POVM (SIC-POVM)~\cite{Gour2014,Fuchs2017,Renes2004}. A set of rank-one SIC-POVM
$\{\frac{1}{d}|\phi_j\rangle\langle\phi_j|\}^{d^2}_{j=1}$ satisfies $|\langle\phi_j|\phi_k\rangle|^2=(d\delta_{jk}+1)/(d+1)$
for any $j$ and $k$ with $|\phi_j\rangle$ being a normalized quantum state and $d$ the dimension of the Hilbert space.
One way to construct a set of SIC-POVM is utilizing the Weyl-Heisenberg operators~\cite{Renes2004,Scott2010}, which is
defined by $D_{ab}=(-e^{i\pi/d})^{ab}A^{a}B^{b}$. The operators $A$ and $B$ satisfy $A|k\rangle=|k+1\rangle$,
$B|k\rangle=e^{i2\pi k/d}|k\rangle$ with $\{|k\rangle\}^{d-1}_{k=0}$ an orthonormal basis in the Hilbert space.
There exists a normalized fiducial vector $|\psi\rangle$ in the Hilbert space such that $\{\frac{1}{d}D_{ab}
|\psi\rangle\langle\psi|D^{\dagger}_{ab}\}^d_{a,b=1}$ is a set of SIC-POVM. In the package, $|\psi\rangle$ is
taken as the one numerically found by Fuchs et al. in Ref.~\cite{Fuchs2017}. If the users want to see the
specific formula of the SIC-POVM, the function {\codefont SIC(n)} can be called. The input {\codefont n} is
the dimension of the density matrix. Currently, the function {\codefont SIC(n)} only valid when $n\leq 151$.

%================================ Figure ============================================
\begin{figure}[tp]
\centering\includegraphics[width=8cm]{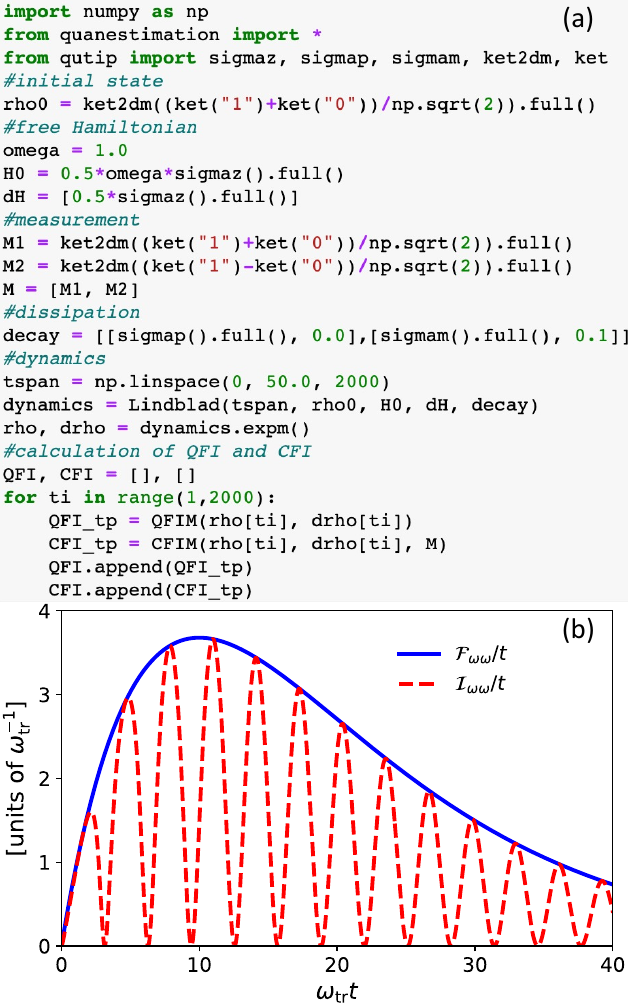}
\caption{(a) The demonstrating code for the calculation of QFI and CFI with
QuanEstimation. (b) The evolution of $\mathcal{F}_{\omega\omega}/t$ (solid blue
line) and $\mathcal{I}_{\omega\omega}/t$ (dashed red line). The initial state is
$|+\rangle$. The true value of $\omega$ ($\omega_{\mathrm{tr}}$) is set to be $1$,
and the decay rates are set to be $\gamma_{+}/\omega_{\mathrm{tr}}=0$ and
$\gamma_{-}/\omega_{\mathrm{tr}}=0.1$. Planck units are applied here.
\label{fig:QFI_code}}
\end{figure}
%====================================================================================

In both functions {\codefont QFIM()} and {\codefont CFIM()}, the outputs are real numbers ($\mathcal{F}_{aa}$
and $\mathcal{I}_{aa}$) in the single-parameter case, namely, when {\codefont drho} only contains one entry, and
they are real symmetric or Hermitian matrices in the multi-parameter scenarios. The basis of QFIM and CFIM are
determined by the order of entries in {\codefont drho}. For example, when {\codefont drho} is
$[\partial_0\rho,\partial_1\rho,\dots]$, the basis of the QFIM and CFIM is $\{x_0,x_1,\dots\}$.

For some specific scenarios, the calculation method in {\codefont QFIM()} may be not efficient enough. Therefore,
we also provide the calculation of QFIM in some specific scenarios. The first one is the calculation in the
Bloch representation. In this case, the function for the calculation of QFIM is of the form
\begin{lstlisting}[breaklines=true,numbers=none,frame=trBL]
QFIM_Bloch(r,dr,eps=1e-8)
\end{lstlisting}
The input {\codefont r} is an array representing a Bloch vector and {\codefont dr} is a list of arrays representing
the derivatives of the Bloch vector on $\bold{x}$. Gaussian states are very commonly used in quantum metrology, and the
corresponding QFIM can be calculated by the function:
\begin{lstlisting}[breaklines=true,numbers=none,frame=trBL]
QFIM_Gauss(R,dR,D,dD)
\end{lstlisting}
The input {\codefont R} is an array representing the first-order moment, i.e., the expected value
$\langle\bold{R}\rangle:=\mathrm{Tr}(\rho\bold{R})$ of the vector $\bold{R}=(q_1,p_1,q_2,p_2,\dots)^{\mathrm{T}}$,
where $q_i=(a_i+a^{\dagger}_i)/\sqrt{2}$ and $p_i=(a_i-a^{\dagger}_i)/(i\sqrt{2})$ are the quadrature operators
with $a_i$ ($a^{\dagger}_i$) the annihilation (creation) operator of $i$th bosonic mode. {\codefont dR} is a list
with $i$th entry also a list containing the derivatives $\partial_{\bold{x}}\langle[\bold{R}]_i\rangle$. Here
$[\cdot]_i$ represents the $i$th entry of the vector. {\codefont D} is a matrix representing the second-order
moment, $D_{ij}=\langle [\bold{R}]_i [\bold{R}]_j+[\bold{R}]_j[\bold{R}]_i\rangle/2$, and {\codefont dD} is a list
of matrices representing the derivatives $\partial_{\bold{x}}D$. Notice that {\codefont QFIM\_Bloch()} and
{\codefont QFIM\_Gauss()} can only compute the SLD-based QFIM.

\emph{Example.} Now we present an example to show the usage of these functions. Consider a single qubit
Hamiltonian $H=\omega\sigma_3/2$ with $\sigma_3$ a Pauli matrix and $\omega$ the frequency. Take $\omega$
as the parameter to be estimated and assume its true value (denoted by $\omega_{\mathrm{tr}}$) is 1.
Planck unit ($\hbar=1$) is applied in the Hamiltonian. The dynamics is governed by the master equation
\begin{eqnarray}
\partial_t\rho&=&-i\left[H, \rho\right]+\gamma_{+}\left(\sigma_+\rho\sigma_{-}-\frac{1}{2}
\left\{\sigma_{-}\sigma_{+}, \rho\right\}\right) \nonumber\\
& &+\gamma_{-}\left(\sigma_-\rho\sigma_{+}-\frac{1}{2}\left\{\sigma_{+}\sigma_{-}, \rho\right\}\right),
\label{eq:ME_spon}
\end{eqnarray}
where $\sigma_{\pm}=(\sigma_1\pm\sigma_2)/2$ with $\sigma_{1}$, $\sigma_{2}$ also Pauli matrices.
$\gamma_{+}$ and $\gamma_{-}$ are the decay rates. The measurement is taken as
$\{|+\rangle\langle+|,|-\rangle\langle-|\}$ with
\begin{equation}
|\pm\rangle:=\frac{1}{\sqrt{2}}(|0\rangle\pm|1\rangle).
\end{equation}
Here $|0\rangle$ ($|1\rangle$) is the eigenstate of $\sigma_3$ with respect to the eigenvalue $1$ ($-1$). The
specific code for the calculation of QFI/CFI are given in Fig.~\ref{fig:QFI_code}(a), and the corresponding
evolution of $\mathcal{F}_{\omega\omega}/t$ (solid blue line) and $\mathcal{I}_{\omega\omega}/t$ (dashed red line)
are shown in Fig.~\ref{fig:QFI_code}(b). The operators such as the density matrix and measurement can either be
generated via QuTiP as in the demonstrating code or direct input.

\subsection{Holevo Cram\'{e}r-Rao bound}
\label{sec:HCRB}

%================================ Figure ============================================
\begin{figure}[tp]
\centering\includegraphics[width=8.0cm]{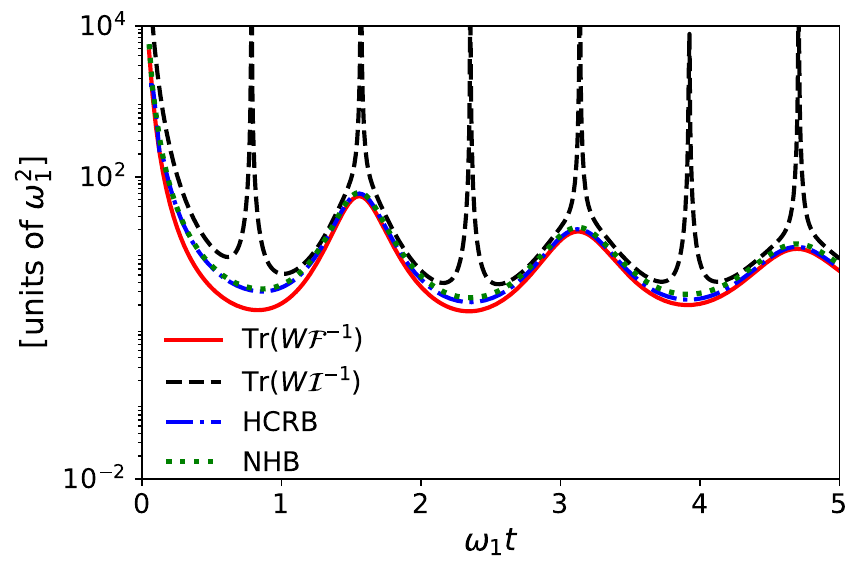}
\caption{Time evolution of $\mathrm{Tr}(W\mathcal{F}^{-1})$ (solid red line),
$\mathrm{Tr}(W\mathcal{I}^{-1})$ (dashed black line), HCRB (dash-dotted blue
line), and NHB (dotted green line) in the case of two-qubit system with the XX
coupling. The probe state is $(|00\rangle+|11\rangle)/\sqrt{2}$. $W=\openone$
and $\omega_1=1$. The true values of $\omega_2$ and $g$ are $1$ and $0.1$,
respectively. The decay rates $\gamma_1=\gamma_2=0.05\omega_1$. The POVM for
$\mathrm{Tr}(W\mathcal{I}^{-1})$ is $\{\Pi_1$, $\Pi_2$, $\openone-\Pi_1-\Pi_2\}$
with $\Pi_1=0.85|00\rangle\langle 00|$ and $\Pi_2=0.4|\!+\!+\rangle\langle+\!+\!|$.
Planck units are applied here. }
\label{fig:HCRB}
\end{figure}
%====================================================================================

Holevo Cram\'{e}r-Rao bound (HCRB) is another useful asymptotic bound in quantum parameter
estimation and tighter than the quantum Cram\'{e}r-Rao bound in general. The HCRB can be expressed
as~\cite{Holevo1973,Rafal2020,Nagaoka1989,Hayashi2008}
\begin{equation}
\mathrm{Tr}(W\mathrm{cov}(\hat{\bold{x}},\{\Pi_y\}))\geq \min_{\bold{X},V} \mathrm{Tr}(WV)
\end{equation}
with $W$ the weight matrix and $V$ a matrix satisfying $V\geq Z(\bold{X})$. Here $Z(\bold{X})$ is a
Hermitian matrix and its $ab$th entry is defined by $[Z(\bold{X})]_{ab}:=\mathrm{Tr}(\rho X_a X_b)$,
where $\bold{X}=[X_0,X_1,\cdots]$ is a vector of operators and its $a$th entry is defined by
$X_a:=\sum_y (\hat{x}_a-x_a)\Pi_y$ with $\hat{x}_a$ the $a$th entry of $\hat{\bold{x}}$.
To let the local estimator $\hat{\bold{x}}$ unbiased, $\bold{X}$ needs to satisfy
$\mathrm{Tr}(X_a\partial_b\rho)=\delta_{ab},\,\forall a, b$. Here $\delta_{ab}$ is the Kronecker
delta function. An equivalent formulation of HCRB is~\cite{Holevo1973,Rafal2020,Nagaoka1989,Hayashi2008}
\begin{equation}
\min_{\bold{X},V}\mathrm{Tr}(WV)\!=\!\min_{\bold{X}}~\!\mathrm{Tr}(W\mathrm{Re}(Z))
\!+\!\Vert\sqrt{W}\mathrm{Im}(Z)\sqrt{W}\Vert,
\end{equation}
where $\mathrm{Re}(Z)$ and $\mathrm{Im}(Z)$ represent the real and imaginary parts of $Z$, and $\Vert\cdot\Vert$
is the trace norm, i.e., $\Vert A\Vert:=\mathrm{Tr}\sqrt{A^{\dagger}A}$ for a matrix $A$. Numerically, in
a specific matrix basis $\{\lambda_i\}$ which satisfies $\mathrm{Tr}(\lambda_i\lambda_j)=\delta_{ij}$, the HCRB
can be solved via the semidefinite programming as it can be reformulated into a linear semidefinite
problem~\cite{Albarelli2019},
\begin{align}
& \min_{\bold{X},V}~\mathrm{Tr}(WV),  \nonumber \\
& \mathrm{subject}~\mathrm{to}~
\begin{cases}
\left(\begin{array}{cc}
V & \Lambda^{\mathrm{T}}R^{\dagger} \\
R\Lambda & \openone\\
\end{array}\right)\geq 0, \\
\sum_i[\Lambda]_{ai}\mathrm{Tr}(\lambda_i\partial_b\rho)=\delta_{ab}.
\end{cases}
\end{align}
Here the $ij$th entry of $\Lambda$ is obtained by decomposing $\bold{X}$ in the basis $\{\lambda_i\}$,
$X_i=\sum_j [\Lambda]_{ij}\lambda_j$, and $R$ satisfies $Z=\Lambda^{\mathrm{T}}R^{\dagger}R\Lambda$.
The semidefinite programming can be solved by the package CVXPY~\cite{Diamond2016,Agrawal2018} in Python
and Convex~\cite{Udell2014} in Julia. In QuanEstimation, the HCRB can be calculated via the function:
\begin{lstlisting}[breaklines=true,numbers=none,frame=trBL]
HCRB(rho,drho,W,eps=1e-8)
\end{lstlisting}
The input {\codefont W} is the weight matrix and {\codefont rho}, {\codefont drho} have been introduced
previously. Since $Z_{aa}$ is equivalent to the variance of the unbiased observable $O:=\sum_y\hat{x}_a\Pi_y$
[unbiased condition is $\mathrm{Tr}(\rho O)=x$], i.e., $Z_{aa}=\mathrm{Tr}(\rho O^2)-[\mathrm{Tr}(\rho O)]^2$,
in the case of single-parameter estimation the optimal $V$ is nothing but $Z_{aa}$ itself. Furthermore, it
can be proved that $Z_{aa}\geq 1/\mathcal{F}_{aa}$ and the equality is attainable asymptotically. Hence,
one can see that $\min_{X_a}Z_{aa}=1/\mathcal{F}_{aa}$, which means the HCRB is equivalent to the quantum
Cram\'{e}r-Rao bound in the single-parameter estimation. Due to better numerical efficiency of QFI computation,
whenever  {\codefont drho} has only one entry, the calling of {\codefont HCRB()} will automatically jump to
{\codefont QFIM()} in the package. Similarly, if $W$ is a rank-one matrix, the HCRB also reduces to
$\mathrm{Tr}(W\mathcal{F}^{-1})$ and thus in this case the calculation of HCRB will also be replaced by
the calculation of QFIM.

\emph{Example.} Now let us take a two-parameter estimation as an example to demonstrate the calculation of HCRB with
QuanEstimation. Consider a two-qubit system with the XX coupling. The Hamiltonian of this system is
\begin{equation}
H=\omega_1\sigma^{(1)}_3+\omega_2\sigma^{(2)}_3+g\sigma^{(1)}_1\sigma^{(2)}_1,
\label{eq:demo_twopara}
\end{equation}
where $\omega_{1}$, $\omega_2$ are the frequencies of the first and second qubit,
$\sigma^{(1)}_{i}=\sigma_{i}\otimes\openone$, and $\sigma^{(2)}_{i}=\openone\otimes\sigma_{i}$ for $i=1,2,3$.
$\openone$ is the identity matrix. Planck units are applied here ($\hbar=1$). The parameters $\omega_2$ and $g$ are
the ones to be estimated. The dynamics is governed by the master equation
\begin{equation}
\partial_t\rho=-i\left[H, \rho\right]+\sum_{i=1,2}\gamma_i\left(\sigma_3^{(i)}\rho\sigma_3^{(i)}-\rho \right)
\label{eq:mq_twopara}
\end{equation}
with $\gamma_i$ the decay rate for $i$th qubit. The time evolutions of quantum Cram\'{e}r-Rao bound
[$\mathrm{Tr}(W\mathcal{F}^{-1})$], classical Cram\'{e}r-Rao bound [$\mathrm{Tr}(W\mathcal{I}^{-1})$], and
HCRB are shown in Fig.~\ref{fig:HCRB}. The POVM for $\mathrm{Tr}(W\mathcal{I}^{-1})$ is
$\{\Pi_1$, $\Pi_2$, $\openone-\Pi_1-\Pi_2\}$ with $\Pi_1=0.85|00\rangle\langle 00|$ and
$\Pi_2=0.4|\!++\rangle\langle++\!|$. The probe state is $(|00\rangle+|11\rangle)/\sqrt{2}$ and the weight
matrix $W=\openone$. As shown in this plot, HCRB (dash-dotted blue line) is tighter than $\mathrm{Tr}(W\mathcal{F}^{-1})$
(solid red line), which is in agreement with the fact  that the HCRB is in general tighter than the quantum Cram\'{e}r-Rao
bound, unless the quantum Cram\'{e}r-Rao bound is attainable, in which case the two bounds coincide~\cite{Rafal2020}.

\subsection{Nagaoka-Hayashi bound}

Apart from the HCRB, the Nagaoka-Hayashi bound (NHB)~\cite{Nagaoka1989,Hayashi999,Conlon2021} is another available
bound for quantum multiparameter estimation, and is tighter than the HCRB in general. The expression of the NHB is
\begin{equation}
\mathrm{Tr}(W\mathrm{cov}(\hat{\bold{x}},\{\Pi_y\}))\geq
\min_{\bold{X},\mathcal{Q}}\mathrm{Tr}\left((W\otimes\rho)\mathcal{Q}\right).
\end{equation}
Here $\mathcal{Q}$ is a symmetric block matrix with each block a Hermitian matrix, and it satisfies
$\mathcal{Q}\geq\bold{X}^{\mathrm{T}}\bold{X}$ with $\bold{X}$ defined in Sec.~\ref{sec:HCRB}, namely,
$\bold{X}=[X_0,X_1,\cdots]$ and $X_a=\sum_y(\hat{x}_a-x_a)\Pi_y$. Similar to the HCRB, the calculation of NHB
can also be reformulated into a linear semidefinite problem~\cite{Conlon2021} as follows:
\begin{align}
& \min_{\bold{X},\mathcal{Q}}~\mathrm{Tr}\left((W\otimes\rho)\mathcal{Q}\right),  \nonumber \\
& \mathrm{subject}~\mathrm{to}~
\begin{cases}
\left(\begin{array}{cc}
\mathcal{Q} & \bold{X}^{\mathrm{T}} \\
\bold{X} & \openone\\
\end{array}\right)\geq 0, \\
\mathrm{Tr}(\rho X_a)=0,\,\forall a, \\
\mathrm{Tr}(X_a\partial_b\rho)=\delta_{ab},\,\forall a, b.\\
\end{cases}
\end{align}
In QuanEstimation, the NHB can be calculated via the function:
\begin{lstlisting}[breaklines=true,numbers=none,frame=trBL]
NHB(rho,drho,W)
\end{lstlisting}
The performance of the NHB is also demonstrated in Fig.~\ref{fig:HCRB} with the Hamiltonian in
Eq.~(\ref{eq:demo_twopara}) and dynamics in Eq.~(\ref{eq:mq_twopara}). In this case the NHB
(dotted green line) is indeed slightly tighter than the HCRB, and thus also tighter than the quantum
Cram\'{e}r-Rao bound [$\mathrm{Tr}(W\mathcal{F}^{-1})$]. However, there still exist a gap between
the classical Cram\'{e}r-Rao bound [$\mathrm{Tr}(W\mathcal{I}^{-1})$] and the NHB, indicating
that the chosen measurement may not be an optimal one.

\subsection{Bayesian estimation}
\label{sec:Bayesian}

Bayesian estimation is another well-used method in parameter estimation, in which the prior distribution is updated
via the posterior distribution obtained by the Bayes' rule
\begin{equation}
p(\bold{x}|y)=\frac{p(y|\bold{x})p(\bold{x})}{\int p(y|\bold{x})p(\bold{x})\mathrm{d}\bold{x}},
\label{eq:Bayes_posterior}
\end{equation}
where $p(\bold{x})$ is the current prior distribution, $y$ is the result obtained in practice, and
$\int\mathrm{d}\bold{x}:=\int\mathrm{d}x_0\int\mathrm{d}x_1\cdots$. The prior distribution is then updated with
$p(\bold{x}|y)$, and the estimated value of $\bold{x}$ is obtained via a reasonable estimator, such as the
expected value $\hat{\bold{x}}=\int\bold{x} p(\bold{x}|y)\mathrm{d}\bold{x}$ or the maximum a posteriori
estimation (MAP), $\hat{\bold{x}}=\mathrm{argmax}_{\bold{x}}\,p(\bold{x}|y)$.

In QuanEstimation, the Bayesian estimation can be performed via the function:
\begin{lstlisting}[breaklines=true,numbers=none,frame=trBL]
pout,xout = Bayes(x,p,rho,y,M=[],
            estimator="mean",savefile=False)
\end{lstlisting}
The input {\codefont x} is a list of arrays representing the regimes of $\bold{x}$, which is the same with the
function {\codefont BayesInput()} discussed in Sec.~\ref{sec:para}. Notice that in the package all the
calculations of the integrals over the prior distributions are performed discretely. Hence, for now the input prior
distribution is required to be an array, instead of a continuous function. {\codefont p} is an array representing the
values of $p(\bold{x})$ with respect to $\bold{x}$. It is multidimensional in the case of multiparameter estimation, i.e.,
the entry number of {\codefont x} are at least two. The input {\codefont rho} is a (multidimensional) list of matrices
representing the values of density matrix with respect to all values of $\bold{x}$, which can be alternatively generated
via the function {\codefont BayesInput()} if specific functions of $H$ and $\partial_{\bold{x}}H$ on $\bold{x}$ can be
provided. {\codefont M=[]} is a list of matrices representing a set of POVM and its default setting is a SIC-POVM.
{\codefont y} is an array representing the results obtained in an experiment. The result corresponds to the POVM operator
input in {\codefont M}, which means it is an integer between 0 and $d-1$ with $d$ the entry number of the set of
POVM. The type of estimator can be set via {\codefont estimator=" "} and currently it has two choices. When
{\codefont estimator="mean"} the estimator is the expected value, and when {\codefont estimator="MAP"} the estimator
is the MAP. The output {\codefont pout} (a multidimensional array) and {\codefont xout} (an array) are the final
posterior distribution and estimated value of $\bold{x}$ obtained via the chosen estimator. When {\codefont savefile=True},
two files "pout.npy" and "xout.npy" will be generated, which include the updated $p(\bold{x})$ and the corresponding optimal
$\bold{x}$ in all rounds. If the users call this function in the full-Julia package, the output files are "pout.csv"
and "xout.csv".

%================================ Figure ============================================
\begin{figure}[tp]
\centering\includegraphics[width=8.5cm]{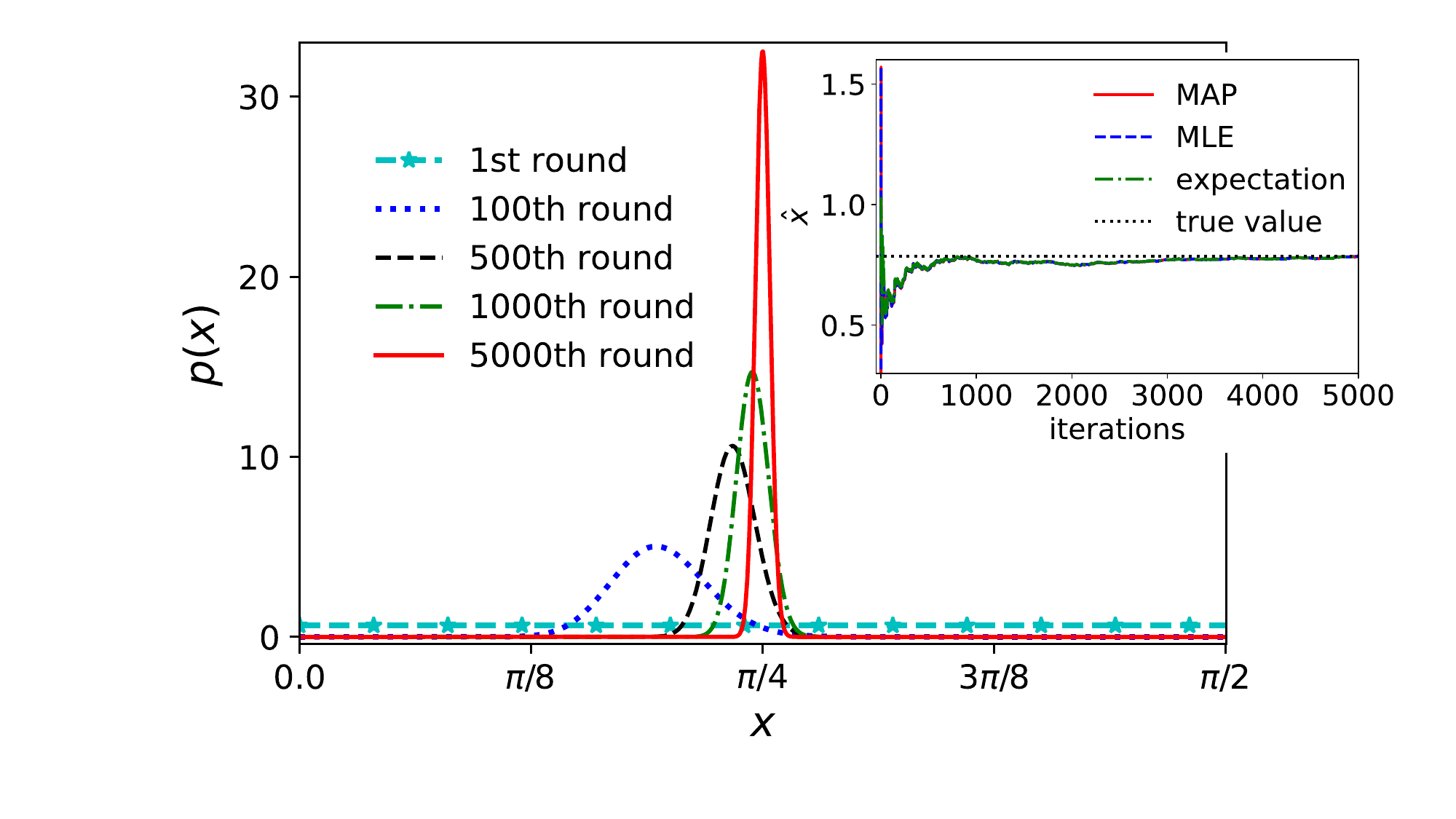}
\caption{Iteration of posterior distribution by the Bayes' rule. The inset shows the change
of estimated value as a function of iteration for MAP (solid red line), MLE (dashed blue line),
and expectation (dash-dotted green line). The dotted black line represents the true value.}
\label{fig:bayes_mle}
\end{figure}
%====================================================================================

\emph{Example.} Now let us consider a simple example with the Hamiltonian
\begin{equation}
H = \frac{\kappa\omega_0}{2}(\sigma_1\cos x + \sigma_3\sin x),
\label{eq:Bayes_demo}
\end{equation}
where $x$, $\kappa$ are two dimensionless parameters and $x$ is taken as the unknown one. Planck units are applied here
($\hbar=1$) and $\omega_0$ is set to be 1. The initial state is taken as $|+\rangle$ and the target time $\omega_0 T=1$.
The prior distribution is assumed to be uniform in the regime $[0,\pi/2]$. The measurement is
$\{|+\rangle\langle +|,|-\rangle\langle-|\}$. The results in experiment are simulated by a random generation according
to the probabilities $p(\pm|x)=\langle\pm|\rho|\pm\rangle$ with respect to the value $x=\pi/4$. As shown in
Fig.~\ref{fig:bayes_mle}, with the growth of iteration number, the deviation decreases monotonously and the estimated
value (center value of the distribution) approaches to $\pi/4$, which can also be confirmed by the convergence of
estimated value (solid red line) shown in the inset. As a matter of fact, here the maximum likelihood estimation (MLE)
can also provide similar performance by taking the likelihood function with the MAP estimator
$\hat{\bold{x}}=\mathrm{argmax}_{\bold{x}}\,\prod_i p(y_i|\bold{x})$ (dashed blue line in the inset). In QuanEstimation,
this MLE can be calculated by the function:
\begin{lstlisting}[breaklines=true,numbers=none,frame=trBL]
Lout,xout = MLE(x,rho,y,M=[],savefile=False)
\end{lstlisting}
When {\codefont savefile=True}, two files "Lout.npy" and "xout.npy" will be generated including all the data in the
iterations.

In Bayesian estimation, another useful tool is the average Bayesian cost~\cite{Robert2007} for the quadratic cost,
which is defined by
\begin{equation}
\bar{C}:=\int p(\bold{x})\sum_y p(y|\bold{x})(\bold{x}-\hat{\bold{x}})^{\mathrm{T}}
W(\bold{x}-\hat{\bold{x}})\,\mathrm{d}\bold{x}
\end{equation}
with $W$ the weight matrix. In QuanEstimation, this average Bayesian cost can be calculated via the function:
\begin{lstlisting}[breaklines=true,numbers=none,frame=trBL]
BayesCost(x,p,xest,rho,M,W=[],eps=1e-8)
\end{lstlisting}
Here {\codefont x} and {\codefont p} are the same with those in {\codefont Bayes()}. {\codefont xest} is a list of arrays
representing the estimator $\hat{\bold{x}}$. The $i$th entry of each array in {\codefont xest} represents the estimator
with respect to $i$th result. In the case of the single-parameter scenario, $W$ is chosen to be 1 regardless of the input.
The average Bayesian cost satisfies the inequality~\cite{Rafal2020}
\begin{equation}
\bar{C}\geq\int p(\bold{x})\left(\bold{x}^{\mathrm{T}}W\bold{x}\right)\mathrm{d}\bold{x}
-\sum_{ab}W_{ab}\mathrm{Tr}\left(\bar{\rho}\bar{L}_a \bar{L}_b\right),
\label{eq:BCB}
\end{equation}
where $\bar{\rho}:=\int p(\bold{x})\rho\,\mathrm{d}\bold{x}$ and the operator $\bar{L}_a$ is determined by the equation
$\int x_a p(\bold{x})\rho\,\mathrm{d}\bold{x}=(\bar{L}_a\bar{\rho}+\bar{\rho}\bar{L}_a)/2$. In the case of the
single-parameter scenario, the inequality above reduces to
\begin{equation}
\bar{C}\geq \int p(x) x^2\,\mathrm{d}x-\mathrm{Tr}(\bar{\rho}\bar{L}^2)
\end{equation}
and represents a bound which is always saturable---the optimal measurement correspond to projection measurement in the
eigenbasis of $\bar{L}$, while the corresponding eigenvalues represent the estimated values of the parameter. If the
mean value $\int p(x) x\,\mathrm{d}x$ is subtracted to zero, then the inequality above can be rewritten into
$\bar{C}\geq \delta^2 x-\mathrm{Tr}(\bar{\rho}\bar{L}^2)$ with $\delta^2 x:=\int p(x) x^2\,\mathrm{d}x-\int p(x)
x\,\mathrm{d}x$ the variance of $x$ under the prior distribution. In QuanEstimation, the bound given in Eq.~(\ref{eq:BCB})
can be calculated via the following function:
\begin{lstlisting}[breaklines=true,numbers=none,frame=trBL]
BCB(x,p,rho,W=[],eps=1e-8)
\end{lstlisting}
Here the inputs {\codefont x} and {\codefont p} are the some with those in {\codefont Bayes()} and {\codefont BayesCost()}.
{\codefont W} represents the weight matrix and the default value is the identity matrix.

\subsection{Bayesian Cram\'{e}r-Rao bounds}

In the Bayesian scenarios, the quantum Cram\'{e}r-Rao Bounds and Holevo Cram\'{e}r-Rao bound are not
appropriate to grasp the the ultimate precision limits
as they are ignorant of the prior information. Still, Bayesian Cram\'{e}r-Rao bounds can be used instead. In these scenarios,
the covariance matrix is redefined as
\begin{equation}
\mathrm{cov}(\hat{\bold{x}},\{\Pi_y\})\!=\!\int \!p(\bold{x})\sum_y\mathrm{Tr}(\rho\Pi_y)
(\hat{\bold{x}}\!-\!\bold{x})(\hat{\bold{x}}\!-\!\bold{x})^{\mathrm{T}}\mathrm{d}\bold{x},
\end{equation}
where the integral $\int\mathrm{d}\bold{x}:=\iiint\mathrm{d}x_0\mathrm{d}x_1\cdots$. In such cases, one version of
the Bayesian Cram\'{e}r-Rao bound (BCRB) is of the form
\begin{equation}
\mathrm{cov}(\hat{\bold{x}},\{\Pi_y\})\geq \int p(\bold{x})
\left(B\mathcal{I}^{-1}B+\bold{b}\bold{b}^{\mathrm{T}}\right)\mathrm{d}\bold{x},
\label{eq:BCRB_type1}
\end{equation}
where $\mathcal{I}$ is the CFIM, and $\bold{b}=(b(x_0),b(x_1),\dots)^{\mathrm{T}}$ is the vector of biases,
i.e., $b(x_a)=\sum_y\hat{x}_a p(y|\bold{x})-x_a$ for each $x_a$ with $p(y|\bold{x})$ the conditional probability.
$B$ is a diagonal matrix with the $a$th entry $B_{aa}=1+[\bold{b}']_{a}$. Here $\bold{b}':=(\partial_0 b(x_0),
\partial_1 b(x_1),\dots)^{\mathrm{T}}$. The quantum correspondence of this bound (BQCRB) reads
\begin{equation}
\mathrm{cov}(\hat{\bold{x}},\{\Pi_y\})\geq\int p(\bold{x})
\left(B\mathcal{F}^{-1}B+\bold{b}\bold{b}^{\mathrm{T}}\right)\mathrm{d}\bold{x},
\label{eq:BQCRB_type1}
\end{equation}
where $\mathcal{F}$ is the QFIM of all types. As a matter of fact, there exists a similar version of
Eq.~(\ref{eq:BCRB_type1}), which can be expressed by
\begin{equation}
\mathrm{cov}(\hat{\bold{x}},\{\Pi_y\})\geq \mathcal{B}\,\mathcal{I}_{\mathrm{Bayes}}^{-1}
\,\mathcal{B}+\int p(\bold{x})\bold{b}\bold{b}^{\mathrm{T}}\mathrm{d}\bold{x},
\label{eq:BCRB_type2}
\end{equation}
where $\mathcal{I}_{\mathrm{Bayes}}=\int p(\bold{x})\mathcal{I}\mathrm{d}\bold{x}$ is the average CFIM with
$\mathcal{I}$ the CFIM defined in Eq.~(\ref{eq:CFIM}). $\mathcal{B}=\int p(\bold{x})B\mathrm{d}\bold{x}$ is the
average of $B$. Its quantum correspondence reads
\begin{equation}
\mathrm{cov}(\hat{\bold{x}},\{\Pi_y\})\geq \mathcal{B}\,\mathcal{F}_{\mathrm{Bayes}}^{-1}
\,\mathcal{B}+\int p(\bold{x})\bold{b}\bold{b}^{\mathrm{T}}\mathrm{d}\bold{x},
\label{eq:BQCRB_type2}
\end{equation}
where $\mathcal{F}_{\mathrm{Bayes}}=\int p(\bold{x})\mathcal{F}\mathrm{d}\bold{x}$ is average QFIM with $\mathcal{F}$
the QFIM of all types.

Another version of the Bayesian Cram\'{e}r-Rao bound is of the form
\begin{equation}
\mathrm{cov}(\hat{\bold{x}},\{\Pi_y\})\geq \int p(\bold{x})
\mathcal{G}\left(\mathcal{I}_p+\mathcal{I}\right)^{-1}\mathcal{G}^{\mathrm{T}}\mathrm{d}\bold{x},
\label{eq:BCRB_type3}
\end{equation}
and its quantum correspondence can be expressed by
\begin{equation}
\mathrm{cov}(\hat{\bold{x}},\{\Pi_y\})\geq \int p(\bold{x})
\mathcal{G}\left(\mathcal{I}_p+\mathcal{F}\right)^{-1}\mathcal{G}^{\mathrm{T}}\mathrm{d}\bold{x},
\label{eq:BQCRB_type3}
\end{equation}
where the entries of $\mathcal{I}_{p}$ and $\mathcal{G}$ are defined by
\begin{equation}
[\mathcal{I}_{p}]_{ab}:=[\partial_a \ln p(\bold{x})][\partial_b \ln p(\bold{x})],
\label{eq:BayesIp}
\end{equation}
and $\mathcal{G}_{ab}:=[\partial_b\ln p(\bold{x})][\bold{b}]_a+B_{aa}\delta_{ab}$. The derivations and thorough
discussions of these bounds will be further discussed in an independent paper, which will be announced in a short time.

The functions in QuanEstimation to calculate $\mathcal{I}_{\mathrm{Bayes}}$ and $\mathcal{F}_{\mathrm{Bayes}}$ are:
\begin{lstlisting}[breaklines=true,numbers=none,frame=trBL]
BCFIM(x,p,rho,drho,M=[],eps=1e-8)
BQFIM(x,p,rho,drho,LDtype="SLD",eps=1e-8)
\end{lstlisting}
And the functions for the calculations of BCRBs and BQCRBs are:
\begin{lstlisting}[breaklines=true,numbers=none,frame=trBL]
BCRB(x,p,dp,rho,drho,M=[],b=[],db=[],
     btype=1,eps=1e-8)
BQCRB(x,p,dp,rho,drho,b=[],db=[],btype=1,
      LDtype="SLD",eps=1e-8)
\end{lstlisting}
The input {\codefont x} and {\codefont p} are the same with those in the function {\codefont Bayes()}. {\codefont dp} is
a (multidimensional) list of arrays representing the derivatives of the prior distribution, which is only essential when
{\codefont btype=3}. In the case that {\codefont btype=1} and {\codefont btype=2}, it could be set as {\codefont []}.
{\codefont rho} and {\codefont drho} are (multidimensional) lists representing the values of $\rho$ and
$\partial_{\bold{x}}\rho$. For example, if the input {\codefont x} includes three arrays, which are the values of $x_0$,
$x_1$, and $x_2$ for the integral, then the $ijk$th entry of {\codefont rho} and {\codefont drho} are a matrix $\rho$ and
a list $[\partial_{0}\rho,\partial_{1}\rho,\partial_{2}\rho]$ with respect to the values $[x_0]_i$, $[x_1]_j$, and $[x_2]_k$.
Here $[x_0]_i$, $[x_1]_j$, and $[x_2]_k$ represent the $i$th, $j$th, and $k$th value in the first, second, and
third array in {\codefont x}. As a matter of fact, if the users can provide specific functions of $H$ and
$\partial_{\bold{x}}H$ on $\bold{x}$, {\codefont rho} and {\codefont drho} can be alternatively generated via the
functions {\codefont BayesInput()} and {\codefont Lindblad()} [or {\codefont Kraus()}]. {\codefont b} and {\codefont db}
are two lists of arrays representing $\bold{b}$ and $\bold{b}'$, and the default settings for both of them are zero vectors
(unbiased). In {\codefont BCRB()} the measurement is input via {\codefont M=[]}, and if it is empty, a set of rank-one
SIC-POVM will be automatically applied, similar to that in {\codefont CFIM()}. Moreover, {\codefont btype=1}, {\codefont btype=2},
and {\codefont btype=3} represent the calculation of Eqs.~(\ref{eq:BCRB_type1}), (\ref{eq:BCRB_type2}), and (\ref{eq:BCRB_type3}).
In the meantime, in {\codefont BQCRB()}, {\codefont btype=1}, {\codefont btype=2}, and {\codefont btype=3} represent the
calculation of Eqs.~(\ref{eq:BQCRB_type1}), (\ref{eq:BQCRB_type2}) and (\ref{eq:BQCRB_type3}). Similar to {\codefont QFIM()},
{\codefont LDtype=" "} here is the type of logarithmic derivatives, including three choices: {\codefont "SLD"}, {\codefont "RLD"},
and {\codefont "LLD"}. Recently, Ref.~\cite{Liu2016} provide an optimal biased bound based on the type-1 BQCRB in the case of
single-parameter estimation, which can be calculated in QuanEstimation via the function:
\begin{lstlisting}[breaklines=true,numbers=none,frame=trBL]
OBB(x,p,dp,rho,drho,d2rho,LDtype="SLD",
    eps=1e-8)
\end{lstlisting}
The input {\codefont dp} is an array containing the derivatives $\partial_x p$. {\codefont d2rho} is a list
containing the second order derivative of the density matrix on the unknown parameter.

%================================ Figure ============================================
\begin{figure}[tp]
\centering\includegraphics[width=8cm]{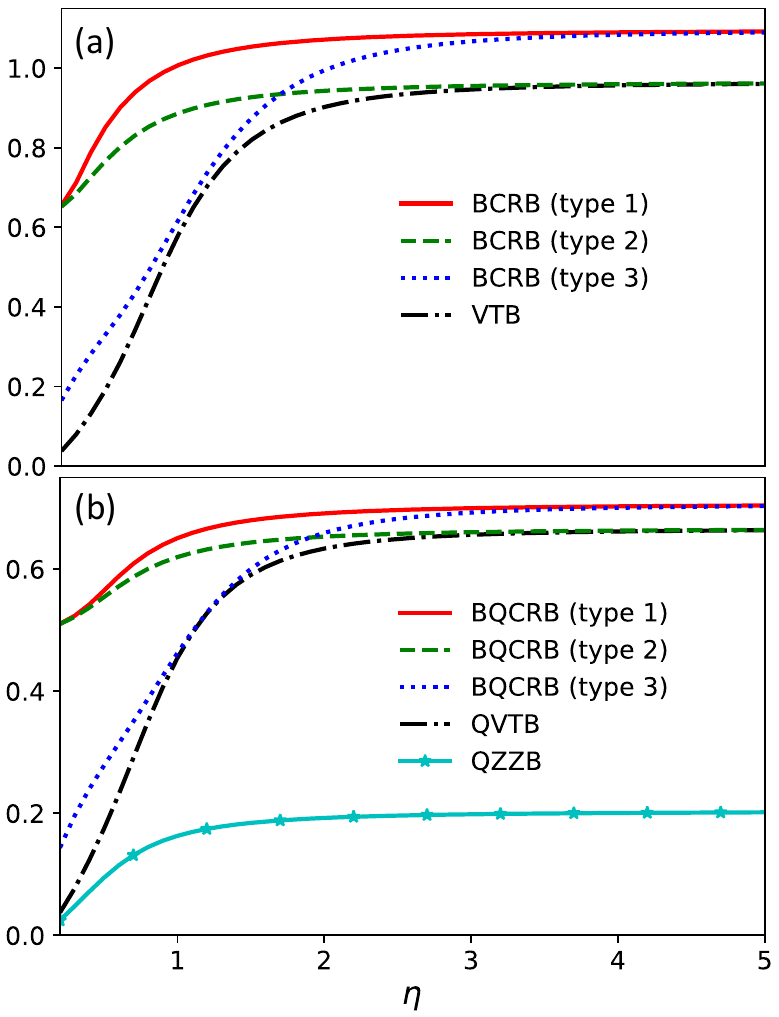}
\caption{(a) The performance of classical Bayesian bounds, including BCRB of type 1 (solid
red line), type 2 (dashed green line), type 3 (dotted blue line), and VTB (dash-dotted
black line). (b) The performance of quantum Bayesian bounds, including BQCRB
of type 1 (solid red line), type 2 (dashed green line), type 3 (dotted blue line),
QVTB (dash-dotted black line), and QZZB (solid cyan pentagram line). The parameters
$\mu=0$ and $\kappa=\pi/2$ in the plots. Planck units are applied here. }
\label{fig:bayes}
\end{figure}
%====================================================================================

Another famous Bayesian version of Cram\'{e}r-Rao bound is introduced by Van Trees in 1968~\cite{vanTrees1968},
which is known as the Van Trees bound (VTB). The VTB is expressed by
\begin{equation}
\mathrm{cov}(\hat{\bold{x}},\{\Pi_y\})\geq \left(\mathcal{I}_{\mathrm{prior}}
+\mathcal{I}_{\mathrm{Bayes}}\right)^{-1},
\end{equation}
where $\mathcal{I}_{\mathrm{prior}}=\int p(\bold{x})\mathcal{I}_{p}\mathrm{d}\bold{x}$ is the CFIM for $p(\bold{x})$
with $\mathcal{I}_p$ defined in Eq.~(\ref{eq:BayesIp}). In the derivation, the assumption
\begin{equation}
\int\partial_{a}\left[b(x_b)p(\bold{x})\right]\mathrm{d}\bold{x}=0
\end{equation}
is applied for all subscripts $a$ and $b$. In 2011, Tsang, Wiseman and Caves~\cite{Tsang2011} provided a quantum
correspondence of the VTB (QVTB). The Tsang-Wiseman-Caves bound is of the form
\begin{equation}
\mathrm{cov}(\hat{\bold{x}},\{\Pi_y\})\geq \left(\mathcal{I}_{\mathrm{prior}}
+\mathcal{F}_{\mathrm{Bayes}}\right)^{-1}.
\end{equation}
The functions in QuanEstimation for the calculation of VTB and QVTB are:
\begin{lstlisting}[breaklines=true,numbers=none,frame=trBL]
VTB(x,p,dp,rho,drho,M=[],eps=1e-8)
QVTB(x,p,dp,rho,drho,LDtype="SLD",eps=1e-8)
\end{lstlisting}
Here {\codefont dp} is a (multidimensional) list of arrays representing the derivatives of the prior distribution.
For example, if {\codefont x} includes 3 arrays, which are the values of $x_0$, $x_1$, and $x_2$ for the integral,
then the $ijk$th entry of {\codefont dp} is an array $(\partial_0 p,\partial_1 p,\partial_2 p)$ with respect to
values $[x_0]_i$, $[x_1]_j$ and $[x_2]_k$.

\emph{Example.} Let us still take the Hamiltonian in Eq.~(\ref{eq:Bayes_demo}) and initial state $|+\rangle$ as
an example. $x$ is still the parameter to be estimated. The prior distribution is taken as a Gaussian distribution
\begin{equation}
p(x)=\frac{1}{c\eta\sqrt{2\pi}}e^{-\frac{(x-\mu)^2}{2\eta^2}}
\label{eq:Bayes_prior}
\end{equation}
in a finite regime $[-\pi/2, \pi/2]$, where $\mu$ is the expectation, $\eta$ is the standard deviation, and
$c=\frac{1}{2}\big[\mathrm{erf}(\frac{\pi-2\mu}{2\sqrt{2}\eta})+\mathrm{erf}(\frac{\pi+2\mu}{2\sqrt{2}\eta})\big]$
is the normalized coefficient. Here $\mathrm{erf}(x):=\frac{2}{\sqrt{\pi}}\int^x_0 e^{-t^2}\mathrm{d}t$ is the error
function. The measurement in the classical bounds is taken as a set of SIC-POVM. The performance of the classical and
quantum Bayesian bounds are given in Figs.~\ref{fig:bayes}(a) and \ref{fig:bayes}(b). As shown in Fig.~\ref{fig:bayes}(a),
in this case BCRB of type 1 (solid red line) and type 2 (dashed green line) are tighter than type 3 (dotted blue
line) and VTB (dash-dotted black line) when the deviation $\eta$ is small. With the increase of $\eta$, BCRB of type 1
and type 3 coincide with each other, so do BCRB of type 2 and VTB. Furthermore, BCRB of type 1 and type 3 are always
tighter than type 2 and VTB in this example. The performance of quantum Bayesian bounds are similar, as shown in
Fig.~\ref{fig:bayes}(b). BQCRB (solid red line for type 1 and dashed green line for type 2) are tighter than type 3
(dotted green line) and QVTB (dash-dotted black line) when $\eta$ is small and BQCRB of type 1 (type 2) and type 3
(QVTB) coincide with each other for a large $\eta$.

\subsection{Quantum Ziv-Zakai bound}

Apart from the Cram\'{e}r-Rao bounds, the Ziv-Zakai bound is another useful bound in Bayesian scenarios. It was
first provided by Ziv and Zakai in 1969~\cite{Ziv1969} for the single-parameter estimation and then extended to
the linear combination of multiple parameters by Bell et al.~\cite{Bell1997}, which is also referred to
as the Bell-Ziv-Zakai bound. In 2012, Tsang provided a quantum correspondence of the Ziv-Zakai bound~\cite{Tsang2012}
(QZZB), and in 2015 Berry et al.~\cite{Berry2015} provided a quantum correspondence of the Bell-Ziv-Zakai bound.
In QZZB, the variance $\mathrm{var}(\hat{x},\{\Pi_y\})$, a diagonal entry of the covariance matrix, satisfies the
following inequality
\begin{eqnarray}
\mathrm{var}(\hat{x},\{\Pi_y\}) &\geq & \frac{1}{2}\int_0^\infty \mathrm{d}\tau\tau
\mathcal{V}\int_{-\infty}^{\infty} \mathrm{d}x\min\!\left\{p(x), p(x+\tau)\right\} \nonumber \\
& & \times\left(1-\frac{1}{2}||\rho(x)-\rho(x+\tau)||\right),
\end{eqnarray}
where $||\cdot||$ is the trace norm. $\mathcal{V}$ is the "valley-filling" operator
satisfying $\mathcal{V}f(\tau)=\max_{h\geq 0}f(\tau+h)$. In the numerical calculations, the prior distribution has
to be limited or truncated in a finite regime $[\alpha,\beta]$, i.e., $p(x)=0$ when $x>\beta$ or $x<\alpha$, and
then the QZZB reduces to
\begin{eqnarray}
\mathrm{var}(\hat{x},\{\Pi_y\}) &\geq & \frac{1}{2}\int_0^{\beta-\alpha}\mathrm{d}\tau\tau
\mathcal{V}\int_{\alpha}^{\beta}\mathrm{d}x\min\left\{p(x), p(x+\tau)\right\} \nonumber \\
& & \times\left(1-\frac{1}{2}||\rho(x)-\rho(x+\tau)||\right).
\end{eqnarray}
The function in QuanEstimation for the calculation of QZZB is:
\begin{lstlisting}[breaklines=true,numbers=none,frame=trBL]
QZZB(x,p,rho,eps=1e-8)
\end{lstlisting}
The performance of QZZB is also demonstrated with the Hamiltonian in Eq.~(\ref{eq:Bayes_demo}) and prior
distribution in Eq.~(\ref{eq:Bayes_prior}), as shown in Fig.~\ref{fig:bayes}(b). In this example, its performance
(solid cyan pentagram line) is worse than BQCRB and QVTB. However, this tightness relation may dramatically change
in other systems or with other prior distributions. Hence, in a specific scenario using QuanEstimation to perform
a thorough comparison would be a good choice to find the tightest tool for the scheme design.

\section{Metrological resources}
\label{sec:resource}

The improvement of precision usually means a higher consumption of resources. For example, the repetition of
experiments will make the deviation of the unknown parameter to scale proportionally to $1/\sqrt{n}$ ($n$ the repetition
number) in theory. The repetition number or the total time is thus the resource responsible for this improvement.
Constraint on quantum resources is an important aspect in the study of quantum parameter estimation, and is crucial
to reveal the quantum advantage achievable in practical protocols. The numerical calculations of some typical resources
have been added in QuTiP, such as various types of entropy and the concurrence. Hence, we do not need to rewrite
them in QuanEstimation. Currently, two additional metrological resources, spin squeezing and the time to reach
a given precision limit are provided in the package. The spin squeezing can be calculated via the function:
\begin{lstlisting}[breaklines=true,numbers=none,frame=trBL]
SpinSqueezing(rho,basis="Dicke",output="KU")
\end{lstlisting}
Here the input {\codefont rho} is a matrix representing the state. The basis of the state can be adjusted
via {\codefont basis=" "}. Two options {\codefont "Dicke"} and {\codefont "Pauli"} represent the Dicke
basis and the original basis of each spin. {\codefont basis="Pauli"} here is equivalent to choose
{\codefont basis="uncoupled"} in the function {\codefont jspin()} in QuTiP. Two types of spin squeezing
can be calculated in this function. {\codefont output="KU"} means the output is the one given by Kitagawa
and Ueda~\cite{Kitagawa1993}, and {\codefont output="WBIMH"} means the output is the one given by Wineland
et al.~\cite{Wineland1992}.

The time to reach a given precision limit can be calculated via the function:
\begin{lstlisting}[breaklines=true,numbers=none,frame=trBL]
TargetTime(f,tspan,func,*args,**kwargs)
\end{lstlisting}
Notice that the dynamics needs to be run first before using this function. For example, it is available to be
called after the calling of both {\codefont Lindblad()} and {\codefont Lindblad.expm()}. The input {\codefont f} is
a float number representing the given value of the precision limit. The time is searched within the regime defined
by the input {\codefont tspan} (an array). {\codefont func} is the handle of a function {\codefont func()} depicting
the precision limit. {\codefont *args} is the corresponding input parameters, in which {\codefont rho} and
{\codefont drho} should be the output of {\codefont Lindblad.expm()} [or {\codefont Lindblad.ode()} and any other
method that may included in the future]. {\codefont **kwargs} is the keyword arguments in {\codefont func()}. The
difference between input parameters and keyword arguments in QuanEstimation is that the keyword arguments have
default values and thus one does not have to assign values to them when calling the function. Currently, all the
asymptotic bounds discussed in Sec.~\ref{sec:tools} are available to be called here.

%==================================== Table =================================
\begin{table}[tp]
% \begin{ruledtabular}
\begin{tabular}{c|c|c|c}
\hline
\hline
Algorithms & method= & \multicolumn{2}{c}{~**kwargs and default values~}\\
\hline
\multirow{6}{*}{auto-GRAPE} & \multirow{6}{*}{"auto-GRAPE"} & "Adam"  & True \\
\multirow{6}{*}{(GRAPE)} & \multirow{6}{*}{("GRAPE")}  & "ctrl0"  & [] \\
  &   & "max\_episode"  & 300 \\
  &   & "epsilon" & 0.01 \\
  &   & "beta1" & 0.90 \\
  &   & "beta2" & 0.99 \\
\hline
\multirow{7}{*}{PSO} & \multirow{7}{*}{"PSO"} & "p\_num" & 10 \\
  &   & "ctrl0"  & [] \\
  &   & "max\_episode"  & [1000,100] \\
  &   & "c0"  & 1.0 \\
  &   & "c1"  & 2.0 \\
  &   & "c2"  & 2.0 \\
  &   & "seed"  & 1234 \\
\hline
\multirow{6}{*}{DE} & \multirow{6}{*}{"DE"} & "p\_num" & 10 \\
  &   & "ctrl0"  & [] \\
  &   & "max\_episode"  & 1000 \\
  &   & "c"  & 1.0 \\
  &   & "cr"  & 0.5 \\
  &   & "seed"  & 1234 \\
\hline
\multirow{5}{*}{DDPG} & \multirow{5}{*}{"DDPG"}   & "ctrl0" & [] \\
  &   & "max\_episode" & 500 \\
  &   & "layer\_num"  & 3 \\
  &   & "layer\_dim"  & 200 \\
  &   & "seed"  & 1234 \\
\hline
\hline
\end{tabular}
% \end{ruledtabular}
\caption{Available control methods in QuanEstimation andcorresponding
default parameter settings. Notice that auto-GRAPE and GRAPE are not
available when {\codefont control.HCRB()} is called.}
\label{table:ctrl_paras}
\end{table}
%============================================================================

\section{Control optimization}
\label{sec:control_opt}

Quantum control is a leading approach in quantum metrology to achieve the improvement of measurement
precision and boost the resistance to decoherence. This is possible thanks to high controllability of
typical quantum metrological setups. A paradigmatic controllable Hamiltonian is of the form
\begin{equation}
H=H_0(\bold{x})+\sum^K_{k=1}u_k(t) H_k,
\end{equation}
where $H_0(\bold{x})$ is the free Hamiltonian containing the unknown parameters $\bold{x}$ and $H_k$ is the
$k$th control Hamiltonian with the corresponding control amplitude $u_k(t)$. In quantum parameter estimation,
the aim of control is to improve the precision of the unknown parameters. Hence, natural choices for the the
objective function $f$ are the various metrological bounds. The quantum Cram\'{e}r-Rao bounds are easiest to
calculate and hence will typically be the first choice. In the single-parameter estimation, the QFI or CFI
can be taken as the objective function, depending whether the measurement can be optimized or is fixed. In
the multiparameter scenario, the objective function can be $\mathrm{Tr}(W\mathcal{F}^{-1})$,
$\mathrm{Tr}(W\mathcal{I}^{-1})$, or the HCRB.

Searching the optimal controls in order to achieve the maximum or minimum values of an objective function is the
core task in quantum control. Most existing optimization algorithms, such as Gradient ascent pulse
engineering~\cite{Khaneja2005,Liu2017a,Liu2017b}, Krotov's method~\cite{Reich2013a,Reich2013b,Goerz2014}, and
machine learning~\cite{Xu2019, Xu2021}, are capable of providing useful control strategies in quantum parameter estimation. The
gradient-based algorithms usually perform well in small-scale systems. For complex problems where the gradient-based
methods are more challenging or even fail to work at all, gradient-free algorithms are a good alternative. Here we
introduce several control algorithms in quantum parameter estimation that have been added into our package and give
some illustrations.

First, we present the specific code in QuanEstimation for the execution of the control optimization,
\begin{lstlisting}[breaklines=true,numbers=none,frame=trBL,mathescape=true]
control = ControlOpt(savefile=False,
                method="auto-GRAPE",**kwargs)
control.dynamics(tspan,rho0,H0,dH,Hc,
                 decay=[],ctrl_bound=[],
                 dyn_method="expm")
control.QFIM(W=[],LDtype="SLD")
control.CFIM(M=[],W=[])
control.HCRB(W=[])
\end{lstlisting}
The input {\codefont tspan} is an array representing the time for the evolution. {\codefont rho0} is a matrix
representing the density matrix of the initial state. {\codefont H0} is a matrix representing the free Hamiltonian
$H_0(\bold{x})$ and {\codefont Hc} is a list containing the control Hamiltonians, i.e., $[H_1,H_2,\dots]$.
{\codefont dH} is a list of matrices representing $\partial_{\bold{x}}H_0$. In the case that only one entry
exists in {\codefont dH}, the objective functions in {\codefont control.QFIM()} and {\codefont control.CFIM()}
are the QFI and CFI, and if more than one entries are input, the objective functions are $\mathrm{Tr}(W\mathcal{F}^{-1})$
and $\mathrm{Tr}(W\mathcal{I}^{-1})$. Different types of QFIM can be selected as the objective function via the
variable {\codefont LDtype=" "}, which includes three options {\codefont "SLD"}, {\codefont "RLD"}, and
{\codefont "LLD"}. The measurement for CFI/CFIM is input via {\codefont M=[]} in {\codefont control.CFIM()} and
the default value is a SIC-POVM. The weight matrix $W$ can be manually input via {\codefont W=[]}, and the default
value is the identity matrix.

In some cases, the control amplitudes have to be limited in a regime, for example $[a,b]$, which can be realized
by input {\codefont ctrl\_bound=[a,b]}. If no value is input, the default regime is $[-\infty,\infty]$.
{\codefont decay=[]} is a list of decay operators and corresponding decay rates for the master equation in
Eq.~(\ref{eq:mastereq}) and its input rule is {\codefont decay=[[Gamma\_1,gamma\_1],...]}. The dynamics
is solved via the matrix exponential by default, which can be switched to ODE by setting {\codefont dyn\_method="ode"}.
The default value for {\codefont savefile} is {\codefont False}, which means only the controls obtained in the final
episode will be saved in the file named "controls.csv", and if it is set to be {\codefont True}, the controls
obtained in all episodes will be saved in this file. The values of QFI, CFI, $\mathrm{Tr}(W\mathcal{F}^{-1})$ or
$\mathrm{Tr}(W\mathcal{I}^{-1})$ in all episodes will be saved regardless of this setting in the file named "f.csv".
Another file named "total\_reward.csv" will also be saved to save the total rewards in all episodes when DDPG is chosen
as the optimization method. Here the word "episode" is referred to as a round of update of the objective function in
the scenario of optimization.

The switch of optimization algorithms can be realized by {\codefont method=" "}, and the corresponding
parameters can be set via {\codefont **kwargs}. All available algorithms in QuanEstimation are given in
Table~\ref{table:ctrl_paras} together with the corresponding default parameter settings. Notice that
in the case that {\codefont method="auto-GRAPE"} is applied, {\codefont dyn\_method="ode"} is not
available for now. In some algorithms maybe more than one set of guessed controls are needed, and if not
enough sets are input then random-value controls will be generated automatically to fit the number. In
the meantime, if excessive number of sets are input, only the suitable number of controls will be used.
{\codefont LDtype="SLD"} is the only choice when {\codefont method="GRAPE"} as the QFIMs based on RLD
and LLD are unavailable to be the objective function for GRAPE in the package. All the aforementioned
algorithms will be thoroughly introduced and discussed with examples in the following subsections.

%================================ Algorithm =================================
\begin{algorithm*}[tp]
%\SetAlgoNoLine
\SetArgSty{<texttt>}
\caption{GRAPE} \label{algorithm:grape}
Initialize the control amplitude $u_k(t)$ for all $t$ and $k$; \\
\For {episode=1, $M$}{
Receive initial state $\rho_{0}$ ($\rho_{\mathrm{in}}$); \\
\For {$t=1, T$}{
Evolve with the control $\rho_t=e^{\Delta t\mathcal{L}_t} \rho_{t-1}$; \\
Calculate the derivatives $\partial_\bold{x}\rho_t=-i\Delta t [\partial_\bold{x} H_0(\bold{x})]^{\times}\rho_t
+e^{\Delta t\mathcal{L}_t} \partial_\bold{x} \rho_{t-1}$; \\
Save $\rho_t$ and $\partial_\bold{x}\rho_t$; \\
\For {$k=1, K$}{
Calculate $\frac{\delta \rho_t}{\delta u_k(t)}=-i\Delta t H^{\times}_k\rho_t$,
$\partial_\bold{x}\!\left(\!\frac{\delta\rho_t}{\delta u_k(t)}\!\right)=
-i\Delta t H^{\times}_k(\partial_{\bold{x}}\rho_t)$; \\
\For {$j=t-1, 1$}{
Calculate $\frac{\delta \rho_t}{\delta u_k(j)}=e^{\Delta t\mathcal{L}_t}
\frac{\delta \rho_{t-1}}{\delta u_k(j)}$,
$\partial_\bold{x}\!\left(\frac{\delta\rho_t}{\delta u_k(j)}\right)=\left(\partial_\bold{x}
e^{\Delta t\mathcal{L}_t}\right)\frac{\delta\rho_{t-1}}{\delta u_k(j)}
+e^{\Delta t\mathcal{L}_t}\partial_\bold{x}\!\left(\frac{\delta \rho_{t-1}}
{\delta u_k(j)}\right)$;}
Save $\frac{\delta \rho_t}{\delta u_k(t)}$, $\partial_\bold{x}\!\left(\frac{\delta\rho_t}
{\delta u_k(t)}\right)$
and all $\frac{\delta \rho_t}{\delta u_k(j)}$,
$\partial_\bold{x}\!\left(\frac{\delta\rho_t}{\delta u_k(j)}\right)$;
}}
Calculate the SLDs for all $\bold{x}$ and the objective function $f(T)$; \\
{\For {$t=1, T$}{
\For {$k=1, K$}{
Calculate the gradient $\frac{\delta f(T)}{\delta u_k(t)}$ with $\frac{\delta\rho_T}
{\delta u_k(t)}$ and $\partial_\bold{x}\!\left(\frac{\delta \rho_T}{\delta u_k(t)}\right)$; \\
Update control $u_k(t)\!\leftarrow\! u_k(t)\!+\!\epsilon\frac{\delta f(T)}{\delta u_k(t)}$.
}}}
}
Save the controls $\{u_k(t)\}$ and corresponding $f(T)$.
\end{algorithm*}
%============================================================================

Apart from the QFIM and CFIM, the HCRB can also be taken as the objective function in the case of multiparameter
estimation, which can be realized by calling {\codefont control.HCRB()}. Notice that auto-GRAPE and GRAPE are
not available in {\codefont method=" "} here as the calculation of HCRB is performed via optimizations (semidefinite
programming), not direct calculations. Due to the equivalence between the HCRB and quantum Cram\'{e}r-Rao bound in
the single-parameter estimation, if {\codefont control.HCRB()} is called in this case, the entire program will
be terminated and a line of reminder will arise to remind the users to invoke {\codefont control.QFIM()} instead.

\subsection{Gradient ascent pulse engineering}

The gradient ascent pulse engineering algorithm (GRAPE) was developed by Khaneja et al.~\cite{Khaneja2005}
in 2005 for the design of pulse sequences in the Nuclear Magnetic Resonance systems, and then applied into the
quantum parameter estimation for the generation of optimal controls~\cite{Liu2017a,Liu2017b}, in which the
gradients of the objective function $f(T)$ at a fixed time $T$ were obtained analytically. In the pseudocode given
in Ref.~\cite{Liu2022}, the propagators between any two time points have to be saved, which would occupy a large
amount of memory during the computation and make it difficult to deal with high-dimensional Hamiltonians or long-time
evolutions. To solve this problem, a modified pseudocode is provided as given in Algorithm~\ref{algorithm:grape}.
In this modified version, after obtaining the evolved state $\rho_t$ and $\partial_{\bold{x}}\rho_t$, the gradient
$\delta\rho_t/\delta u_k(t)$ and its derivatives with respect to $\bold{x}$ are then calculated via the equations
\begin{equation}
\frac{\delta\rho_t}{\delta u_k(t)}=-i\Delta t H^{\times}_k(\rho_t)
\end{equation}
with $\Delta t$ a small time interval, $H^{\times}_k(
\cdot)=[H_k,\cdot]$ the commutator between $H_{k}$ and other
operators, and
\begin{equation}
\partial_\bold{x}\!\left(\frac{\delta\rho_t}{\delta u_k(t)}\!\right)
=-i\Delta t H^{\times}_k(\partial_\bold{x}\rho_t),
\end{equation}
The gradients $\delta\rho_t/\delta u_k(j)$ ($j<t$) are calculated adaptively according to the equation
\begin{equation}
\frac{\delta \rho_t}{\delta u_k(j)}=e^{\Delta t\mathcal{L}_t}\frac{\delta \rho_{t-1}}{\delta u_k(j)}
\end{equation}
and its derivatives are obtained via
\begin{equation}
\partial_\bold{x}\!\left(\frac{\delta\rho_t}{\delta u_k(j)}\right)\!=\!\left(\partial_\bold{x}
e^{\Delta t\mathcal{L}_t}\right)\!\frac{\delta\rho_{t-1}}{\delta u_k(j)}\!+\!e^{\Delta t\mathcal{L}_t}
\partial_\bold{x}\!\left(\frac{\delta\rho_{t-1}}{\delta u_k(j)}\right)\!.
\end{equation}
In this process, only the gradients $\delta\rho_t/\delta u_k(t)$, $\delta\rho_t/\delta u_k(j)$ and their
derivatives need to be saved for the further use in the next round, and will be deleted after that.
This operation avoids the usage of propagators and thus saves a lot of memory during the computation.
In this way, the gradients $\delta\rho_T/\delta u_k(t)$ ($t$ is any time here) are obtained adaptively and
$\delta f(T)/\delta u_k(t)$ can then be calculated accordingly. The specific expression of
$\delta f(T)/\delta u_k(t)$ can be found in Refs.~\cite{Liu2017a,Liu2017b,Liu2020}. Adam~\cite{Kingma2014}
is also applied in this algorithm for the further improvement of the computational efficiency.

Consider the dynamics governed by Eq.~(\ref{eq:ME_spon}) with parameter settings in
Fig.~\ref{fig:QFI_code}, and the control Hamiltonian
\begin{equation}
u_1(t)\sigma_1+u_2(t)\sigma_2+u_3(t)\sigma_3.
\label{eq:ctrl_demo}
\end{equation}
In the case of $\omega_{\mathrm{tr}}T=5$ with zeros as the initial guess of the controls, the QFI converges
in 35 rounds with Adam, yet it takes 780 rounds to converge when Adam is not applied.

\subsection{Auto-GRAPE}

In the multiparameter estimation, the gradients of $\mathrm{Tr}(W\mathcal{F}^{-1})$ and
$\mathrm{Tr}(W\mathcal{I}^{-1})$ are very difficult to obtain analytically when the length of $\bold{x}$ is large.
This is because the analytical calculation of $\mathcal{F}^{-1}$ and $\mathcal{I}^{-1}$ are difficult, if not
completely impossible. Hence, the functions $\sum_a W_{aa}/\mathcal{F}_{aa}$ and $\sum_a W_{aa}/\mathcal{I}_{aa}$,
which are lower bounds of $\mathrm{Tr}(W\mathcal{F}^{-1})$ and $\mathrm{Tr}(W\mathcal{I}^{-1})$, or their inverse
functions, are taken as the superseded objective functions in GRAPE~\cite{Liu2017b}. Although it has been proved
that these superseded functions show positive performance on the generation of controls, it is still possible
that the direct use of $\mathrm{Tr}(W\mathcal{F}^{-1})$ and $\mathrm{Tr}(W\mathcal{I}^{-1})$ might bring better
results. To investigate it, hereby we provide a new GRAPE algorithm based on the automatic differentiation
technology. This algorithm is referred to as the auto-GRAPE in this paper.

Automatic differentiation (AD) is an emerging numerical technology in machine learning to evaluate the exact
derivatives of an objective function~\cite{Baydin2018}. Recently, Song et al.~\cite{Song2022} used AD to
generate controls with complex and optional constraints. AD decomposes the calculation of the objective
function into some basic arithmetic and apply the chain rules to calculate the derivatives. AD not only provides
high precision results of the derivatives, but its computing complexity is no more than the calculation
of the objective function. Hence, it would be very useful to evaluate the gradient in GRAPE. Here we use a
Julia package Zygote~\cite{Innes2019} to implement our auto-GRAPE algorithm. auto-GRAPE is available for all
types of QFIM in the package. In the following we only discuss the SLD-based QFIM for simplicity. The pseudocode
of auto-GRAPE for SLD-based QFIM is given in Algorithm~\ref{algorithm:autogrape}. In Zygote, "array mutation"
operations should be avoided as the evaluation of differentiation for such operations are not supported currently.
However, the general numerical calculation of the SLD is finished via the calculation of its each entry, as shown
in Eq.~(\ref{eq:SLD_eigen}). This entry-by-entry calculation would inevitably cause mutations of the array, and
cannot directly apply automatic differentiation for now. Here we introduce two methods to realize AD in our
package. One method is to avoid the entry-by-entry calculation directly. Luckily, \v{S}afr\'{a}nek provided
a method for the calculation of SLD and QFIM in the Liouville space~\cite{Safranek2018}, which just avoids
the entry-by-entry calculation. Denote $\mathrm{vec}(A)$ as the column vector with respect to a $d$-dimensional
matrix $A$ in Liouville space and $\mathrm{vec}(A)^{\dagger}$ as the conjugate transpose of $\mathrm{vec}(A)$.
The entry of $A$ is defined by $[\mathrm{vec}(A)]_{id+j}:=A_{ij}$ ($i,j\in[0,d-1]$ and the subscript of
$\mathrm{vec}(A)$ starts from 0). Then the SLD can be calculated via the equation~\cite{Safranek2018}
\begin{equation}
\mathrm{vec}(L_a)=2(\rho\otimes\openone+\openone\otimes\rho^{*})^{-1}\mathrm{vec}(\partial_a\rho),
\label{eq:SLD_auto}
\end{equation}
where $\rho^{*}$ is the conjugate of $\rho$. The above calculation procedure treats the array as an entirety and
only contains basic linear algebra operations on the array. Hence, it is available to be used for the implementation
of automatic differentiation with the existing tools like Zygote. In QuanEstimation, the vectorization of matrices
are performed with the aforementioned method, i.e., $[\mathrm{vec}(A)]_{id+j}:=A_{ij}$, in all Python scripts, yet
in the Julia scripts, the vectorization is taken as $[\mathrm{vec}(A)]_{i+jd}:=A_{ij}$ for the calculation convenience.
Since all the outputs are converted back to the matrices, this difference would not affect the user experience. This
method is easy to be implemented in coding, yet the dimension growth of the calculation in the Liouville space would
significantly affect the computing efficiency and memory occupation. Therefore, we introduce the second method as
follows.

%================================ Algorithm =================================
\begin{algorithm}[tp]
%\SetAlgoNoLine
\SetArgSty{<texttt>}
\caption{auto-GRAPE} \label{algorithm:autogrape}
Initialize the control amplitude $u_k(t)$ for all $t$ and $k$; \\
\For {episode=1, $M$}{
Receive initial state $\rho_{0}$ ($\rho_{\mathrm{in}}$); \\
\For {$t=1, T$}{
Evolve with the control $\rho_t=e^{\Delta t\mathcal{L}_t} \rho_{t-1}$; \\
Calculate the derivatives $\partial_\bold{x} \rho_t=-i\Delta t [\partial_\bold{x} H_0(\bold{x})]^{\times}\rho_t
+e^{\Delta t\mathcal{L}_t} \partial_\bold{x} \rho_{t-1}$; \\
Save $\rho_t$ and $\partial_\bold{x} \rho_t$;\\
}
Calculate the SLD and objective function $f(T)$. \\
Calculate the gradient $\frac{\delta f(T)}{\delta u_k(t)}$ with the automatic differentiation method
for all $t$ and $k$.\\
{\For {$t=1, T$}{
\For {$k=1, K$}{
Update control $u_k(t)\!\leftarrow\! u_k(t)\!+\!\epsilon\frac{\delta f(T)}{\delta u_k(t)}$.
}}}
}
Save the controls $\{u_k(t)\}$ and corresponding $f(T)$.
\end{algorithm}
%============================================================================

%================================ Figure ============================================
\begin{figure}[bp]
\centering\includegraphics[width=8.5cm]{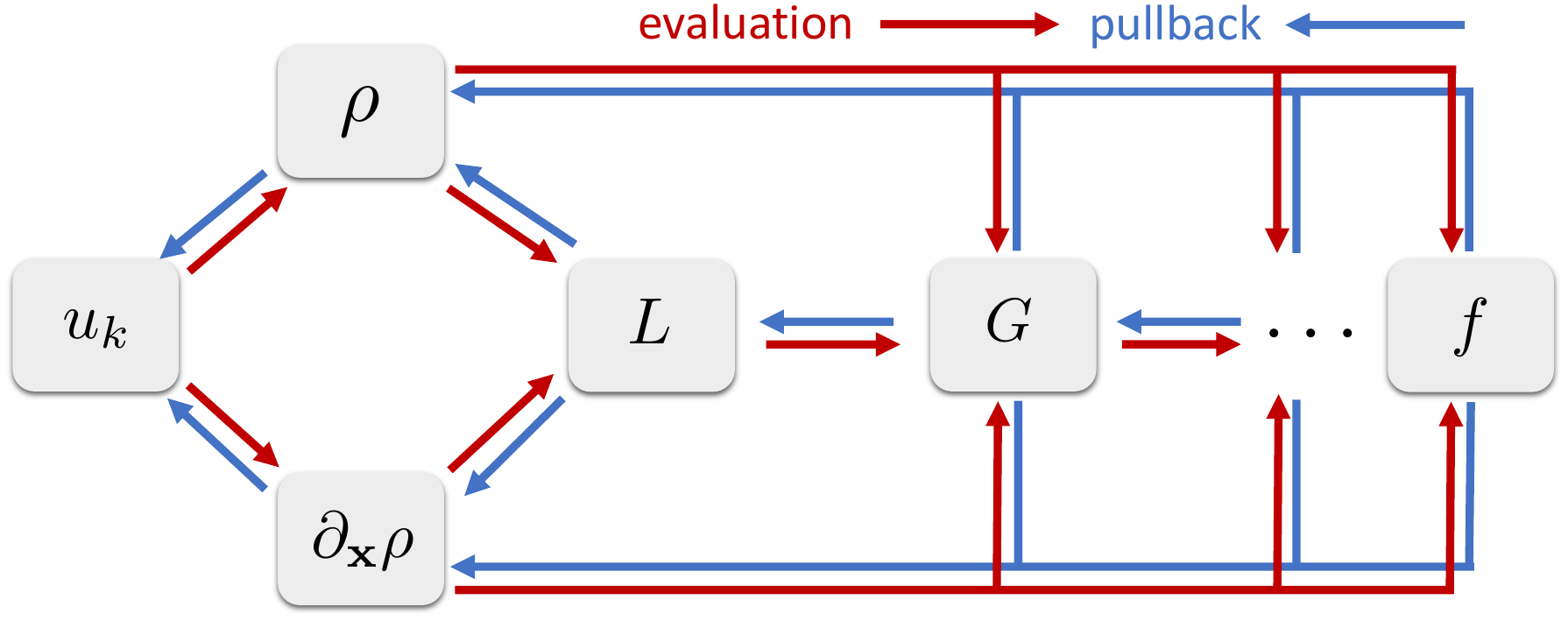}
\caption{The schematic of chain rules in automatic differentiation with the logarithmic
derivative related functions as the objective function.}
\label{fig:AD_illus}
\end{figure}
%====================================================================================

%================================ Table =====================================
\begin{table}[tp]
\def\arraystretch{1.15}
\begin{tabular}{c|c|c|c|c}
\hline
\hline
\multirow{3}{*}{$N$} & \multicolumn{2}{c|}{M1} & \multicolumn{2}{c}{M2}\\
\cline{2-5}
& ~~Computing~~& ~~~~Memory~~~~& ~~Computing~~ &~~~~Memory~~~~\\
&  time  &  allocation  &  time  &   allocation \\
\hline
$2$      &  4.46\,$\mu$s   &  2.99\,KB   & 5.14\,$\mu$s   &  2.24\,KB   \\
$2^{2}$  &  18.09\,$\mu$s  &  17.01\,KB  & 11.17\,$\mu$s  &  5.46 \,KB  \\
$2^{3}$  &  257.65\,$\mu$s &  217.63\,KB & 35.84\,$\mu$s  &  18.79\,KB  \\
$2^{4}$  &  4.55\,ms       &  3.34\,MB   & 151.51\,$\mu$s &  90.18\,KB  \\
$2^{5}$  &  174.61\,ms     &  53.01\,MB  & 962.17\,$\mu$s &  501.85\,KB \\
$2^{6}$  &  9.45\,s        &  846.18\,MB & 11.05\,ms      &  3.31 \,MB  \\
$2^{7}$  &  6151.51\,s     &  137.95\,GB & 45.70\,ms      &  230.98\,MB \\
$2^{8}$  &  -              &  -          & 347.50\,ms     &  1.73\,GB   \\
$2^{9}$  &  -              &  -          & 3.29\,s        &  13.36\,GB  \\
$2^{10}$ &  -              &  -          & 41.51\,s       &  105.08\,GB \\
\end{tabular}
\begin{ruledtabular}
\begin{tabular}{ccccccc}
$\omega T$ & 5 & 10 & 15 & 20 & 30 & 40\\\specialrule{0.05em}{0pt}{3pt}
GRAPE & 5.23\,s & 21.75\,s & 44.95\,s & 71.00\,s &178.56\,s  &373.89\,s \\
auto-GRAPE & 0.32\,s & 0.77\,s & 1.45\,s & 2.19\,s & 4.14\,s & 7.00\,s \\
\end{tabular}
\end{ruledtabular}
\caption{Upper table: Comparison of the average computing time and memory
allocation for the calculation of the gradient of QFI between two realization
methods of AD. M1 and M2 represent the first and second methods. $N$ is the
dimension of the density matrix. The density matrix and its derivative are
generated randomly in the test. Lower table: Comparison of the average
computing time per episode between GRAPE and auto-GRAPE with different target
time $T$. Parallel computing is not applied here. KB, MB, and GB represent
Kilobyte, Megabyte, and Gigabyte, respectively.}
\label{table:auto}
\end{table}
%============================================================================

The core of AD is utilizing the chain rules to evaluate the derivatives of the objective function. As illustrated in
Fig.~\ref{fig:AD_illus}, in AD the value of the objective function $f$ is evaluated from left to right (red arrows),
and the derivatives are calculated backwards (blue arrows), which is also called pullback in the language of
AD. In our case, the differentiation of $f$ on a control amplitude $u_k$ needs to be evaluated through all three
paths, from $f$ to $\rho$, from $f$ to $\partial_{\bold{x}}\rho$ (if $f$ is a function of $\partial_{\bold{x}}\rho$)
and from $f$ to $G$ to $L$. Here $L$ represents the SLDs of all parameters and $G:=G(L)=G(\rho,\partial_{\bold{x}}\rho)$
could be any intermediate function. For example, the contribution of the path from $f$ to $\rho$ to the derivative
$\mathrm{d}f/\mathrm{d}u_k$ is $\frac{\partial f}{\partial \rho}\frac{\partial \rho}{\partial u_k}$. Notice that
here $\partial f/\partial \rho$ is a formal derivative. The paths to $\rho$ and $\partial_{\bold{x}}\rho$ can be
routinely solved in Zygote, however, the path to $L$ cannot be solved due to the entry-by-entry calculation of
SLD in Eq.~(\ref{eq:SLD_eigen}), which causes the difficulty to generate $\partial L/\partial \rho$
and $\partial L/\partial(\partial_{\bold{x}}\rho)$, and therefore $\partial G/\partial\rho$
and $\partial G/\partial(\partial_{\bold{x}}\rho)$ cannot be obtained. The chain rules in AD cannot be applied then.
Hence, we need to manually provide $\partial G/\partial \rho$ and $\partial G/\partial (\partial_{\bold{x}}\rho)$ to
let AD work in our case. To do it, one should first know that the total differentiation $\mathrm{d}G_{\alpha\beta}$
(the $\alpha\beta$th entry of $\mathrm{d}G$) can be evaluated via the equation
\begin{equation}
\mathrm{d}G_{\alpha\beta}=\sum_{ij}\frac{\partial G_{\alpha\beta}}{\partial L_{ij}}\mathrm{d} L_{ij}
+\frac{\partial G_{\alpha\beta}}{\partial (L_{ij})^{*}}\mathrm{d} (L_{ij})^{*},
\end{equation}
which can be written into a more compact matrix form
\begin{equation}
\mathrm{d}G_{\alpha\beta}=\mathrm{Tr}\!\left(\left(\frac{\partial G_{\alpha\beta}}
{\partial L}\right)^{\mathrm{T}}\mathrm{d}L+\left(\frac{\partial G_{\alpha\beta}}
{\partial L^{*}}\right)^{\mathrm{T}}\mathrm{d}L^{*}\right).
\end{equation}
Due to the fact that the SLD is a Hermitian matrix, one can have $dL^{*}=dL^{\mathrm{T}}$, and the equation above
reduces to
\begin{align}
\mathrm{d}G_{\alpha\beta}&=\mathrm{Tr}\!\left(\left(\frac{\partial G_{\alpha\beta}}
{\partial L}\right)^{\mathrm{T}}\mathrm{d}L+\frac{\partial G_{\alpha\beta}}
{\partial L^{\mathrm{T}}}\mathrm{d}L\right) \nonumber \\
&= 2\mathrm{Tr}\!\left(\left(\frac{\partial G_{\alpha\beta}}{\partial L}\right)^{\mathrm{T}}\mathrm{d}L\right).
\label{eq:dG}
\end{align}
Now we introduce an auxiliary function $h$ which satisfies
\begin{equation}
\left(\frac{\partial G_{\alpha\beta}}{\partial L}\right)^{\mathrm{T}}=\rho h^{\mathrm{T}}+h^{\mathrm{T}}\rho.
\end{equation}
This equation is a typical Lyapunov equation and can be numerically solved. Substituting the equation above into
the expression of $\mathrm{d}G_{\alpha\beta}$, one can find that
\begin{equation}
\mathrm{d}G_{\alpha\beta}=2\mathrm{Tr}\left(h^{\mathrm{T}}\mathrm{d}L\rho+h^{\mathrm{T}}\rho\mathrm{d}L\right).
\end{equation}
Due to the fact that $\partial_{\bold{x}}\rho=(\rho L+L\rho)/2$, we have
$\rho\mathrm{d}L+(\mathrm{d}L)\rho=2\mathrm{d}(\partial_{\bold{x}}\rho)-(\mathrm{d}\rho) L-L\mathrm{d}\rho$,
which means
\begin{equation}
\mathrm{d}G_{\alpha\beta}=2\mathrm{Tr}\!\left(2h^{\mathrm{T}}\mathrm{d}(\partial_{\bold{x}}\rho)\right)
-2\mathrm{Tr}\!\left(\left(Lh^{\mathrm{T}}+h^{\mathrm{T}}L\right)\mathrm{d}\rho\right).
\label{eq:dG_h}
\end{equation}
Next, since $G=G(\rho,\partial_{\bold{x}}\rho)$, $\mathrm{d}G_{\alpha\beta}$ can also be expressed by
\begin{equation}
\mathrm{d}G_{\alpha\beta}=2\mathrm{Tr}\!\left(\!\left(
\frac{\partial G_{\alpha\beta}}{\partial\rho}\right)^{\!\mathrm{T}}
\!\mathrm{d}\rho\!+\!\left(\frac{\partial G_{\alpha\beta}}
{\partial(\partial_{\bold{x}}\rho)}\right)^{\!\mathrm{T}}
\!\mathrm{d}(\partial_{\bold{x}}\rho)\!\right).
\end{equation}
This equation is derived through a similar calculation procedure for Eq.~(\ref{eq:dG}). Comparing this equation
with Eq.~(\ref{eq:dG_h}), one can see that
\begin{align}
\frac{\partial G_{\alpha\beta}}{\partial\rho}&=2h, \\
\frac{\partial G_{\alpha\beta}}{\partial (\partial_{\bold{x}}\rho)} &=-hL^{\mathrm{T}}-L^{\mathrm{T}}h.
\end{align}
With these expressions, $\partial G/\partial\rho$ and $\partial G/\partial(\partial_{\bold{x}}\rho)$ can be obtained
correspondingly. In this way, the entire path from $f$ to $L$ is connected. Together with the other two paths, AD can
be fully applied in our case. The performance of computing time and memory allocation for the calculation of the
gradient of QFI between these two realization methods of AD are compared with different dimensional density matrices.
The dimension is denoted by $N$. As shown in the upper table in Table~\ref{table:auto}, the computing time and memory
allocation of the second method are better than the first one except for the case of $N=2$, and this advantage becomes
very significant when $N$ is large. Moreover, the computing time and memory allocation of the first method grow fast
with the increase of dimension, which is reasonable as the calculations, especially the diagonalization, in the first
method are performed in the $N^2$-dimensional space. There is no data of the first method when $N$ is larger than 7
as the memory occupation has exceeded our computer's memory. From this comparison, one can see that the second method
performs better than the first one in basically all aspects and hence is chosen as the default auto-GRAPE method in
QuanEstimation.

\begin{figure*}[tp]
\centering\includegraphics[width=17.5cm]{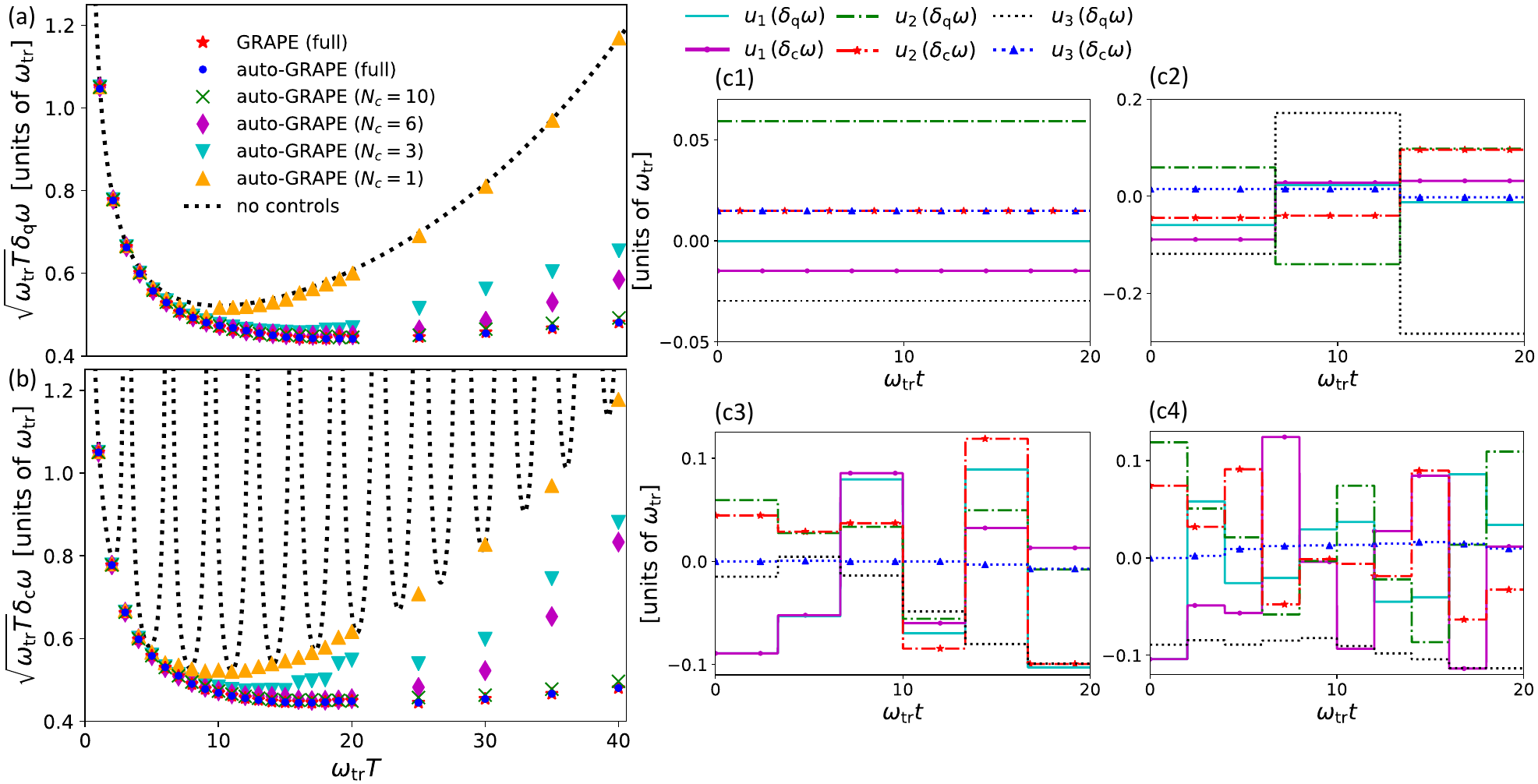}
\caption{The performance of control-enhanced (a) $\sqrt{\omega_{\mathrm{tr}}T}\delta_{\mathrm{q}}\omega$
and (b) $\sqrt{\omega_{\mathrm{tr}}T}\delta_{\mathrm{c}}\omega$ with different $N_{\mathrm{c}}$. The
optimal controls are generated via GRAPE and auto-GRAPE with the dynamics in Eq.~(\ref{eq:ME_spon}) and
control Hamiltonian in Eq.~(\ref{eq:ctrl_demo}). The dotted black lines in (a) and (b) represent
$\sqrt{\omega_{\mathrm{tr}}T}\delta_{\mathrm{q}}\omega$ and $\sqrt{\omega_{\mathrm{tr}}T}\delta_{\mathrm{c}}\omega$
without control. The red pentagrams are those obtained via GRAPE with a full $N_{\mathrm{c}}$, i.e., $N_{\mathrm{c}}$
equals to the number of time steps. The blue circles, green crosses, purple diamonds, cyan downward triangles and
orange upward triangles represent those obtained via auto-GRAPE with $N_{\mathrm{c}}$ being full, 10, 6, 3 and 1,
respectively. Other parameters are set to be the same with those in Fig.~\ref{fig:QFI_code}. (c1-c4) The optimal
controls in the case of $\omega_{\mathrm{tr}}T=20$ with $N_{\mathrm{c}}$ being (c1) 1, (c2) 3, (c3) 6 and (c4) 10,
respectively. The true value $\omega_{\mathrm{tr}}$ is set to be $1$. Planck units are applied here.}
\label{fig:SPara_ctrl}
\end{figure*}

\emph{Example.} Consider the dynamics in Eq.~(\ref{eq:ME_spon}) and control Hamiltonian in Eq.~(\ref{eq:ctrl_demo}).
Now define
\begin{eqnarray}
\delta_{\mathrm{c}}\omega &:=& 1/\sqrt{\mathcal{I}_{\omega\omega}}, \label{eq:c_deviation} \\
\delta_{\mathrm{q}}\omega &:=& 1/\sqrt{\mathcal{F}_{\omega\omega}} \label{eq:q_deviation}
\end{eqnarray}
as the theoretical optimal deviations with and without fixed measurement. The corresponding performance of controls generated via
GRAPE and auto-GRAPE are shown in Figs.~\ref{fig:SPara_ctrl}(a) and \ref{fig:SPara_ctrl}(b), which are obtained by
300 episodes in general. In QuanEstimation, the number of episodes can be set via the variable {\codefont max\_episode=300}
in {\codefont **kwargs} in Table~\ref{table:ctrl_paras}. As shown in these plots, the values of
$\sqrt{\omega_{\mathrm{tr}}T}\delta_{\mathrm{q}}\omega$ in (a) and $\sqrt{\omega_{\mathrm{tr}}T}\delta_{\mathrm{c}}\omega$
in (b) obtained via GRAPE (red pentagrams) and auto-GRAPE (blue circles) basically coincide with each other, which is
reasonable as they are intrinsically the same algorithm, just with different gradient calculation methods. However,
auto-GRAPE shows a significant improvement on the computing time consumption, as given in the lower table in
Table~\ref{table:auto}, especially for a large target time $T$. The growth of average computing time per episode with
the increase of $T$ in auto-GRAPE is quite insignificant compared to that in GRAPE. Adam can be applied by setting
{\codefont Adam=True} in {\codefont **kwargs}. For the sake of a good performance, one can set appropriate Adam parameters
in {\codefont **kwargs}, including the learning rate {\codefont epsilon}, the exponential decay rate for the first (second)
moment estimates {\codefont beta1} ({\codefont beta2}). The default values of these parameters in the package are 0.01 and
0.90 (0.99). If {\codefont Adam=False}, the controls are updated with the constant step {\codefont epsilon}. Due to the
convergence problem of Adam in some cases, several points in the figure are obtained by a second running of the code with
a constant step, which takes the optimal control obtained in the first round (with Adam) as the initial guess.

In some scenarios, the time resolution of the control amplitude could be limited if the dynamics is
too fast or the target time is too short. Hence, in the numerical optimization in such cases, the time steps
of control cannot equal to that of the dynamics. Here we use the total control amplitude number $N_{\mathrm{c}}
=T/\Delta t_{\mathrm{c}}$ with $\Delta t_{\mathrm{c}}$ the control time step, to represent the time resolution
of the control and we assume $\Delta t_{\mathrm{c}}$ is fixed in the dynamics. A full $N_{\mathrm{c}}$ in
Figs.~\ref{fig:SPara_ctrl}(a) and \ref{fig:SPara_ctrl}(b) means $\Delta t_{\mathrm{c}}$ equals to the dynamical
time step $\Delta t$. In the numerical calculation, it is possible that quotient of $\Delta t_{\mathrm{c}}$ by
$\Delta t$ is not an integer, indicating that the existing time of all control amplitudes cannot be equivalent.
To avoid this problem, in QuanEstimation the input number ($N_{t}$) of dynamical time steps is automatically
adjusted to $kN_{\mathrm{c}}$ with $k$ the smallest integer to let $kN_{\mathrm{c}}>N_t$, if it is not already
an integer multiple of $N_{\mathrm{c}}$. For example, if $N_{\mathrm{c}}=3$ and $N_{\mathrm{t}}=100$,
then $N_{\mathrm{t}}$ is adjusted to 102. Notice that in the package GRAPE is not available to deal with a non-full
$N_{\mathrm{c}}$ scenario for a technical reason. If GRAPE is invoked in this case, it would automatically go
back to auto-GRAPE. As a matter of fact, auto-GRAPE outperforms GRAPE in most aspects, therefore, we strongly
suggest the users choose auto-GRAPE, instead of GRAPE, in practice.

The performance of controls with limited $N_{\mathrm{c}}$ is also demonstrated in Figs.~\ref{fig:SPara_ctrl}(a)
and \ref{fig:SPara_ctrl}(b) with the dynamics in Eq.~(\ref{eq:ME_spon}) and control Hamiltonian in
Eq.~(\ref{eq:ctrl_demo}). It can be seen that the constant-value controls ($N_{\mathrm{c}}=1$, orange upward
triangles) cannot reduce the values of $\delta_{\mathrm{c}}\omega$ and $\delta_{\mathrm{q}}\omega$. In the case
of fixed measurement it can only suppress the oscillation of $\delta_{\mathrm{c}}\omega$. The performance improves
with the increase of $N_{\mathrm{c}}$ and when $N_{\mathrm{c}}=10$, the values of $\delta_{\mathrm{q}}\omega$ and
$\delta_{\mathrm{c}}\omega$ are very close to those with a full $N_{\mathrm{c}}$. This fact indicates that inputting
10 control amplitudes is good enough in this case and a full $N_{\mathrm{c}}$ control is unnecessary. A limited
$N_{\mathrm{c}}$ here could be easier to realize in practice and hence benefit the experimental realization.

%================================ Figure ====================================
%============================================================================

\subsection{Particle swarm optimization}

%================================ Figure ====================================
\begin{figure*}[tp]
\centering\includegraphics[width=17.5cm]{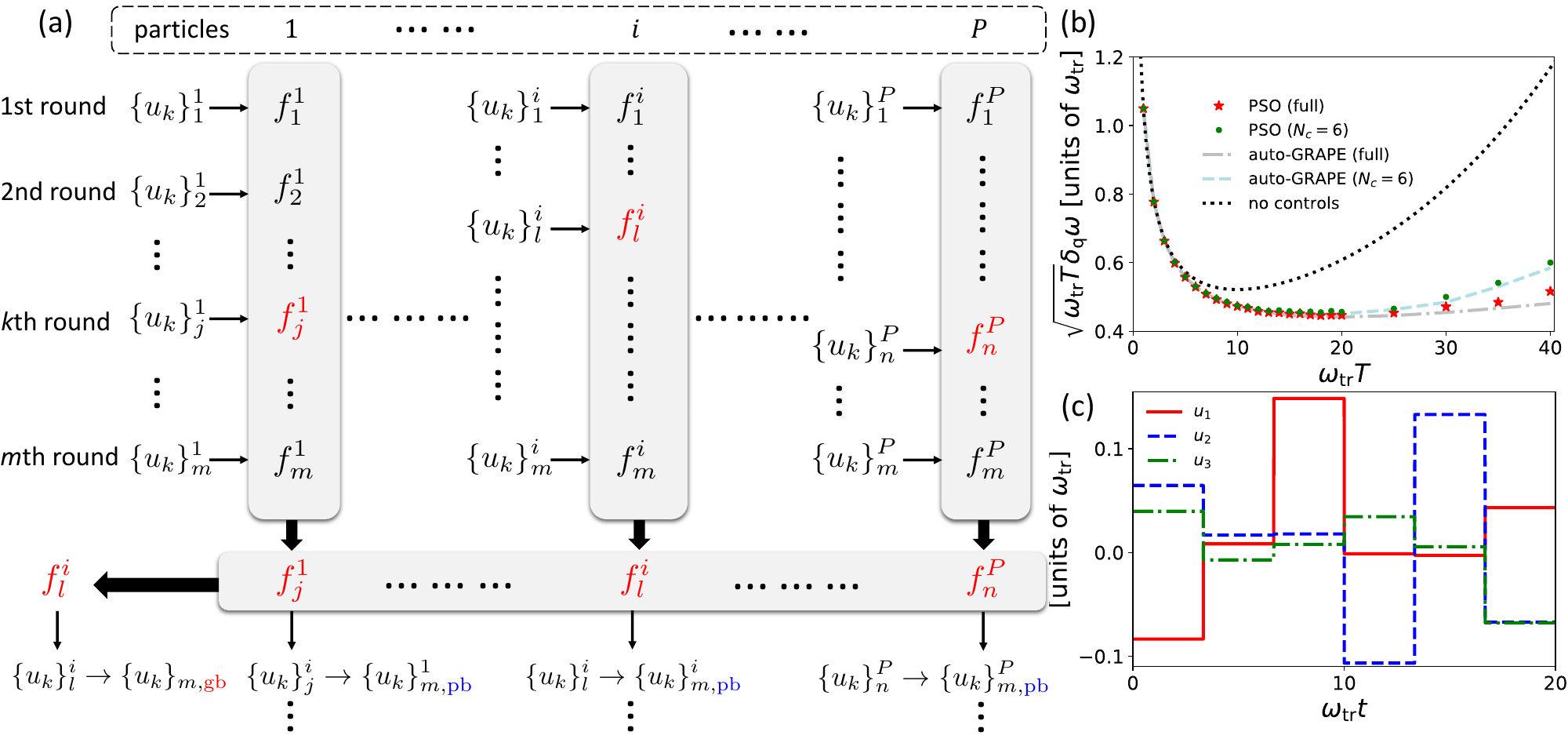}
\caption{(a) Illustration of the basic operation of PSO in $m$th round of episode.
The personal best (with the blue subscript pb) for each particle in this
round is obtained by comparing all the values of $f$ of this particle in
all previous rounds including the current one. The global best (with the red
subscript gb) is obtained by comparing the values of $f$ of all personal bests in
this round. The light gray areas represent the process of comparison, which takes
the values of $\{u_k\}$ with respect to the maximum value of $f$. (b) The
control-enhanced values of $\sqrt{\omega_{\mathrm{tr}}T}\delta_{\mathrm{q}}\omega$
with a full $N_{\mathrm{c}}$ (red pentagrams) and $N_{\mathrm{c}}=6$ (green circles),
where the controls are generated via PSO. (c) The optimal controls for $N_{\mathrm{c}}=6$
in the case of $\omega_{\mathrm{tr}}T=20$. The true value $\omega_{\mathrm{tr}}$ is set
to be $1$. Planck units are applied here.}
\label{fig:PSO}
\end{figure*}
%============================================================================

%================================ Algorithm =================================
\begin{algorithm*}[tp]
%\SetAlgoNoLine
\SetArgSty{<texttt>}
\caption{PSO} \label{algorithm:pso}
Initialize the control $\left\{u_k\right\}^i_1$ for each $i \in \left[1,P\right]$; \\
Initialize $\left\{\delta u_k\right\}^i_1=0$ for each $i \in \left[1,P\right]$;\\
Assign $f(\left\{u_k\right\}^i_{0,\mathrm{pb}})=0$ for each $i \in \left[1,P\right]$; \\
\For {$m=1, M$\ \do}{
\For {$i=1, P$\ \do}{
Receive the control $\left\{u_k\right\}^i_m$;\\
Evolve the state with $\left\{u_k\right\}^i_m$ and calculate the  objective function
$f(\left\{u_k\right\}^i_m)$ at the target time $T$; \\
Compare $f(\left\{u_k\right\}^i_m)$ with value of the personal best in last episode
$f(\left\{u_k\right\}^i_{m-1,\mathrm{pb}})$ and assign the new personal best
$\left\{u_k\right\}^i_{m,\mathrm{pb}}=\mathrm{arg}
\left(\max\left\{f(\left\{u_k\right\}^i_{m-1,\mathrm{pb}}),
f(\left\{u_k\right\}^i_m)\right\}\right)$;}
Compare all $f(\left\{u_k\right\}^i_{m,\mathrm{pb}})$ with $i\in[1,P]$ and assign the global best
$\left\{u_k\right\}_{m, \mathrm{gb}}=\mathrm{arg}\left(\max\limits_{i\in\left[1,P\right]}
f(\left\{u_k\right\}^i_{m, \mathrm{pb}})\right)$;\\
\For {$i=1, P$\ \do}
{Calculate $\left\{\delta u_k\right\}^i_m= c_0 \left\{\delta u_k\right\}^i_{m-1} +
\mathrm{rand}() \cdot c_1\big(\left\{u_k\right\}^i_{m, \mathrm{pb}}-\left\{u_k\right\}^i_m\big) +
\mathrm{rand}() \cdot c_2\big(\left\{u_k\right\}_{m,\mathrm{gb}}-\left\{u_k\right\}^i_m\big)$;\\
Update the control $\left\{u_k\right\}^i_{m+1} = \left\{u_k\right\}^i_m+\left\{\delta u_k\right\}^i_m$.
}}
Save the global best $\{u_k\}_{M,\mathrm{gb}}$ and corresponding $f$.
\end{algorithm*}
%============================================================================

Particle swarm optimization (PSO) is a well-used gradient-free method in
optimizations~\cite{Kennedy1995,Eberhart2001}, and has been applied in the detection
of gravitational waves~\cite{Michimura2018}, the characterization of open systems~\cite{Stenberg2016},
the prediction of crystal structure~\cite{Wang2010}, and in quantum metrology it has been used
to generate adaptive measurement schemes in phase estimations~\cite{Hentschel2010,Hentschel2011}.

A typical version of PSO includes a certain number (denoted by $P$) of parallel particles. In quantum
control, these particles are just $P$ sets of controls $\{u_k\}$ labelled by $\{u_k\}^i$ for $i=1,\dots,P$.
The value of $\{u_k\}$ of $i$th particle in $m$th round of episode is further denoted by $\{u_k\}^i_m$.
The basic optimization  philosophy of PSO is given in Fig.~\ref{fig:PSO}(a) and the pseudocode is given
in Algorithm~\ref{algorithm:pso}. In the pseudocode, $\left\{u_k\right\}^i_{0,\mathrm{pb}}$ and
$f(\left\{u_k\right\}^i_{0,\mathrm{pb}})$ are just formal notations representing the initialization
of the personal bests. There exist two basic concepts in PSO, the personal best and global best. In
the $m$th round of episode, the personal best of $i$th particle ($\{u_k\}^i_{m,\mathrm{pb}}$) is
assigned by the $\{u_k\}$ with respect to the maximum value of $f$ among all previous episodes of
this particle, namely,
\begin{equation}
\{u_k\}^i_{m,\mathrm{pb}}=\mathrm{arg}\left(\max \limits_{n\in\left[1,m\right]}
f(\left\{u_k\right\}^i_n)\right)
\end{equation}
with $\mathrm{arg}(\cdot)$ the argument. For example, as illustrated in Fig.~\ref{fig:PSO}, if $f^1_j$ is
the maximum in $\{f^1_1,f^1_2,\dots,f^1_m\}$, then $\{u_k\}^1_{m,\mathrm{pb}}$ is assigned by $\{u_k\}^1_j$.
Once the personal bests are obtained for all particles, the global best is assigned by the $\{u_k\}$ with
respect to the maximum value of $f$ among all personal bests, i.e.,
\begin{equation}
\{u_k\}_{m, \mathrm{gb}}=\mathrm{arg}\left(\max\limits_{i\in\left[1,P\right]}
f(\{u_k\}^i_{m,\mathrm{pb}})\right).
\end{equation}
With all personal bests and the global best, the velocity $\{\delta u_k\}^i_m$ for the $i$th particle is
calculated by
\begin{align}
\{\delta u_k\}^i_m =& c_0 \{\delta u_k\}^i_{m-1}
\!+\!\mathrm{rand}()\cdot c_1\left(\{u_k\}^i_{m,\mathrm{pb}}-\{u_k\}^i_m\right) \nonumber \\
& +\mathrm{rand}()\cdot c_2\left(\{u_k\}_{m,\mathrm{gb}}-\{u_k\}^i_m\right),
\end{align}
where rand() represents a random number within $[0,1]$ and $c_0$, $c_1$, $c_2$ are three positive
constant numbers. In the package, these parameters can be adjusted in {\codefont **kwargs}, shown in
Table~\ref{table:ctrl_paras}, via the variables {\codefont c0}, {\codefont c1} and {\codefont c2}. A
typical choice for these constants is $c_0=1$, $c_1=c_2=2$, which are also the default values in the package.
{\codefont max\_episode} in {\codefont **kwargs} represents the episode number to run. If it is only set to be
a number, for example {\codefont max\_episode=1000}, the program will continuously run 1000 episodes. However,
if it is a list, for example {\codefont max\_episode=[1000,100]}, the program will also run 1000 episodes
in total but replace $\{u_k\}$ of all particles with the current global best every 100 episodes.
{\codefont p\_num} represents the particle number and is set to be 10 in default. The initial guesses
of control can be input via {\codefont ctrl0} and the default choice {\codefont ctrl0=[]} means all the guesses
are randomly generated. In the case that the number of input guessed controls is less than the particle number,
the algorithm will generate the remaining ones randomly. On the other hand, if the number is larger than the
particle number, only the suitable number of controls will be used. The optimization result can be realized
repeatedly by fixing the value of the variable {\codefont seed}, and its default value is 1234 in the package.

%================================ Algorithm =================================
\begin{algorithm*}[tp]
%\SetAlgoNoLine
\SetArgSty{<texttt>}
\caption{DE}
Initialize the control $\{u_k\}^i$ for $i\in[1,P]$; \\
Evolve the state with $\{u_k\}^i$ and calculate the objective function $f(\{u_k\}^i)$
at the target time $T$ for $i\in[1,P]$; \\
\For {episode=1, $M$}{
\For {$i=1,P$}{
Randomly generate three integers $p_1$, $p_2$, $p_3$ in the regime $[1,P]$; \\
Generate $\{G_k\}$ via the equation $\{G_k\}=\{u_k\}^{p_1}+c(\{u_k\}^{p_2}-\{u_k\}^{p_3})$; \\
\For {$k=1, K$}{
Generate a random integer $a\in[1, N_{\mathrm{c}}]$; \\
\For {$j=1, N_{\mathrm{c}}$}{
Generate a random number $r\in[0,1]$ and assign
$ [Q_k]_j=
\begin{cases}
[G_k]_j, & {\mathrm{if}~r\leq c_r~\mathrm{or}~j=a}, \\
[u_k]_j, & {\mathrm{if}~r>c_r~\mathrm{and}~j\neq a};
\end{cases}$
}}
Evolve the state with the control $\{Q_k\}$ and calculate $f(\{Q_k\})$ at time $T$; \\
\If {$f(\{u_k\}^i)<f(\{Q_k\})$}{
Replace $\{u_k\}^i$ with $\{Q_k\}$.
}
}}
Compare all $f(\{u_k\}^i)$ and save $\{u_k\}^i$ (and corresponding $f$) with
respect to the largest $f$.
\label{algorithm:DE}
\end{algorithm*}
%===============================================================================

\emph{Example.} Here we also illustrate the performance of controls generated via PSO with the dynamics in
Eq.~(\ref{eq:ME_spon}) and control Hamiltonian in Eq.~(\ref{eq:ctrl_demo}). $\delta_{\mathrm{q}}\omega$ is defined
in Eq.~(\ref{eq:q_deviation}). The performance of controls with a full $N_{\mathrm{c}}$ (red pentagrams) and
$N_{\mathrm{c}}=6$ (green circles) are shown in Fig.~\ref{fig:PSO}(b), and the corresponding optimal controls for
$N_{\mathrm{c}}=6$ are given in Fig.~\ref{fig:PSO}(c). Compared to the result obtained via auto-GRAPE (dash-dotted
gray line for a full $N_{\mathrm{c}}$ and dashed light-blue line for $N_{\mathrm{c}}=6$), the performance of PSO
is worse than that of auto-GRAPE, especially in the case of a large target time $T$ with a full $N_{\mathrm{c}}$.
This is due to the fact that the search space is too large for PSO in such cases as the time step $\Delta t$ is
fixed in the calculation and a larger $T$ means a larger value of $N_{\mathrm{c}}$. For example, in the case of
$\omega_{\mathrm{tr}}T=40$ and $N_{\mathrm{c}}=10000$, the total parameter number in the optimization is 30000.
PSO can provide a good performance when the dimension of the search space is limited. In the case of $N_{\mathrm{c}}=6$,
the result of PSO basically coincides with that of auto-GRAPE. Hence, for those large systems that the calculation of
gradient is too time-consuming or the search space is limited, the gradient-free methods like PSO would show their powers.

\subsection{Differential evolution}

%================================ Figure ====================================
\begin{figure*}[tp]
\centering\includegraphics[width=14cm]{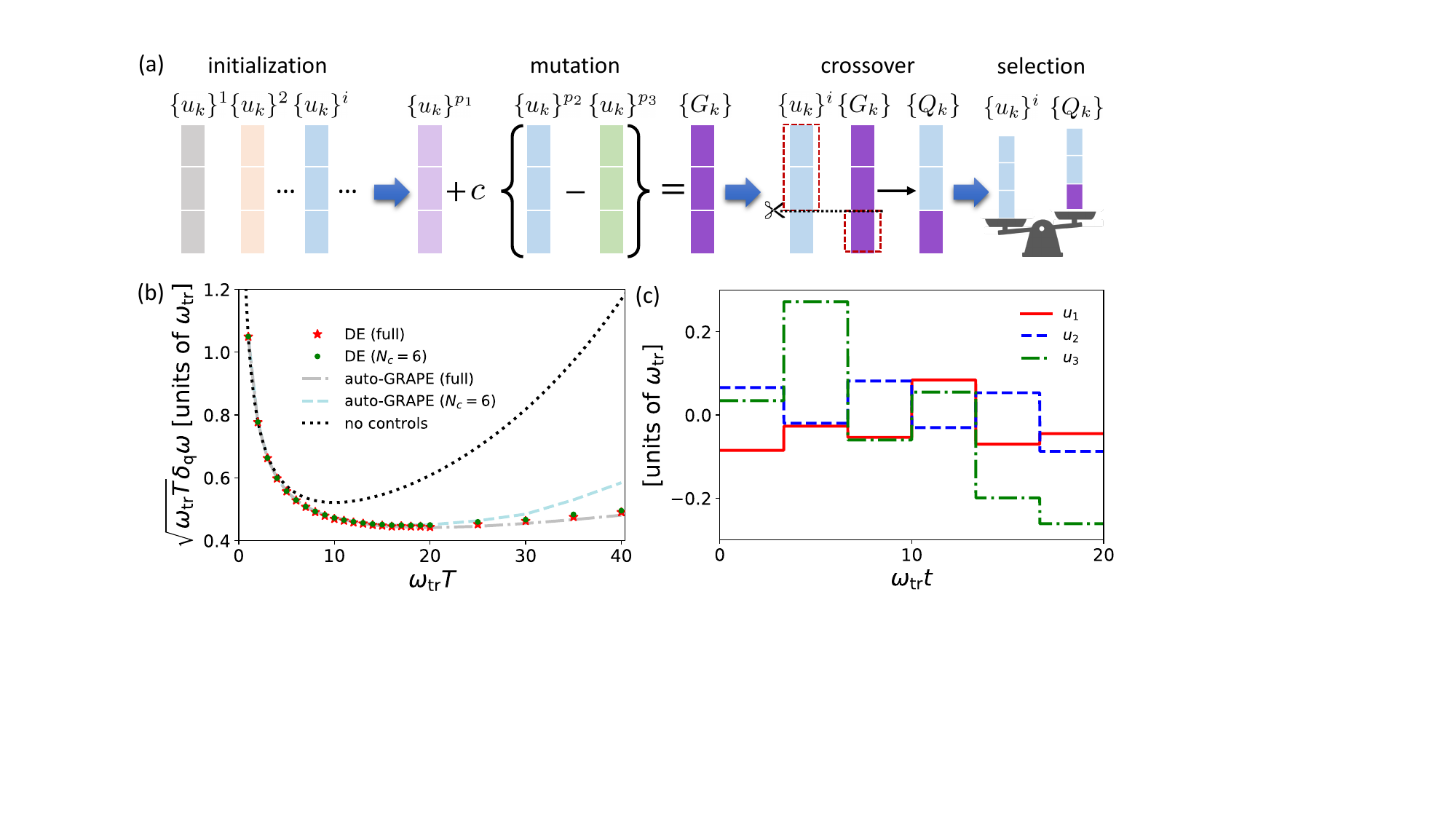}
\caption{(a) Illustration of the optimization process of DE, which includes four steps:
initialization, mutation, crossover and selection. (b) The control-enhanced values
of $\sqrt{\omega_{\mathrm{tr}}T}\delta_{\mathrm{q}}\omega$ with a full $N_{\mathrm{c}}$
(red pentagrams) and $N_{\mathrm{c}}=6$ (green circles), where the controls are generated
via DE. (c) The optimal controls for $N_{\mathrm{c}}=6$ in the case of $\omega_{\mathrm{tr}}T=20$.
The true value $\omega_{\mathrm{tr}}$ is set to be $1$. Planck units are applied here.}
\label{fig:DE}
\end{figure*}
%============================================================================

Differential evolution (DE) is another useful gradient-free algorithm in optimizations~\cite{Storn1997}.
It has been used to design adaptive measurements in quantum phase estimation~\cite{Lovett2013,Palittapongarnpim2017},
high-quality control pulses in quantum information~\cite{Yang2019,Yang2020}, and help to improve the
learning performance in quantum open systems~\cite{Ma2017,Dong2019}. Different with PSO, DE would not
converge prematurely in general and its diversification is also better since the best solution does not
affect other solutions in the population~\cite{Kachitvichyanukul2012}.

A typical DE includes a certain number (denoted by $P$) of $\{u_k\}^i$, which is usually referred to
as the populations in the language of DE. In QuanEstimation, the population number and the guessed
controls can be set via the variables {\codefont p\_num} and {\codefont ctrl0} in
{\codefont **kwargs}, as shown in Table~\ref{table:ctrl_paras}. The rule for the usage of
{\codefont ctrl0} here is the same as {\codefont ctrl0} in PSO. After the initialization
of all $\{u_k\}^i$ ($i\in[1,P]$), two important processes in DE, mutation and crossover, are performed,
as illustrated in Fig.~\ref{fig:DE}(a) with the pseudocode in Algorithm~\ref{algorithm:DE}. In the step
of mutation, three populations $\{u_k\}^{p_1}$, $\{u_k\}^{p_2}$ and $\{u_k\}^{p_3}$ are randomly picked
from all $\{u_k\}^i$, and used to generate a new population $\{G_k\}$ via the equation
\begin{equation}
\{G_k\}=\{u_k\}^{p_1}+c(\{u_k\}^{p_2}-\{u_k\}^{p_3})
\end{equation}
with $c\in[0,1]$ (or $[0,2]$) a constant number. The next step is the crossover. At the beginning of this step,
a random integer $a$ is generated in the regime $[1,N_{\mathrm{c}}]$, which is used to make sure the crossover
happens definitely. Then another new population $\{Q_k\}$ is generated for each $\{u_k\}^i$ utilizing $\{G_k\}$.
Now we take $j$th entry of $Q_k$ ($[Q_k]_j$) as an example to show the generation rule. In the first, a random
number $r$ is picked in the regime $[0,1]$. Then $[Q_k]_j$ is assigned via the equation
\begin{equation}
[Q_k]_j=\begin{cases}
[G_k]_j, & {\mathrm{if}~r\leq c_r~\mathrm{or}~j=a}, \\
[u_k]_j, & {\mathrm{if}~r>c_r~\mathrm{and}~j\neq a},
\end{cases}
\end{equation}
where $[G_k]_j$ is the $j$th entry of $G_k$ and $[u_k]_j$ is the $j$th entry of a $u_k$ in $\{u_k\}^i$. This
equation means if $r$ is no larger than a given constant $c_r$ (usually called crossover constant in DE),
then assign $[G_k]_j$ to $[Q_k]_j$, otherwise assign $[u_k]_j$ to $[Q_k]_j$. In the meantime, the $a$th
entry of $Q_k$ always takes the value of $[G_k]_j$ regardless the value of $r$ to make sure at least one
point mutates. After the crossover, the values of  objective functions $f(\{u_k\}^i)$ and $f(\{Q_k\})$ are compared,
and $\{u_k\}^i$ is replaced by $\{Q_k\}$ if $f(\{Q_k\})$ is larger. In the package, $c$ and $c_r$ can be adjusted
via the variables {\codefont c} and {\codefont cr} in {\codefont **kwargs}, and the default values are 1.0 and 0.5.

\emph{Example.} The performance of controls generated via DE is also illustrated with the dynamics in Eq.~(\ref{eq:ME_spon})
and control Hamiltonian in Eq.~(\ref{eq:ctrl_demo}). $\delta_{\mathrm{q}}\omega$ is defined in Eq.~(\ref{eq:q_deviation}).
As shown in Fig.~\ref{fig:DE}(b), different with PSO, the performance of DE with a full $N_{\mathrm{c}}$ (red pentagrams)
is very close to that of auto-GRAPE (dash-dotted gray line), even for a large target time $T$, which indicates that DE works
better than PSO in this example. More surprisingly, in the case of $N_{\mathrm{c}}=6$, DE (green circles) not only outperforms
PSO, but also significantly outperforms auto-GRAPE (dashed light-blue line). This result indicates that no algorithm has the
absolute advantage in general. Comparison and combination of different algorithms are a better approach to design optimal
controls in quantum metrology, which can be conveniently finished via QuanEstimation. The optimal controls obtained via DE
for $N_{\mathrm{c}}=6$ are given in Fig.~\ref{fig:DE}(c) in the case of $\omega_{\mathrm{tr}}T=20$. The results above are
obtained with 1000 episodes, which can be adjusted via {\codefont max\_episode=1000} in {\codefont **kwargs}.

\subsection{Deep Deterministic Policy Gradients}

%================================ Figure ====================================
\begin{figure}[bp]
\centering\includegraphics[width=8.5cm]{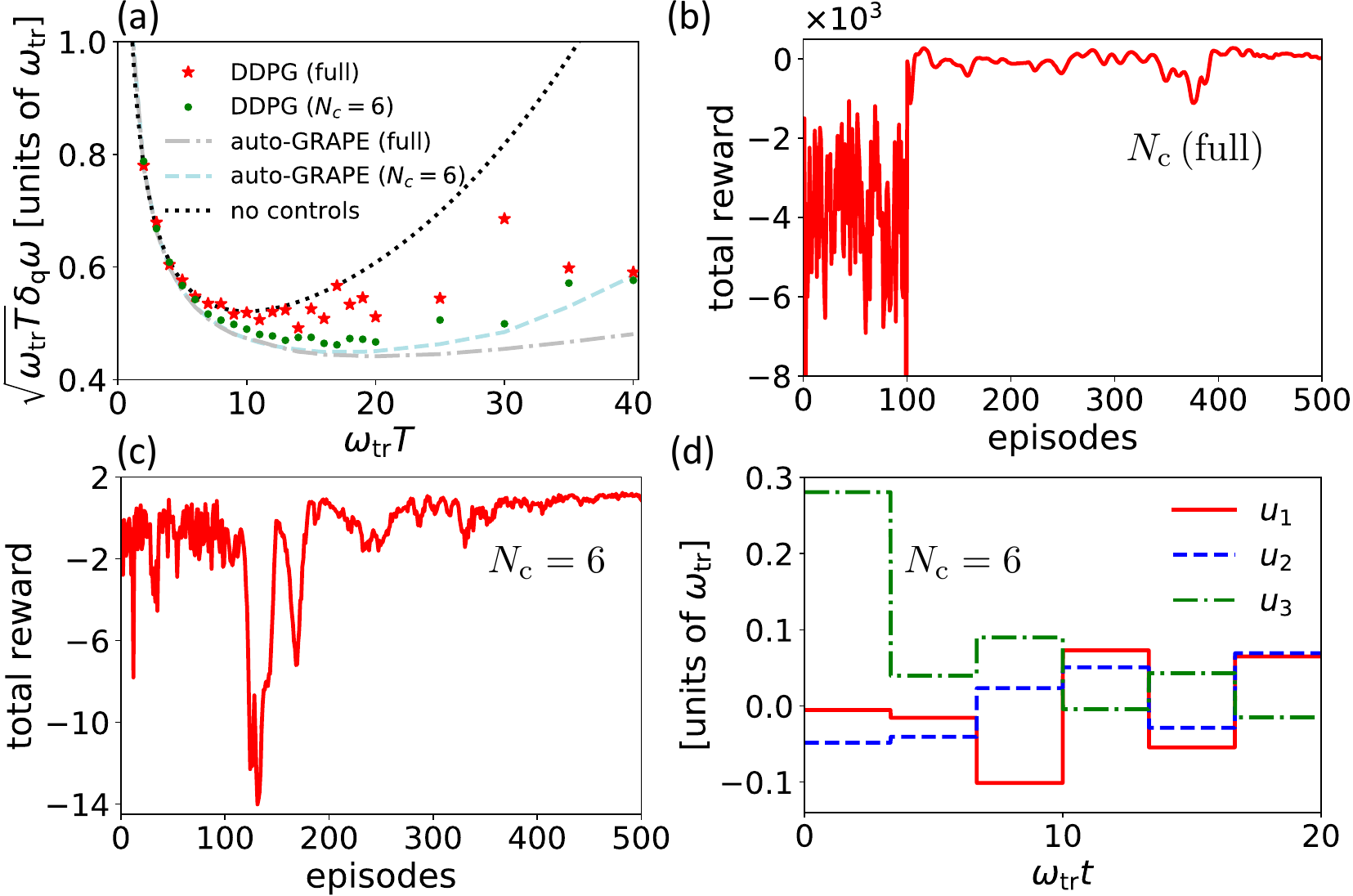}
\caption{(a) The control-enhanced values of $\sqrt{\omega_{\mathrm{tr}}T}\delta_{\mathrm{q}}\omega$
with a full $N_{\mathrm{c}}$ (red pentagrams) and $N_{\mathrm{c}}=6$ (green circles),
where the controls are generated via DDPG. (b-c) The change of total reward in the
episodes in the case of (b) a full $N_{\mathrm{c}}$ and (c) $N_{\mathrm{c}}=6$.
(d) The controls obtained via DDPG for $N_{\mathrm{c}}=6$ in the case of
$\omega_{\mathrm{tr}}T=20$. The true value $\omega_{\mathrm{tr}}$ is set to be $1$.
Planck units are applied here.}
\label{fig:DDPG}
\end{figure}
%============================================================================

Deep deterministic policy gradients (DDPG) is a powerful tool in machine learning~\cite{Lillicrap2015}
and has already been applied in quantum physics to perform quantum multiparameter estimation~\cite{Xu2021}
and enhance the generation of spin squeezing~\cite{Tan2021}. The pseudocode of DDPG for quantum estimation
and the corresponding flow chart can be found in Ref.~\cite{Liu2022}, and the details will not be repeatedly
addressed herein.

\emph{Example.} The performance of controls generated via DDPG in the case of single-parameter estimation is also
illustrated with the dynamics in Eq.~(\ref{eq:ME_spon}) and control Hamiltonian in Eq.~(\ref{eq:ctrl_demo}), as shown
in Fig.~\ref{fig:DDPG}(a). $\delta_{\mathrm{q}}\omega$ is defined in Eq.~(\ref{eq:q_deviation}). The reward is taken
as the logarithm of the ratio between the controlled and non-controlled values of the QFI at time $t$. It can be seen
that the performance of DDPG with a full $N_{\mathrm{c}}$ (red pentagrams) shows a significant disparity with that of
auto-GRAPE (dash-dotted gray line). A more surprising fact is that it is even worse than the performance of both
auto-GRAPE (dashed light-blue line) and DDPG (green circles) with $N_{\mathrm{c}}=6$. And the performance of DDPG with
$N_{\mathrm{c}}=6$ also presents no advantage compared to PSO and DE. However, we cannot rashly say that PSO and DE
outperform DDPG here as DDPG involves way more parameters and maybe a suitable set of parameters would let its performance
comparable or even better than PSO and DE. Nevertheless, we can still safely to say that PSO and DE, especially DE, are
easier to find optimal controls in this example and DDPG does not present a general advantage here. The total reward in
the case of $\omega_{\mathrm{tr}}T=20$ with a full $N_{\mathrm{c}}$ and $N_{\mathrm{c}}=6$ are given in Figs.~\ref{fig:DDPG}(b)
and \ref{fig:DDPG}(c), respectively. The total reward indeed increases and converges for a full $N_{\mathrm{c}}$, but the
final performance is only sightly better than the non-controlled value [dotted black line in Fig.~\ref{fig:DDPG}(a)].
For $N_{\mathrm{c}}=6$, the total reward does not significantly increase, which means the corresponding performance
of $\delta_{\mathrm{q}}\omega$ basically comes from the average performance of random controls. The controls obtained
via DDPG for $N_{\mathrm{c}}=6$ are shown in Fig.~\ref{fig:DDPG}(d).

\subsection{Performance of the convergence speed}

\begin{figure}[tp]
\centering\includegraphics[width=8.5cm]{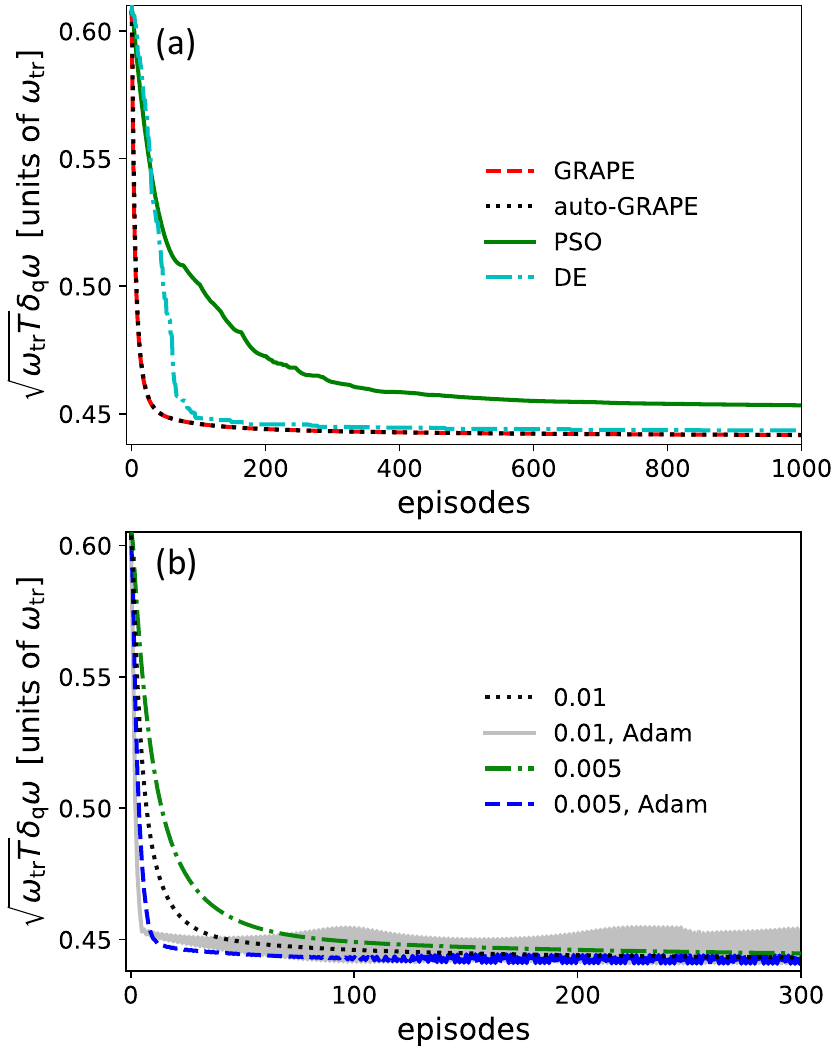}
\caption{(a) The convergence performance of different algorithms, including
GRAPE (dashed red line), auto-GRAPE (dotted black line), PSO (solid green line)
and DE (dash-dotted cyan line). (b) The convergence performance of auto-GRAPE
with constant step $\epsilon=0.01$ (dotted black line), $\epsilon=0.005$
(dash-dotted green line), and with Adam (solid gray line for $\epsilon=0.01$
and dashed blue line for $\epsilon=0.005$). The target time $\omega_{\mathrm{tr}}T=20$,
and the true value $\omega_{\mathrm{tr}}$ is set to be 1. Planck units are
applied here. }
\label{fig:converg}
\end{figure}

Apart from the improvement of the  objective function, the convergence speed is also an important aspect of an
algorithm to evaluate its performance. Here we illustrate the convergence performance of different algorithms
in Fig.~\ref{fig:converg} in the single-parameter scenario discussed previously, namely, the dynamics in
Eq.~(\ref{eq:ME_spon}) and control Hamiltonian in Eq.~(\ref{eq:ctrl_demo}) with a full $N_{\mathrm{c}}$. As
shown in Fig.~\ref{fig:converg}(a), GRAPE (dashed red line) and auto-GRAPE (dotted black line) show higher
convergence speed than PSO (solid green line) and DE (dash-dotted cyan line). This phenomenon coincides with
the common understanding that the gradient-based methods converge faster than gradient-free methods in general.
DE converges slower than GRAPE and auto-GRAPE, but the final performance of QFI basically coincides with them.
PSO presents the slowest speed in this example and the final result of QFI is also worse than others. DDPG is
not involved in this figure as its improvement on the QFI is not as significant as others.

The effect of Adam in auto-GRAPE is also illustrated in Fig.~\ref{fig:converg}(b). Denote $\epsilon$ as the
learning rate in Adam. In the case of constant-step update, auto-GRAPE with $\epsilon=0.01$ (dotted black line)
converges faster than that with $\epsilon=0.005$ (dash-dotted green line), which is common and reasonable as a
large step usually implies a higher convergence speed. However, when Adam is invoked, this difference becomes
very insignificant and both lines (solid gray line for $\epsilon=0.01$ and dashed blue line for $\epsilon=0.005$)
converge faster than constant-step updates. However, it should be noticed that a large $\epsilon$ in Adam may
result in a strong oscillation of $\delta_{\mathrm{q}}\omega$ in the episodes, and it should be adjusted to smaller
values if one wants to avoid this phenomenon.

%================================ Figure ====================================
%============================================================================

\subsection{Multiparameter estimation}
\label{sec:multi}

%================================ Figure ====================================
\begin{figure}[tp]
\centering\includegraphics[width=8.5cm]{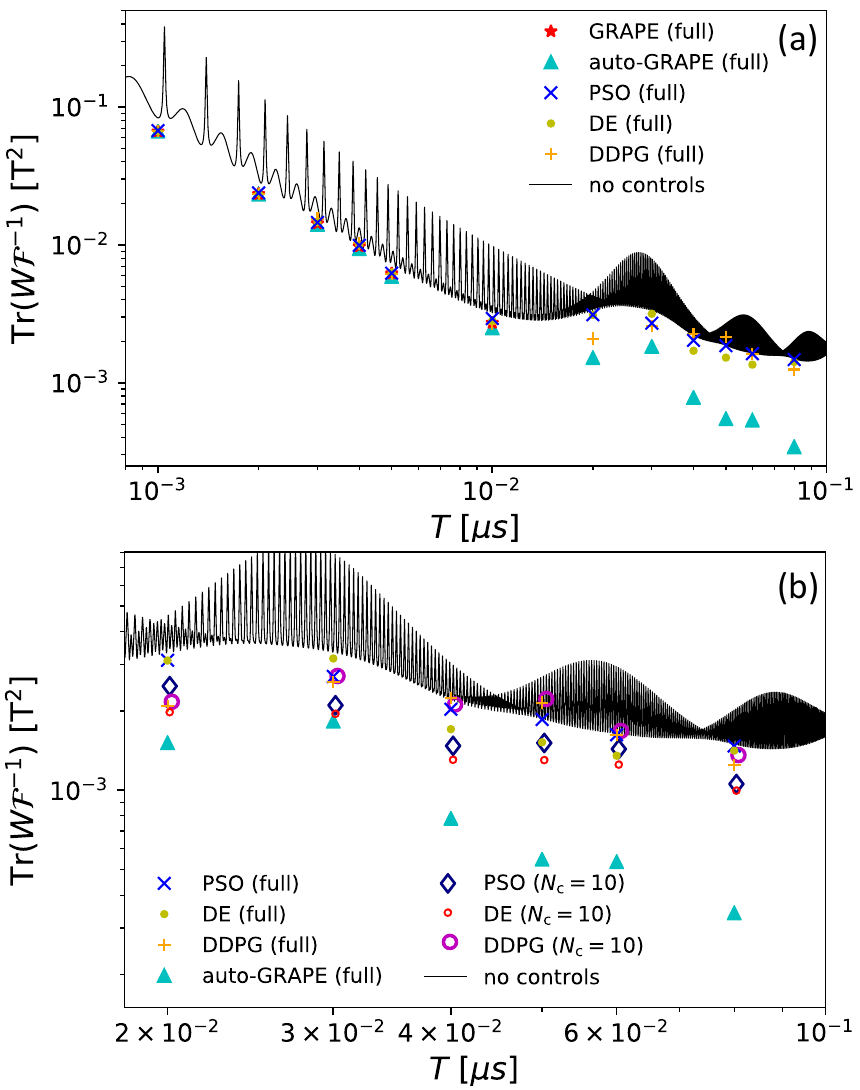}
\caption{(a) The performance of controls generated via different algorithms,
including GRAPE (red pentagrams), auto-GRAPE (cyan triangles) with full
$N_{\mathrm{c}}$, PSO (blue crosses) with full $N_{\mathrm{c}}$, DE (yellow
circles) with full $N_{\mathrm{c}}$ and DDPG (orange pluses) with full
$N_{\mathrm{c}}$. (b) The performance of controls generated via PSO (dark
blue diamonds), DE (small red hollow circles) and DDPG (large purple hollow
circles) with limited $N_{\mathrm{c}}$ ($N_{\mathrm{c}}=10$). $W$ is chosen
to be $\openone$.}
\label{fig:multipara}
\end{figure}
%============================================================================

Compared to the single-parameter estimation, multiparameter estimation is a more challenging problem in
quantum metrology. In this case, $\mathrm{Tr}(W\mathcal{F}^{-1})$ cannot be used as the objective function
in the implementation of GRAPE as the analytical calculation of $\mathcal{F}^{-1}$ is very difficult,
if not fully impossible, when the number of parameter is large. Hence, in GRAPE when $W=\openone$,
$\sum_{a}1/\mathcal{F}_{aa}$, a lower bound of $\mathrm{Tr}(\mathcal{F}^{-1})$, is taken as
the superseded objective function~\cite{Liu2020,Liu2017b,Liu2022}. Unfortunately, $\sum_{a}W_{aa}/\mathcal{F}_{aa}$
fails to be a valid lower bound for a general $W$. In this case, to keep $\sum_{a}W_{aa}/\mathcal{F}_{aa}$
a valid lower bound, the parameters for estimation have to be reorganized by the linear combination of the
original ones to let $W$ be diagonal, which causes the inconvenience to implement GRAPE insuch cases.
Different with GRAPE, this problem naturally vanishes in auto-GRAPE as the inverse matrix $\mathcal{F}^{-1}$
is calculated automatically and so does the gradient. In the meantime, PSO and DE would also not face such
problems as they are gradient-free.

\emph{Example.} Here we take an electron-nuclear spin system, which can be readily realized in the nitrogen-vacancy
centers, as an example to demonstrate and compare the performance of different algorithms included in QuanEstimation.
The Hamiltonian of this system reads~\cite{Barry2020,Schwartz2018,Rembold2020}
\begin{equation}
H_0/\hbar=DS^2_3+g_{\mathrm{S}}\vec{B}\cdot\vec{S}+g_{\mathrm{I}}\vec{B}\cdot\vec{I}
+\vec{S}^{\,\mathrm{T}}\mathcal{A}\vec{I},
\label{eq:NV_H}
\end{equation}
where $S_i=s_i\otimes\openone$ and $I_i=\openone\otimes\sigma_i$ ($i=1,2,3$) represent the
electron and nuclear ($^{15}\mathrm{N}$) operators with $s_1$, $s_2$ and $s_3$ spin-1 operators.
Their specific expressions are
\begin{eqnarray}
s_1 = \frac{1}{\sqrt{2}}\left(\begin{array}{ccc}
0 & 1 & 0 \\
1 & 0 & 1 \\
0 & 1 & 0
\end{array}\right),
s_2 = \frac{1}{\sqrt{2}}\left(\begin{array}{ccc}
0 & -i & 0\\
i & 0 & -i\\
0 & i & 0
\end{array}\right)\!\!, \nonumber
\end{eqnarray}
and $s_3=\mathrm{diag}(1,0,-1)$. The vectors $\vec{S}=(S_1,S_2,S_3)^{\mathrm{T}}$,
$\vec{I}=(I_1,I_2,I_3)^{\mathrm{T}}$ and $\mathcal{A}$ is the hyperfine tensor. In this case,
$\mathcal{A}=\mathrm{diag}(A_1,A_1,A_2)$ with $A_1$ and $A_2$ the axial and transverse magnetic hyperfine
coupling coefficients. The hyperfine coupling between the magnetic field and electron are approximated to be
isotopic. The coefficients $g_{\mathrm{S}}=g_\mathrm{e}\mu_\mathrm{B}/\hbar$ and
$g_{\mathrm{I}}=g_\mathrm{n}\mu_\mathrm{n}/\hbar$. Here $g_\mathrm{e}$ ($g_\mathrm{n}$) is the $g$ factor
of the electron (nuclear), $\mu_\mathrm{B}$ ($\mu_\mathrm{n}$) is the Bohr (nuclear) magneton and $\hbar$ is
the Plank's constant. The control Hamiltonian is
\begin{equation}
H_{\mathrm{c}}/\hbar=\sum^3_{i=1}\Omega_i(t)S_i,
\label{eq:NV_c}
\end{equation}
where $\Omega_i(t)$ is a time-varying Rabi frequency. In practice, the electron suffers from the noise of
dephasing, which means the dynamics of the full system is described by the master equation
\begin{equation}
\partial_t\rho=-i[H_0+H_{\mathrm{c}},\rho]+\frac{\gamma}{2}(S_3\rho S_3-S^2_3\rho-\rho S^2_3),
\label{eq:NV_ME}
\end{equation}
with $\gamma$ the dephasing rate, which is usually inverse proportional to the dephasing time $T^{*}_2$.

Now we use this system as a controllable magnetometer to estimate the magnetic field $\vec{B}$, which is
a three-parameter estimation problem. The optimal controls can be obtained via different algorithms.
In this case, the initial state is taken as $(|1\rangle+|\!-\!1\rangle)\otimes|\!\!\uparrow\rangle/\sqrt{2}$,
where $(|1\rangle+|\!-\!1\rangle)/\sqrt{2}$ is an electron state with $|1\rangle$ ($|\!-\!1\rangle$) the
eigenstate of $s_3$ corresponding to the eigenvalue $1$ ($-1$). $|\!\!\uparrow\rangle$ is a nuclear state and
the eigenstate of $\sigma_3$ corresponding to the eigenvalue 1. $W$ is chosen to be $\openone$. The systematic
parameters are chosen as $D=2\pi\times 2.87$\,GHz, $g_{\mathrm{S}}=2\pi\times 28.03$\,GHz/T,
$g_{\mathrm{I}}=2\pi\times 4.32$\,MHz/T, $A_1=2\pi\times 3.65$\,MHz, $A_2=2\pi\times 3.03$\,MHz, and the
true values of $\vec{B}$ are $B_1=B_2=B_3=0.50$\,mT. The dephasing rate $\gamma=2\pi\times 1$\,MHz. All the
parameter values are selected according to Refs.~\cite{Barry2020,Felton2009}.

The performance of controls given by different algorithms is given in Fig.~\ref{fig:multipara}. The control
amplitude is limited in the regime $[-20\,\mathrm{MHz},20\,\mathrm{MHz}]$. In the case of a full $N_{\mathrm{c}}$
[$N_{\mathrm{c}}=2000T/(0.01\,\mathrm{\mu s})$], as shown in Fig.~\ref{fig:multipara}(a), the performance of GRAPE
(red pentagrams),
auto-GRAPE (cyan triangles), PSO (blue crosses), DE (yellow circles) and DDPG (orange pluses) basically coincide
for small target time ($T\leq 0.01\,\mathrm{\mu s}$), and the reduction of $\mathrm{Tr}(W\mathcal{F}^{-1})$
is limited compared to the non-controlled values (solid black line). In the regime of large target time
($T> 0.01\,\mathrm{\mu s}$), auto-GRAPE shows the best performance. GRAPE is not applied in these points as its
time consumption is too heavy for our computers. PSO and DE only find controls that provide slightly enhancement
on $\mathrm{Tr}(W\mathcal{F}^{-1})$ in this regime. The different behaviors of the performance are due to the
large search space in this case. For example, the total control number for $T=0.08\,\mathrm{\mu s}$ is 48000
including all three controls $\Omega_{1}$, $\Omega_{2}$ and $\Omega_{3}$. In such a large parameter space, different
with the gradient-based methods, the gradient-free methods cannot promise to find optimal values. Hence, the
gradient-based methods would be a good choice in such cases. However, one should notice that the gradient-based
methods like auto-GRAPE could be more memory consuming than gradient-based methods. In the case that the computer
memory is limited, one may have to choose gradient-free methods.

In the case of a small search space, for example $N_{\mathrm{c}}=10$, the performance of PSO and DE improve
significantly, as shown in Fig.~\ref{fig:multipara}(b). Both PSO (dark blue diamonds) and DE (small red hollow
circles) with $N_{\mathrm{c}}=10$ outperform the full $N_{\mathrm{c}}$ cases, yet
DDPG with $N_{\mathrm{c}}=10$ (large purple hollow circles) does not show this behavior. Similar to the
single-parameter scenario, DE provides a better performance than PSO and DDPG when the control number
$N_{\mathrm{c}}$ is limited. A more interesting fact is that for some target time, like $T=0.03\,\mathrm{\mu s}$,
PSO and DE even provide comparable performance with auto-GRAPE with a full $N_{\mathrm{c}}$, indicating that
the control schemes given by PSO and DE in this case not only meet the best precision limit, but are also simpler
to implement in experiments than the full-$N_{\mathrm{c}}$ scheme given by auto-GRAPE.

\subsection{Minimum parameterization time optimization}

The control optimizations discussed in the previous subsections are performed with a fixed target time $T$. In some
scenarios, the goal is not to achieve the highest precision within a fixed time, but to reach a given precision as
soon as possible. This problem requires the search of minimum time to reach a given value of the objective function,
which can be realized in QuanEstimation in the class {\codefont ControlOpt()}. After the callings of
{\codefont control=ControlOpt()} and {\codefont control.dynamics()}, one can use the following code to solve this problem:
\begin{lstlisting}[breaklines=true,numbers=none,frame=trBL,mathescape=true]
control.mintime(f,W=[],M=[],method="binary",
                target="QFIM",LDtype="SLD")
\end{lstlisting}
Here the input {\codefont f} is a float number representing the given value of the objective function. The type of
objective function can be adjusted via {\codefont target=" "}, which includes three options {\codefont "QFIM"}
(default), {\codefont "CFIM"}, and {\codefont "HCRB"}. The measurement can be input via {\codefont M=[]} if
necessary, and in this case the objective function will be chosen as the CFIM regardless of the setting in
{\codefont target=" "}. In the case of {\codefont target="QFIM"}, the type of QFIM can be changed via
{\codefont LDtype=" "}. The choices include {\codefont "SLD"}, {\codefont "RLD"}, and {\codefont "LLD"}.
{\codefont method="binary"} represents the binary search (logarithmic search) and {\codefont method="forward"}
represents the forward search from the beginning of time. Choosing a suitable method may help to improve the
calculation efficiency. For example, if the users already know that the minimum time is very small compared to $T$,
the forward search would be more efficient than the binary search. Notice that the search is restricted in the regime
$[0,T]$ where $T$ is given by the array {\codefont tspan} input in {\codefont ControlOpt()}, and the current code
can only deal with full-$N_{\mathrm{c}}$ controls. The outputs are two files "mtspan.csv" and "controls.csv" containing
the array of time from the start to the searched minimum time and the corresponding length of optimal controls,
respectively.

\section{State optimization}
\label{sec:state_opt}

%==================================== Table =================================
\begin{table}[tp]
% \begin{ruledtabular}
\begin{tabular}{c|c|c|c}
\hline
\hline
~~Algorithms~~ & ~~method=~~ & \multicolumn{2}{c}{~~**kwargs and default values~~}\\
\hline
\multirow{6}{*}{AD} & \multirow{6}{*}{"AD"} & "Adam"  & False \\
  &   & "psi0"  & [] \\
  &   & "max\_episode"  & 300 \\
  &   & "epsilon" & 0.01 \\
  &   & "beta1" & 0.90 \\
  &   & "beta2" & 0.99 \\
\hline
\multirow{7}{*}{PSO} & \multirow{7}{*}{"PSO"} & "p\_num" & 10 \\
  &   & "psi0"  & [] \\
  &   & "max\_episode"  & [1000,100] \\
  &   & "c0"  & 1.0 \\
  &   & "c1"  & 2.0 \\
  &   & "c2"  & 2.0 \\
  &   & "seed"  & 1234 \\
\hline
\multirow{6}{*}{DE} & \multirow{6}{*}{"DE"} & "p\_num" & 10 \\
  &   & "psi0"  & [] \\
  &   & "max\_episode"  & 1000 \\
  &   & "c"  & 1.0 \\
  &   & "cr"  & 0.5 \\
  &   & "seed"  & 1234 \\
\hline
\multirow{9}{*}{NM} & \multirow{9}{*}{"NM"} & "p\_num" & 10 \\
  &   & "psi0"  & [] \\
  &   & "max\_episode"  & 1000 \\
  &   & "ar"  & 1.0 \\
  &   & "ae"  & 2.0 \\
  &   & "ac"  & 0.5 \\
  &   & "as0"  & 0.5 \\
  &   & "seed"  & 1234 \\
\hline
\multirow{3}{*}{RI} & \multirow{3}{*}{"RI"}   & "psi0" & [] \\
  &   & "max\_episode" & 300 \\
  &   & "seed"  & 1234 \\
\hline
\multirow{5}{*}{DDPG} & \multirow{5}{*}{"DDPG"}   & "psi0" & [] \\
  &   & "max\_episode" & 500 \\
  &   & "layer\_num"  & 3 \\
  &   & "layer\_dim"  & 200 \\
  &   & "seed"  & 1234 \\
\hline
\hline
\end{tabular}
% \end{ruledtabular}
\caption{Available methods for state optimization in QuanEstimation and
corresponding default parameter settings. Notice that AD is not available
when the HCRB are taken as the objective function.}
\label{table:StateOpt_paras}
\end{table}
%============================================================================

Quantum resources like entanglement and squeezing are key in quantum parameter estimation to demonstrate a quantum
enhanced precision. In contrast to the dynamical resources like time or control, entanglement and squeezing are
usually embedded in the probe state, indicating that different probe states would present dramatically different
performance on the precision. The search of optimal probe states is thus an essential step in the design of optimal
schemes. Various methodologies, including direct analytical calculations~\cite{Caves1981,Liu2013,Jarzyna2012,
Lang2013,Lang2014,Modi2011,Monras2006,Fiderer2019,Safranek2016,Knysh2014,Fujiwara2001,Safranek2019,Pal2021,
Trenyi2022}, semi-analytical~\cite{Dorner2009,Rafal2009,Forsgren2002,Maccone2009,Knysh2011,Yuan2017,Toth2018,
Toth2020,Lukacs2022} and full numerical approaches~\cite{Frowis2014,Knott2016,Rafal2020a,Basilewitsch2020,Larrouy2020},
have been proposed and discussed. More advances of the state optimization in quantum metrology can be found in a
recent review~\cite{Liu2022}. QuanEstimation includes the process of state optimization with various methods,
including both gradient-based and gradient-free methods. The specific code in QuanEstimation for the execution of
state optimization are as follows:
\begin{lstlisting}[breaklines=true,numbers=none,frame=trBL,mathescape=true]
state = StateOpt(savefile=False,method="AD",
                 **kwargs)
state.dynamics(tspan,H0,dH,Hc=[],ctrl=[],
               decay=[],dyn_method="expm")
state.QFIM(W=[],LDtype="SLD")
state.CFIM(M=[],W=[])
\end{lstlisting}
In the case that the parameterization is described by the Kraus operators, replace {\codefont state.dynamics()}
with the code {\codefont state.Kraus(K,dK)}. The parameters above have already been introduced previously. The
default settings {\codefont W=[]} and {\codefont M=[]} means $W=\openone$ and the measurement is a SIC-POVM.
The optimization method can be adjusted via {\codefont method=" "} and corresponding parameters can be set via
{\codefont **kwargs}. The available optimization methods and corresponding default parameter settings are given
in Table~\ref{table:StateOpt_paras}. The dynamics can also be solved with ODE by setting {\codefont dyn\_method="ode"}.
One exception is that when {\codefont method="AD"} is applied. In this case ODE is not available for now.
Two files "f.csv" and "states.csv" will be generated at the end of the program, which include the values of the
objective function in all episodes and the final obtained optimal probe state. When {\codefont savefile=True}, the
states obtained in all episodes will be saved in "states.csv". In the multiparameter estimation, the HCRB can also
be chosen as the objective function by calling the code:
\begin{lstlisting}[breaklines=true,numbers=none,frame=trBL,mathescape=true]
state.HCRB(W=[])
\end{lstlisting}
Notice that if {\codefont method="AD"}, {\codefont state.HCRB()} is not available. Similar to the control
optimization, if the users invoke {\codefont state.HCRB()} in the single-parameter scenario, a warning will
arise to remind them to call {\codefont state.QFIM()} instead.

%================================ Algorithm =================================
\begin{algorithm}[tp]
%\SetAlgoNoLine
\SetArgSty{<texttt>}
\caption{AD for pure states} \label{algorithm:AD}
Receive the guessed probe state $\rho_{0}=|\psi_{\mathrm{in}}\rangle\langle\psi_{\mathrm{in}}|$; \\
\For {episode=1, $M$}{
%\For {t=1, $T$}{
%Evolve and obtain the density matrix $\rho_t=e^{\Delta t\mathcal{L}_t} \rho_{t-1}$; \\
%Calculate the derivatives $\partial_\bold{x}\rho_t=-i\Delta t [\partial_\bold{x} H_0(\bold{x})]^{\times}\rho_t
%+e^{\Delta t\mathcal{L}_t} \partial_\bold{x} \rho_{t-1}$; \\
%}
Calculate the density matrix $\rho_{T}$ and its derivative $\partial_{\bold{x}}\rho_{T}$ at the target time $T$; \\
Calculate the objective function $f(T)$ with $\rho_{T}$ and $\partial_{\bold{x}}\rho_{T}$; \\
Calculate the gradients $\big\{\frac{\delta f(T)}{\delta c_{i}}\big\}$ with the automatic differentiation; \\
Update the coefficients $\{c_{i}\} \leftarrow \{c_{i}\}+\epsilon\big\{\frac{\delta f(T)}{\delta c_{i}}\big\}$; \\
Normalize the coefficients $\{c_i\}\leftarrow \frac{1}{\sqrt{\sum_j |c_j|^2}}\{c_i\}$.
}
Save the coefficients and corresponding $f(T)$, and reconstruct the state.
\end{algorithm}
%============================================================================

In the previous section, we already showed the power of automatic differentiation (AD) in the construction of
auto-GRAPE. Similarly, it can also be used in the state optimization. Due to the convexity of the QFI and
QFIM~\cite{Toth2014,Liu2020}, the optimal probe states are pure states in most scenarios. Hence, we first
consider the state optimization within the set of pure states. The pseudocode of AD in state optimization
for pure states is given in Algorithm~\ref{algorithm:AD}. In a specific basis $\{|i\rangle\langle i|\}$, a probe
state could be expanded by $|\psi\rangle=\sum_i c_i|i\rangle$, and the search of optimal probe states is equivalent
to the search of a set of normalized complex coefficients $\{c_i\}$. In AD, a guessed probe state is first given or
generated and evolved to the target time $T$ according to the given dynamics, during which the density matrices
and corresponding derivatives with respect to $\bold{x}$ are calculated and saved. Then after calculating the objective
function $f(T)$ at time $T$, all gradients $\{\delta f(T)/\delta c_i\}$ are evaluated via the automatic
differentiation, and the coefficients $\{c_i\}$ are updated accordingly with the step $\epsilon$. This step can
be adjusted via {\codefont epsilon} in {\codefont **kwargs}. Finally, the updated coefficients are normalized
as required by the quantum mechanics. In the package, Adam is not applied by default in AD and it can be turned
on by setting {\codefont Adam=True} in {\codefont **kwargs}.

Regarding the gradient-free methods, apart from the PSO, DE and DDPG, QuanEstimation also contains the Nelder-Mead
algorithm (NM)~\cite{Nelder1965}, which has already been used by Fr\"{o}wis et al.~\cite{Frowis2014} to perform
the state optimization in the case of collective spins. The detailed flow chart of NM to locate the minimum value
of an objective function can be found in Ref.~\cite{Liu2022}. For the sake of self-consistency of the paper, here
we present its pseudocode in Algorithm~\ref{algorithm:NM} for the search of the maximum value of $f$ at the target
time $T$.

%================================ Algorithm =================================
\begin{algorithm}[tp]
%\SetAlgoNoLine
\SetArgSty{<texttt>}
\caption{NM for pure states} \label{algorithm:NM}
Receive a set of guessed states $|\psi_1\rangle,\cdots,|\psi_{n+1}\rangle$;\\
\For {episode=1, $M$}{
Evolve all states according to the given dynamics and calculate the objective function $f$
at time $T$; \\
Sort the states and reassign the indices to let
$f(|\psi_1\rangle)\geq f(|\psi_2\rangle)\geq\cdots\geq f(|\psi_{n+1}\rangle)$;\\
Calculate the average state $|\psi_{\mathrm{a}}\rangle=\frac{1}{\sqrt{\mathcal{N}_{\mathrm{a}}}}
\sum^n_{k=1}|\psi_{k}\rangle$; \\
Calculate the reflected state
$|\psi_{\mathrm{r}}\rangle=\frac{1}{\sqrt{\mathcal{N}_{\mathrm{r}}}}
[|\psi_{\mathrm{a}}\rangle+a_{\mathrm{r}}(|\psi_{\mathrm{a}}\rangle-|\psi_{n+1}\rangle)]$; \\
\uIf {$f(|\psi_{\mathrm{r}}\rangle)>f(|\psi_1\rangle)$}
{Calculate the expanded state
$|\psi_{\mathrm{e}}\rangle=\frac{1}{\sqrt{\mathcal{N}_{\mathrm{e}}}}
[|\psi_{\mathrm{a}}\rangle+a_{\mathrm{e}}(|\psi_{\mathrm{r}}\rangle-|\psi_{\mathrm{a}}\rangle)]$; \\
\eIf {$f(|\psi_{\mathrm{r}}\rangle)\geq f(|\psi_{\mathrm{e}}\rangle)$}
{Replace $|\psi_{n+1}\rangle$ with $|\psi_{\mathrm{r}}\rangle$;}
{Replace $|\psi_{n+1}\rangle$ with $|\psi_{\mathrm{e}}\rangle$;}}
\uElseIf {$f(|\psi_1\rangle) \geq f(|\psi_{\mathrm{r}}\rangle) > f(|\psi_n\rangle)$}
{Replace $|\psi_{n+1}\rangle$ with $|\psi_{\mathrm{r}}\rangle$;}
\uElseIf {$f(|\psi_n\rangle) \geq f(|\psi_{\mathrm{r}}\rangle) > f(|\psi_{n+1}\rangle)$}
{Calculate the outside contracted state
$|\psi_{\mathrm{oc}}\rangle=\frac{1}{\sqrt{\mathcal{N}_{\mathrm{oc}}}}
[|\psi_{\mathrm{a}}\rangle+a_{\mathrm{c}}(|\psi\rangle_{\mathrm{r}}-|\psi_{\mathrm{a}}\rangle)]$;\\
\eIf{$f(|\psi_{\mathrm{oc}}\rangle) \geq f(|\psi_{\mathrm{r}}\rangle)$}
{Replace $|\psi_{n+1}\rangle$ with $|\psi_{\mathrm{oc}}\rangle$;}
{Replace all $|\psi_k\rangle$ for $k\in[2,n+1]$ with
$\frac{1}{\sqrt{\mathcal{N}_k}}[|\psi_1\rangle+a_{\mathrm{s}}(|\psi_k\rangle-|\psi_1\rangle)]$;}
}
\Else {Calculate the inside contracted state
$|\psi_{\mathrm{ic}}\rangle=\frac{1}{\sqrt{\mathcal{N}_{\mathrm{ic}}}}
[|\psi_{\mathrm{a}}\rangle-a_{\mathrm{c}}(|\psi_{\mathrm{a}}\rangle-|\psi_{n+1}\rangle)]$;\\
\eIf {$f(|\psi_{\mathrm{ic}}\rangle) > f(|\psi_{n+1}\rangle)$}
{Replace $|\psi_{n+1}\rangle$ with $|\psi_{\mathrm{ic}}\rangle$;}
{Replace all $|\psi_k\rangle$ for $k\in[2,n+1]$ with
$\frac{1}{\sqrt{\mathcal{N}_k}}[|\psi_1\rangle+a_{\mathrm{s}}(|\psi_k\rangle-|\psi_1\rangle)]$;}
}}
Return the optimal state $|\psi_1\rangle$ and corresponding $f$.
\end{algorithm}
%============================================================================

In NM, $n+1$ guessed states are input and sorted descendingly according to the corresponding values of $f$, namely,
$f(|\psi_1\rangle)\geq\cdots\geq f(|\psi_{n+1}\rangle)$. In one episode of optimization, the average state
$|\psi_{\mathrm{a}}\rangle:=\frac{1}{\sqrt{\mathcal{N}_{\mathrm{a}}}}\sum^n_{k=1}|\psi_k\rangle$ and reflected state
$|\psi_{\mathrm{r}}\rangle:=\frac{1}{\sqrt{\mathcal{N}_{\mathrm{r}}}}[|\psi_{\mathrm{a}}\rangle+a_{\mathrm{r}}
(|\psi_{\mathrm{a}}\rangle-|\psi_{n+1}\rangle)]$ are first calculated. In the case that the reflected state is
better than $|\psi_1\rangle$, i.e., $f(|\psi_{\mathrm{r}}\rangle)$ is larger than $f(|\psi_1\rangle)$, the expanded
state $|\psi_{\mathrm{e}}\rangle:=\frac{1}{\sqrt{\mathcal{N}_{\mathrm{e}}}}[|\psi_{\mathrm{a}}\rangle+a_{\mathrm{e}}
(|\psi_{\mathrm{r}}\rangle-|\psi_{\mathrm{a}}\rangle)]$ is then calculated and compared to the reflected state. If
the reflected state is still better, then replace $|\psi_{n+1}\rangle$ with $|\psi_{\mathrm{r}}\rangle$, otherwise
replace $|\psi_{n+1}\rangle$ with $|\psi_{\mathrm{e}}\rangle$. In the case that the performance of the reflected
state is in the middle of $|\psi_1\rangle$ and $|\psi_n\rangle$, just replace $|\psi_{n+1}\rangle$ with it. If its
performance is between $|\psi_n\rangle$ and $|\psi_{n+1}\rangle$, then the outside contracted state
$|\psi_{\mathrm{oc}}\rangle:=\frac{1}{\sqrt{\mathcal{N}_{\mathrm{oc}}}}[|\psi_{\mathrm{a}}\rangle+a_{\mathrm{c}}
(|\psi\rangle_{\mathrm{r}}-|\psi_{\mathrm{a}}\rangle)]$ is calculated and compared to the reflected state.
$|\psi_{n+1}\rangle$ is replaced with $|\psi_{\mathrm{oc}}\rangle$ if $|\psi_{\mathrm{oc}}\rangle$ outperforms the
reflected state, otherwise all states $\{|\psi_k\rangle\}$, except the best one $|\psi_1\rangle$, are replaced with
the states $\frac{1}{\sqrt{\mathcal{N}_k}}[|\psi_1\rangle+a_{\mathrm{s}}(|\psi_k\rangle-|\psi_1\rangle)]$ and the
program goes to the next episode. In the case that $|\psi_{\mathrm{r}}\rangle$ is no better than any state in
$\{|\psi_k\rangle\}$, the inside contracted state
$|\psi_{\mathrm{ic}}\rangle:=\frac{1}{\sqrt{\mathcal{N}_{\mathrm{ic}}}}
[|\psi_{\mathrm{a}}\rangle-a_{\mathrm{c}}(|\psi_{\mathrm{a}}\rangle-|\psi_{n+1}\rangle)]$ is then calculated and
compared to $|\psi_{n+1}\rangle$. If it is better than $|\psi_{n+1}\rangle$, replace $|\psi_{n+1}\rangle$ with it,
otherwise perform the same replacement operation to all states as done previously. At the beginning of next round,
all states are sorted in descending order again.
$\mathcal{N}_{\mathrm{a}}:=\langle\psi_{\mathrm{a}}|\psi_{\mathrm{a}}\rangle$
is the normalization coefficient, same as $\mathcal{N}_{\mathrm{r}}$, $\mathcal{N}_{\mathrm{e}}$,
$\mathcal{N}_{\mathrm{oc}}$ and $\mathcal{N}_{\mathrm{ic}}$. A general setting of the coefficients are
$a_{\mathrm{r}}=1.0$, $a_{\mathrm{e}}=2.0$, $a_{\mathrm{c}}=a_{\mathrm{s}}=0.5$, which are also the default
values in the package. These coefficients can be adjusted in {\codefont **kwargs} (shown in
Table~\ref{table:StateOpt_paras}) via {\codefont ar}, {\codefont ae}, {\codefont ac} and {\codefont as0}.
In the meantime, {\codefont p\_num} in {\codefont **kwargs} represents the state number $n+1$.

%================================ Algorithm =================================
\begin{algorithm}[tp]
%\SetAlgoNoLine
\SetArgSty{<texttt>}
\caption{RI}
Receive the guessed probe state $\rho_0$; \\
\For {$m$=1, $M$}{
Evolve the state with $\rho=\sum_i K_i \rho_0 K^{\dagger}_i$;\\
Calculate the derivative $\partial_{a}\rho = \sum_i(\partial_{a}K_i)\rho_0 K^{\dagger}_i
+ K_i\rho_0(\partial_{a}K^{\dagger}_i)$; \\
Calculate the QFI and the SLD $L$ with $\rho$ and $\partial_{\bold{x}}\rho$; \\
Calculate the matrix $\mathcal{M}$; \\
Find the eigenvector ${|\psi_{m}\rangle}$ of $\mathcal{M}$ corresponding to
its largest eigenvalue; \\
Replace $\rho_0$ with $|\psi_{m}\rangle\langle\psi_{m}|$.}
Return the optimal state ${|\psi_{M}\rangle}$ and the QFI.
\label{algorithm:Iter}
\end{algorithm}
%===============================================================================

%================================ Figure ====================================
\begin{figure*}[tp]
\centering\includegraphics[width=17.5cm]{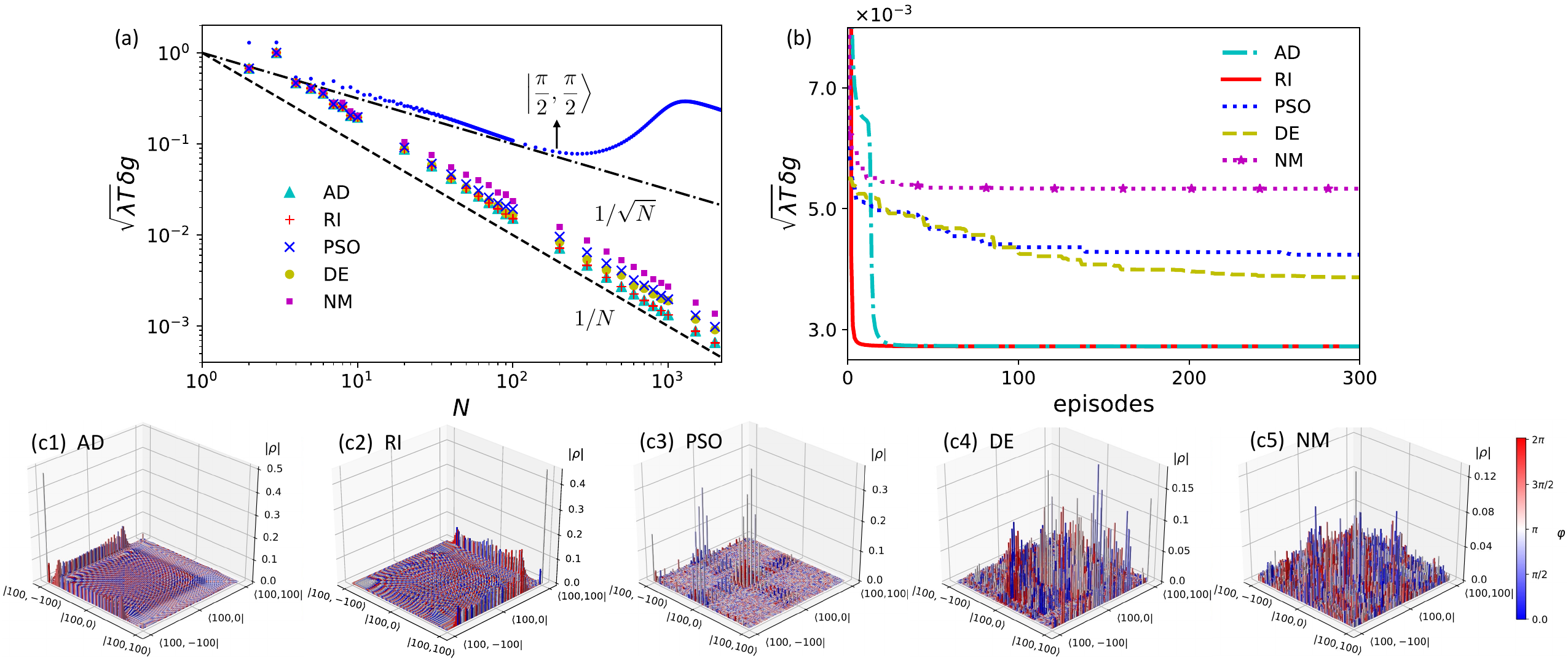}
\caption{(a) The performance of the optimal probe states searched via AD (cyan triangles),
RI (red pluses), PSO (blue crosses), DE (yellow circles) and NM (purple squares) in the
Lipkin-Meshkov-Glick model in the absence of noise. The blue dots represents the
value of $\sqrt{\lambda T}\delta g$ for the coherent spin state $|\pi/2,\pi/2\rangle$,
and the dash-dotted black and dashed black lines represent $1/\sqrt{N}$ and $1/N$,
respectively. (b) The convergence performance of AD (dash-dotted cyan line), RI (solid
red line), PSO (dotted blue line), DE (dashed yellow line), and NM (dotted star purple
line) in the case of $N=500$. (c1-c5) The searched optimal states with different algorithms
in the case of $N=100$. The target time is chosen as $\lambda T=10$. The true value of $g$
is 0.5, and the value of $h/\lambda$ is set to be $0.1$. Planck units are applied here.}
\label{fig:StateOpt_unitary}
\end{figure*}
%============================================================================

%================================ Figure ====================================
\begin{figure*}[tp]
\centering\includegraphics[width=17.5cm]{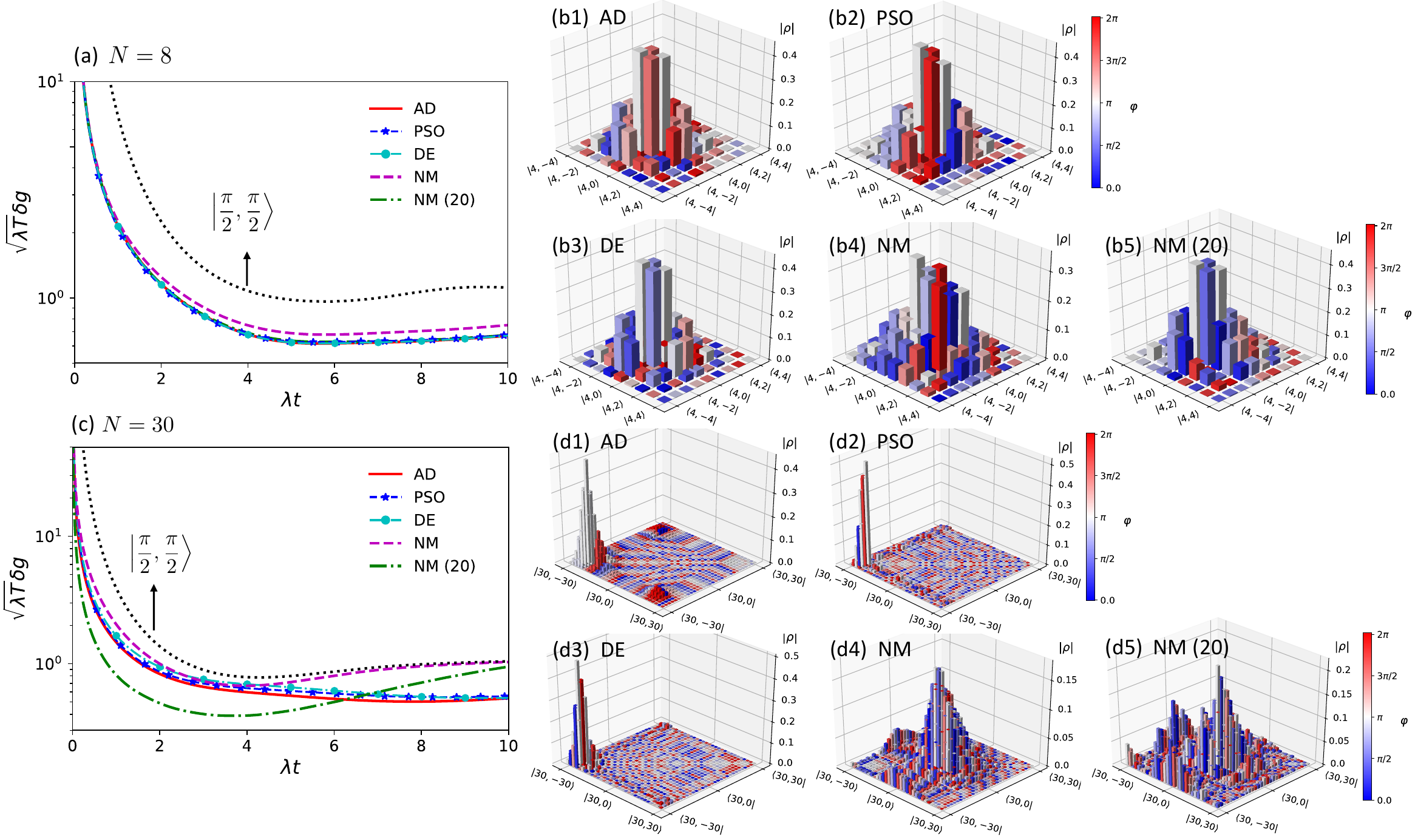}
\caption{The performance of probe states obtained via different algorithms for
(a) $N=8$ and (c) $N=30$ when the collective dephasing exists. The solid red line,
dashed star blue line, dash-dotted circle cyan line, dashed purple line represent
the values of $\sqrt{\lambda T}\delta g$ for the searched states obtained via AD,
PSO, DE, and NM, respectively. The dash-dotted green line represents that of NM
with 20 parallel sets. The dotted black line represent the result of
$|\pi/2,\pi/2\rangle$. (b1-b5) The searched optimal states for $N=8$. (d1-d5) The
searched optimal states for $N=30$. The target time $\lambda T=10$, and the true
values of $g$ is 0.5. The value of $h/\lambda$ is set to be $0.1$ and the decay
rate $\gamma/\lambda=0.1$. Planck units are applied here.}
\label{fig:StateOpt_noise}
\end{figure*}
%============================================================================

Apart from the aforementioned algorithms, there also exist dedicated algorithms for the state optimization
in quantum parameter estimation. Here we introduce a reverse iterative algorithm (RI), which was first proposed
in Refs.~\cite{Demkowicz2011,Macieszczak2014} in the Bayesian estimation context, and then applied to the QFI
in Ref.~\cite{Macieszczak2013a}. In the case of single-parameter estimation, the QFI can be rewritten into
\begin{equation}
\label{eq:qfisup}
\mathcal{F}_{aa} = \sup_{A} \left[2\mathrm{Tr}(A \partial_a\rho)-\mathrm{Tr}(\rho A^2)\right].
\end{equation}
This form is equivalent to the standard definition of the QFI as can be seen by solving the maximization
problem $2\mathrm{Tr}(A\partial_a\rho)-\mathrm{Tr}(\rho A^2)$ with respect to $A$, which is formally a
quadratic function in matrix $A$ and the resulting extremum condition yields the standard linear equation
for $\partial_a\rho=\frac{1}{2}(A\rho+\rho A)$, i.e., the optimal $A=L_a$ is just the SLD operator. When this
solution is plugged into the formula and it yields $\mathrm{Tr}(\rho L^2_a)$, which is in agreement with
the standard definition of the QFI. Consider the parameterization process described by the Kraus operators
given in Eq.~(\ref{eq:kraus_opt}), $\rho=\sum_i K_i(x)\rho_0 K_i^\dagger(x)$. Taking into account
Eq.~(\ref{eq:qfisup}), we see that the problem of identifying the optimal input state $\rho_0$ that maximizes
the QFI, can be written as a double maximization problem,
\begin{equation}
 \sup_{\rho_0}\mathcal{F}_{aa} = \sup_{A,\rho_0}
 \left[2\mathrm{Tr}(A \partial_a\rho)-\mathrm{Tr}(\rho A^2)\right].
\end{equation}
This observation leads to an effective iterative protocol, where for a fixed $\rho_0$ we find the optimal $A$
that maximizes the above expression, and then fixing the optimal $A$ found in the previous step we look for the
optimal $\rho_0$. In order to implement the procedure, note that the QFI can be rewritten in the `Heisenberg
picture' form, where the Kraus operators effectively act on the $L_a$ operators, as
\begin{equation}
\mathcal{F}_{aa}=\mathrm{Tr}\left(\rho_0 \mathcal{M}\right)
\end{equation}
with
\begin{equation}
\mathcal{M}\!=\!\sum_i 2\!\left[(\partial_a K^{\dagger}_i)L_a K_i
\!+\!K^{\dagger}_i L_a(\partial_a K_i)\right]\!-\!K_i^\dagger L^2_a K_i.
\end{equation}
This equation indicates that for a fixed $\mathcal{M}$ (i.e.,~fixed $A=L_a$), the optimal probe state is nothing
but the eigenvector corresponding to the maximum eigenvalue of $\mathcal{M}$. The pseudocode of this algorithm
is given in Algorithm~\ref{algorithm:Iter}. In one round of the optimization, $\mathcal{M}$ is calculated and
its eigenvector with respect to the maximum eigenvalue of $\mathcal{M}$ is calculated and used as the probe
state in the next round. In the package, this method can be invoked via {\codefont method="RI"}. The number
of episodes and the seed can be adjusted in {\codefont **kwargs} (shown in Table~\ref{table:StateOpt_paras})
via {\codefont max\_episode} and {\codefont seed}. Notice that this method is only available when
{\codefont state.Kraus()} is invoked, and in the current version of the package, it only works for the
single-parameter quantum estimation, i.e., the objective function is the QFI. The extension to the CFI and
the case of multiparameter estimation will be thoroughly discussed in an independent paper.

\emph{Example.} Here we use the Lipkin-Meshkov-Glick model as an example to show the state optimization with
QuanEstimation. The Hamiltonian of this model is~\cite{Lipkin1965}
\begin{equation}
H_{\mathrm{LMG}}=-\frac{\lambda}{N}(J_1^2+gJ_2^2)-hJ_3,
\end{equation}
where $J_i=\frac{1}{2}\sum_{j=1}^N \sigma_i^{(j)}$ ($i=1,2,3$) is the collective spin operator with $\sigma_i^{(j)}$
the $i$th Pauli matrix for the $j$th spin. $N$ is the total number of spins, $\lambda$ is the spin-spin interaction
strength, $h$ is the strength of the external field and $g$ is the anisotropic parameter. All searches with
different algorithms start from the coherent spin state $|\theta=\pi/2,\phi=\pi/2\rangle$, which is defined
by~\cite{Ma2011}
\begin{equation}
|\theta,\phi\rangle=\exp\left(-\frac{\theta}{2}e^{-i\phi} J_{+}+\frac{\theta}{2}e^{i\phi}J_-\right)|J,J\rangle,
\end{equation}
where $|J,J\rangle$ is a Dicke state with $J=N/2$ and $J_{\pm}=J_1\pm iJ_2$. Here we consider the case
that the search is constrained to pure states with fixed $J=N/2$, which can be expressed as
$|\psi\rangle=\sum^J_{m=-J}c_m|J,m\rangle$ with $|J,m\rangle$ a general Dicke state and $c_m$ a complex
coefficient. Let us first study the single-parameter scenario with $g$ the parameter to be estimated.

%================================ Figure ====================================
\begin{figure*}[tp]
\centering\includegraphics[width=15.5cm]{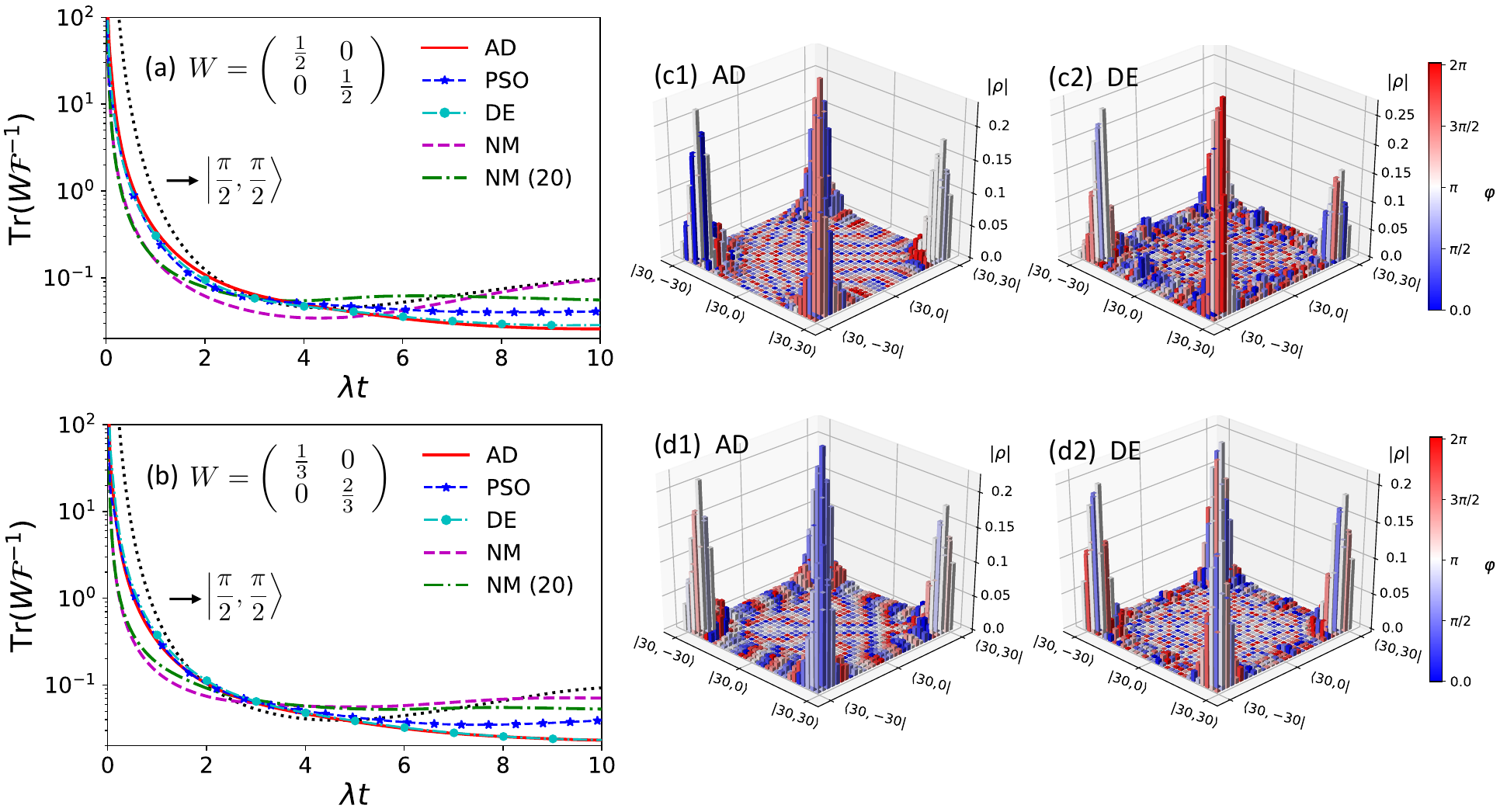}
\caption{The performance of different algorithms for the weight matrix (a)
$W=\mathrm{diag}(1/2,1/2)$ and (b) $W=\mathrm{diag}(1/3,2/3)$. The solid red
line, dashed star blue line, dash-dotted circle cyan line, dashed purple line
and dash-dotted green line represent the results obtained via AD, PSO, DE, NM,
and NM with 20 parallel sets, respectively. The dotted black line represent the
result of $|\pi/2,\pi/2\rangle$. (c1-c2) The optimal states obtained from AD and
DE for $W=\mathrm{diag}(1/2,1/2)$. (d1-d2) The optimal states obtained from AD
and DE for $W=\mathrm{diag}(1/3,2/3)$. The target time $\lambda T=10$. The true
values of $g$ and $h/\lambda$ are set to be 0.5 and $0.1$. Planck units are
applied here.}
\label{fig:StateOpt_multipara}
\end{figure*}
%============================================================================

The performance of the optimal probe states searched via AD (cyan triangles), RI (red pluses), PSO (blue crosses),
DE (yellow circles) and NM (purple squares) in the absence of noise are given in Fig.~\ref{fig:StateOpt_unitary}(a).
Here $\delta g=1/\sqrt{\mathcal{F}_{gg}}$ is the theoretical optimal deviation for $g$. The target time is taken
as $\lambda T=10$ (Planck units are applied). The performance of DDPG is not good enough and thus not shown in
the figure. For a very small $N$, the searched optimal states do not show an obvious advantage than the state
$|\pi/2,\pi/2\rangle$ (blue dots). However, when $N$ is large the advantage becomes significant, and the performance
of all searched states outperform $|\pi/2,\pi/2\rangle$ and $1/\sqrt{N}$ (dash-dotted black line) in the case that
$N$ is larger than around 6. For a large $N$, the performance of the states obtained via AD and RI are the best and
very close to $1/N$ (dashed black line). The performance of DE and PSO basically coincide with each other (more
accurately to say, the performance of DE is slightly better than that of PSO), but is worse than AD and RI. The
performance of NM is the worst in this example. Please note that we cannot rashly say that the general performance
of NM is worse than DE or PSO in the state optimization just based on this plot as different parameter settings in
the algorithms sometimes could dramatically affect the behaviors, yet we basically use the general recommended settings
in all algorithms. Nevertheless, different sensitivities of the parameter settings on the final result still indicates
that DE and PSO are easier to locate optimal states than NM at least in this example.

Regarding the convergence performance in this example, as shown in Fig.~\ref{fig:StateOpt_unitary}(b), RI shows
the fastest convergence speed and the best optimized value. AD is slightly slower than RI but still way faster than the
gradient-free methods. However, the disadvantage of AD is that occupation of memory grows very fast with the increase
of $N$. Hence, RI would be the best choice to try first for the state optimization in the case of unitary parameterization.
In the last, as a demonstration, the searched optimal states via different algorithms in the case of $N=100$ are shown
in Figs.~\ref{fig:StateOpt_unitary}(c1-c5).

\emph{Example.} When the collective dephasing is involved, the dynamics of this system is governed by the following
master equation
\begin{equation}
\partial_t\rho = -i[H_{\mathrm{LMG}},\rho]+\gamma \left(J_3\rho J_3-\frac{1}{2}\left\{\rho, J^2_3\right\}\right)
\label{eq:dephasing_LMG}
\end{equation}
with $\gamma$ the decay rate. The performance of optimal probe states searched via AD (solid red line), PSO (dashed
star blue line), DE (dash-dotted circle cyan line) and NM (dashed purple line) are illustrated with $N=8$ and $N=30$
in Figs.~\ref{fig:StateOpt_noise}(a) and \ref{fig:StateOpt_noise}(c), respectively. The corresponding optimal probe
states are given in Figs.~\ref{fig:StateOpt_noise}(b1-b4) for $N=8$ and Figs.~\ref{fig:StateOpt_noise}(d1-d4) for
$N=30$. In both cases, the states obtained via AD, PSO and DE basically present coincidental performance at time $T$,
and outperform $|\pi/2,\pi/2\rangle$ (dotted black lines). Similar to the unitary scenario, the state obtained via
NM shows a worse performance at time $T$, and it even fails to find a better state than $|\pi/2,\pi/2\rangle$ in the
case of $N=30$. In this figure, the number of parallel sets (also called particles in PSO and populations in DE) are
10 for all NM, DE and PSO. After increasing the number of parallel sets from 10 to 20 [labelled by NM (20) in the
plot], the performance of NM (dash-dotted green line) improves in the case of $N=8$, which basically coincides with
others. However, it still fails to find a better state when $N=30$. More number of parallel sets may be requires for
NM in this case. The states obtained via NM (20) are shown in Figs.~\ref{fig:StateOpt_noise}(b5) and
\ref{fig:StateOpt_noise}(d5) for $N=8$ and $N=30$, respectively.

%==================================== Table =================================
\begin{table}[bp]
\begin{tabular}{c|c|c|c}
\hline
\hline
~~Algorithms~~ & ~~method=~~ & \multicolumn{2}{c}{~~**kwargs and default values~~}\\
\hline
\multirow{7}{*}{PSO} & \multirow{7}{*}{"PSO"} & "p\_num" & 10 \\
  &   & "measurement0"  & [] \\
  &   & "max\_episode"  & [1000,100] \\
  &   & "c0"  & 1.0 \\
  &   & "c1"  & 2.0 \\
  &   & "c2"  & 2.0 \\
  &   & "seed"  & 1234 \\
\hline
\multirow{6}{*}{DE} & \multirow{6}{*}{"DE"} & "p\_num" & 10 \\
  &   & "measurement0"  & [] \\
  &   & "max\_episode"  & 1000 \\
  &   & "c"  & 1.0 \\
  &   & "cr"  & 0.5 \\
  &   & "seed"  & 1234 \\
\hline
\multirow{5}{*}{AD} & \multirow{7}{*}{"AD"} & "Adam"  & False \\
\multirow{5}{*}{(available when}  &   & "measurement0"  & [] \\
\multirow{5}{*}{{\codefont mtype="input"})}  &   & "max\_episode"  & 300 \\
  &   & "epsilon" & 0.01 \\
  &   & "beta1" & 0.90 \\
  &   & "beta2" & 0.99 \\
\hline
\hline
\end{tabular}
\caption{Available methods for measurement optimization in QuanEstimation and
corresponding default parameter settings. Notice that AD is only available
when {\codefont mtype="input"}. Here {\codefont measurement0} is the initial
guess of the measurement.}
\label{table:MeasOpt_paras}
\end{table}
%============================================================================

Next we discuss the state optimization in multiparameter estimation. Consider the simultaneous estimation
of $g$ and $h/\lambda$ in the Lipkin-Meshkov-Glick model with the dynamics in Eq.~(\ref{eq:dephasing_LMG}).
Figures~\ref{fig:StateOpt_multipara}(a) and \ref{fig:StateOpt_multipara}(b) show the performance of optimal
states obtained via different algorithms for $W=\mathrm{diag}(1/2,1/2)$ and $W=\mathrm{diag}(1/3,2/3)$, respectively.
In both cases AD (solid red line) and DE (dash-dotted circle cyan line) present the best performance at the target
time $\lambda T=10$, and DE even slightly outperform AD in the case of $W=\mathrm{diag}(1/2,1/2)$. The performance
of PSO (dashed star blue line) is worse than AD and DE, yet still better than NM (dashed purple line) and NM with
20 parallel sets (dash-dotted green line). The performance of NM does not even outperform the coherent spin state
$|\pi/2,\pi/2\rangle$ (dotted black line) in the case of $W=\mathrm{diag}(1/2,1/2)$. Hence, apart from gradient-based
algorithm like AD, PSO and DE would also be good choices for state optimizations. The optimal states obtained from AD
and DE for $W=\mathrm{diag}(1/2,1/2)$ and $W=\mathrm{diag}(1/3,2/3)$ are demonstrated in
Figs.~\ref{fig:StateOpt_multipara}(c1-c2) and Figs.~\ref{fig:StateOpt_multipara}(d1-d2), respectively. Although the
performance on $\mathrm{Tr}(W\mathcal{F}^{-1})$ are basically the same for these states, they may still have gaps on
other properties like the difficulties of preparation, the robustness to the imperfect preparation and so on. Hence,
in practice one needs to compare these optimal states comprehensively case by case to make wise choices.

\section{Measurement optimization}
\label{sec:measurement_opt}

%================================ Figure ====================================
\begin{figure*}[tp]
\centering\includegraphics[width=17.5cm]{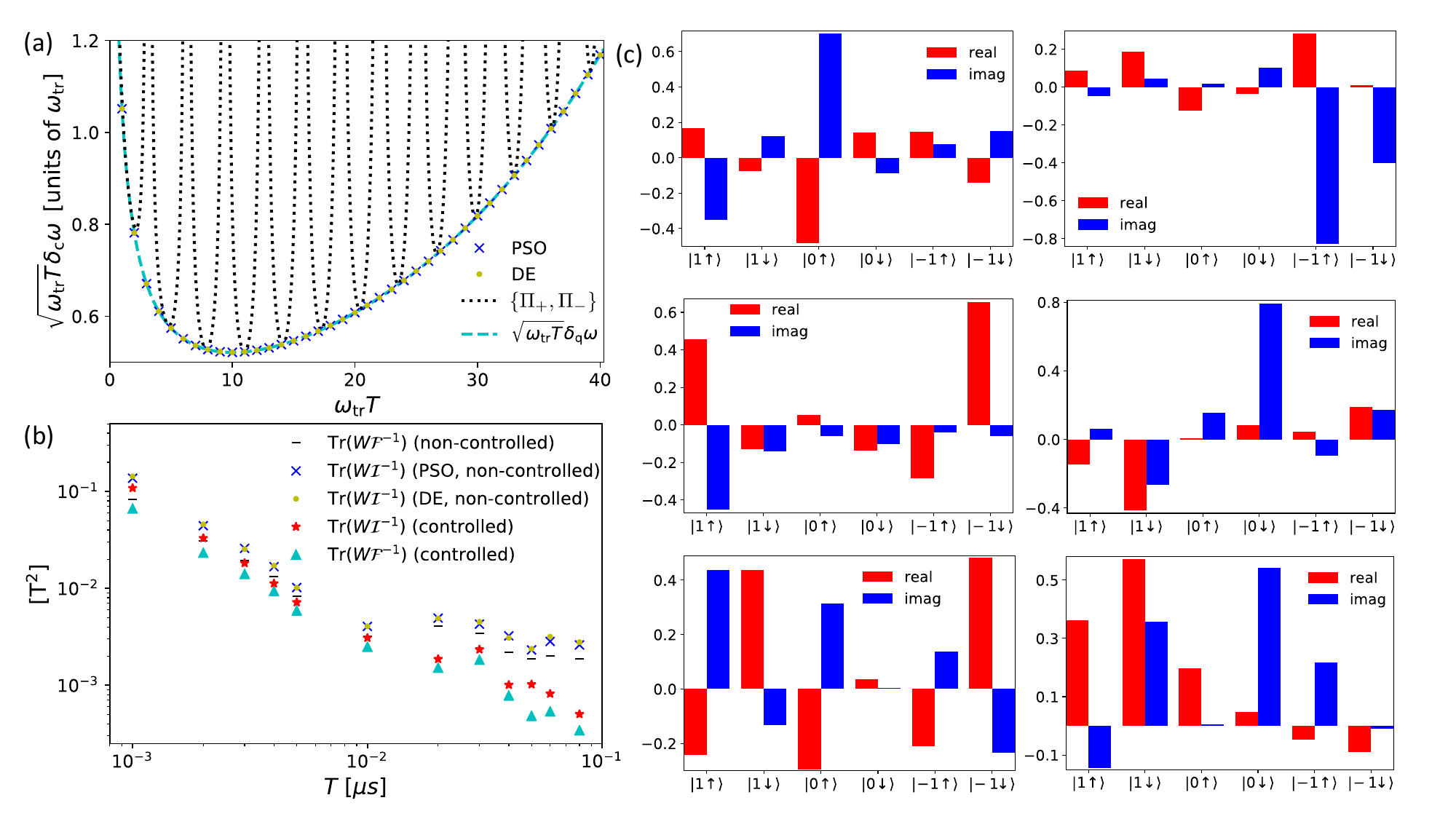}
\caption{(a) The performance of optimal projective measurements obtained via PSO
(blue crosses) and DE (yellow circles) in the case of single-parameter estimation.
The dashed cyan line represents the values of $\sqrt{\omega_{\mathrm{tr}}T}\delta_{\mathrm{q}}\omega$
and the dotted black line represents the values of $\sqrt{\omega_{\mathrm{tr}}T}\delta_{\mathrm{c}}\omega$
with respect to the projective measurement $\{\Pi_{+}\!=\!|+\rangle\langle+|,
\Pi_{-}\!=\!|-\rangle\langle-|\}$. The true value $\omega_{\mathrm{tr}}=1$. Planck units
are applied in this plot. (b) The performance of optimal projective measurements obtained
via PSO (blue crosses) and DE (yellow circles) in the case of multiparameter estimation in
the absence of control. The black underlines and cyan triangles represent the values of
$\mathrm{Tr}(W\mathcal{F}^{-1})$ without and with optimal control. The red pentagrams
represent the controlled values of $\mathrm{Tr}(W\mathcal{I}^{-1})$ with the optimal
measurements obtained in the non-controlled scenario. (c) Demonstration of the optimal
projective measurement obtained by DE in the multiparameter estimation at the target time
$T=0.04\,\mu$s. The red and blue bars represent the real and imaginary parts of the coefficients
of the optimal measurement in the basis $\{|1\!\uparrow\rangle,|1\!\downarrow\rangle,
|0\!\uparrow\rangle,|0\!\downarrow\rangle,|\!-\!1\!\uparrow\rangle,|\!-\!1\!\downarrow\rangle\}$.}
\label{fig:Mopt}
\end{figure*}
%============================================================================

Measurement is critical in quantum parameter estimation~\cite{Yu2021,Rath2021,Zhang2020,XuL2021}. On one hand,
all asymptotic bounds require some optimal measurements to attain if it is attainable, and hence the search
of optimal measurements is a natural requirement in theory to approach the ultimate precision limit. On the
other hand, the choice of measurements is usually limited in practice, and how to find conditioned optimal
measurements with the practical measurements in hand is an important step towards the design of a realizable
scheme. QuanEstimation includes the optimization of measurements for several scenarios. The first one is the
optimization of rank-one projective measurements. A set of projective measurements $\{\Pi_i\}$ satisfies
$\Pi_i\Pi_j=\Pi_i\delta_{ij}$ and $\sum_i\Pi_i=\openone$, and it can be rewritten into
$\{|\phi_i\rangle\langle\phi_i|\}$ with $\{|\phi_i\rangle\}$ an orthonormal basis in the Hilbert space. In
this way, the optimization of rank-one projective measurement is equivalent to identifying the optimal basis,
which can be realized using PSO and DE in QuanEstimation. In this case the automatic differentiation is not
working very well due to the Gram-Schmidt orthogonalization procedure after the update of $\{|\phi_i\rangle\}$
according to the gradients. In some cases, the realizable measurement has to be limited in
the linear combination of a given set of POVM, hence, the second scenario is to find the optimal linear
combination of an input measurement. Moreover, in some cases the measurement $\{\Pi_i\}$ has to be fixed, but
an arbitrary unitary operation can be invoked before performing the measurement, which is equivalent to a new
measurement $\{U\Pi_i U^{\dagger}\}$. Based on this, the third scenario is to find the optimal rotated
measurement of an input measurement.

The code in QuanEstimation for the execution of measurement optimization are as follows:
\begin{lstlisting}[breaklines=true,numbers=none,frame=trBL,mathescape=true]
m = MeasurementOpt(mtype="projection",
           minput=[],savefile=False,
           method="DE",**kwargs)
m.dynamics(tspan,rho0,H0,dH,Hc=[],ctrl=[],
           decay=[],dyn_method="expm")
m.CFIM(W=[])
\end{lstlisting}
In the case that the parameterization is described by the Kraus operators, replace {\codefont m.dynamics()} with
the code {\codefont m.Kraus(rho0,K,dK)}. The optimization method can be adjusted via {\codefont method=" "}
and corresponding parameters can be set via {\codefont **kwargs}. The available optimization methods and
corresponding default parameter settings are given in Table~\ref{table:MeasOpt_paras}. {\codefont dyn\_method="ode"}
is also available here to invoke ODE for soloving the dynamics, except the case that {\codefont method="AD"} is
applied. Two files "f.csv" and "measurements.csv" will be generated at the end of the program. When
{\codefont savefile=True}, the measurements obtained in all episodes will be saved in "measurements.csv".

The variable {\codefont mtype=" "} defines the type of scenarios for the optimization, and currently it includes
two options: {\codefont mtype="projection"} and {\codefont mtype="input"}. The first one means the optimization
is performed in the first scenario, i.e., within the set of projective measurements. In this case,
{\codefont minput=[]} should keep empty. Since $|\phi_i\rangle$ in a rank-one projective measurement
$\{|\phi_i\rangle\langle\phi_i|\}$ can be expended as $|\phi_i\rangle=\sum_j C_{ij}|j\rangle$ in a given
orthonormal basis $\{|j\rangle\}$, the optimization of the rank-one projective measurement is equivalent to
the optimization of a complex matrix $C$. When the gradient-free methods are applied, all entries in $C$ are
updated via the given algorithm in each episode, then adjusted via the Gram-Schmidt orthogonalization
procedure to make sure $\{|\phi_i\rangle\langle\phi_i|\}$ is a legitimate projective measurement, i.e.,
$\langle\phi_i|\phi_j\rangle=\delta_{ij},~\forall i,j$ and $\sum_i|\phi_i\rangle\langle\phi_i|=\openone$. The
second option {\codefont mtype="input"} means the optimization is performed in the second and third scenarios.
The input rule of {\codefont minput} for the second scenario is {\codefont minput=["LC", [Pi1,Pi2,...], m]} and
for the third one is {\codefont minput=["rotation", [Pi1,Pi2,...]]}. Here {\codefont [Pi1,Pi2,...]} is a list of
matrices representing the input measurement $[\Pi_1,\Pi_2,\dots]$. The variable {\codefont m} in the second
scenario is an integer representing the number of operators of the output measurement, and thus should be no
larger than that of the input measurement. For example, assume the input measurement is $\{\Pi_i\}^6_{i=1}$ and
input 4 in the position of {\codefont m} means the the output measurement is $\{\Pi^{\prime}_{i}\}^4_{i=1}$ where
$\Pi^{\prime}_i=\sum^{6}_{j=1}B_{ij}\Pi_j$. The optimization is to find an optimal real matrix $B$ for the optimal
CFI or $\mathrm{Tr}(W\mathcal{I}^{-1})$. To make sure the updated measurement in each episode is still a legitimate
POVM, all entries of $B$ are limited in the regime $[0,1]$ and $\sum_{i}B_{ij}$ is required to be 1, which is
realized by the normalization process. In this scenario, apart from PSO and DE, AD can also be implemented. In
the third scenario, the unitary operation is expressed by $U=\prod_k \exp(i s_k\lambda_k)$ where $\lambda_k$ is
a SU($N$) generator and $s_k$ is a real number in the regime $[0,2\pi]$. The optimization is to find an optimal
set of $\{s_k\}$ for the optimal CFI or $\mathrm{Tr}(W\mathcal{I}^{-1})$, and similar to the second scenario, AD
is also available here besides PSO and DE. In the case that {\codefont mtype="projection"}, each entry of
{\codefont measurement0} in {**kwargs} is a list of arrays, and in the case that {\codefont mtype="input"}, each
entry is an array.

\emph{Example.} Now we consider two models to demonstrate the measurement optimizations in the first scenario. The
first one is a single-parameter case with the single-qubit Hamiltonian $H=\omega\sigma_3/2$ and dynamics in
Eq.~(\ref{eq:ME_spon}). $\delta_{\mathrm{c}}\omega$ and $\delta_{\mathrm{q}}\omega$ are defined in
Eqs.~(\ref{eq:c_deviation}) and (\ref{eq:q_deviation}). As shown in Fig.~\ref{fig:Mopt}(a), $\delta_{\mathrm{c}}\omega$
for the projective measurement $\{\Pi_{+}\!=\!|+\rangle\langle+|,\Pi_{-}\!=\!|-\rangle\langle-|\}$ (dotted black line) can
only reach $\delta_{\mathrm{q}}\omega$ (dashed cyan line) at some specific time points, which has already been shown in
Sec.~\ref{sec:QCRB}. However, utilizing the optimal projective measurements obtained via PSO (blue crosses) and DE (yellow
circles), $\delta_{\mathrm{c}}\omega$ saturates $\delta_{\mathrm{q}}\omega$ for all target time. This performance coincides
with the common understanding that the QFI can be theoretically attained by certain optimal measurements.

In the case of multiparameter estimation, we use the Hamiltonian in Eq.~(\ref{eq:NV_H}) and dynamics in
Eq.~(\ref{eq:NV_ME}) to demonstrate the performance of the optimal projective measurements. The magnetic field
$\vec{B}$ is still the quantity to be estimated. Different with the single-parameter case, the values of
$\mathrm{Tr}(W\mathcal{I}^{-1})$ for the optimal measurements found by PSO (blue crosses) and DE (yellow circles)
cannot attain $\mathrm{Tr}(W\mathcal{F}^{-1})$ (black underlines) in the absence of control, as shown in
Fig.~\ref{fig:Mopt}(b). The gap between $\mathrm{Tr}(W\mathcal{F}^{-1})$ and $\mathrm{Tr}(W\mathcal{I}^{-1})$ is
due to the fact that the quantum Cram\'{e}r-Rao bound is not attainable here. Next, together with the optimal
measurement which gives the lowest $\mathrm{Tr}(W\mathcal{I}^{-1})$, the control is also invoked to further evaluate
the reduction of $\mathrm{Tr}(W\mathcal{I}^{-1})$. Utilizing the optimal controls obtained via auto-GRAPE, the values
of $\mathrm{Tr}(W\mathcal{I}^{-1})$ (red pentagrams) continue to reduce compared to the non-controlled case, yet it
is still unable to attain the controlled values of $\mathrm{Tr}(W\mathcal{F}^{-1})$ (cyan triangles) in general due
to the attainability problem. Nevertheless, their differences are very insignificant for some target time, indicating
that the combined performance of the optimal measurement and optimal control approaches to the ultimate precision limit.
The optimal measurement $\{|\phi_1\rangle\langle\phi_1|,\cdots,|\phi_6\rangle\langle\phi_6|\}$ obtained by DE in the
absence of control are demonstrated in Fig.~\ref{fig:Mopt}(c). The red and blue bars represent the real and imaginary
parts of the coefficients of $|\phi_1\rangle$ to $|\phi_6\rangle$ in the basis
$\{|1\!\uparrow\rangle,|1\!\downarrow\rangle,|0\!\uparrow\rangle,|0\!\downarrow\rangle,|\!-\!1\!\uparrow\rangle,
|\!-\!1\!\downarrow\rangle\}$.

%================================ Figure ====================================
\begin{figure}[tp]
\centering\includegraphics[width=8.5cm]{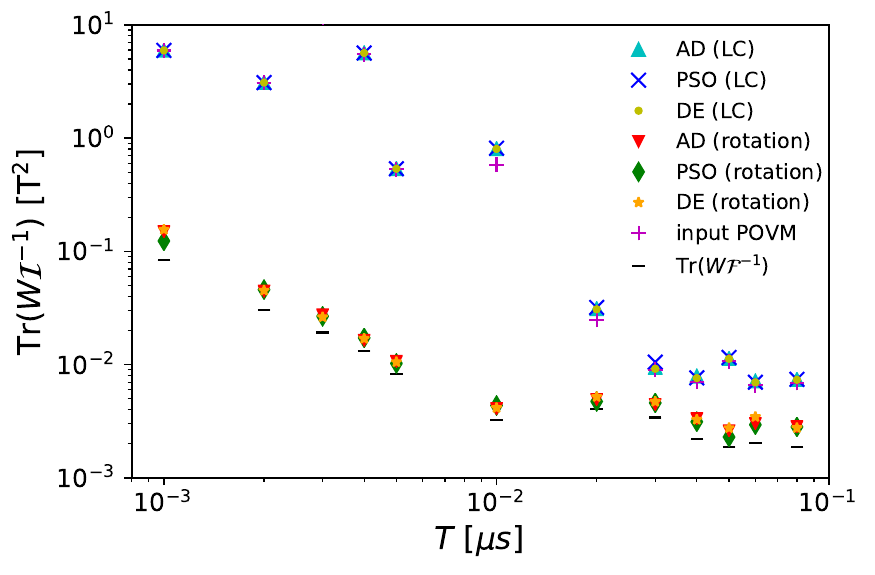}
\caption{Demonstration of the measurement optimization in the second (LC) and
third scenarios (rotation). The cyan upward triangles, blue crosses and
yellow circles represent the performance of optimal measurements found by AD, PSO,
and DE, respectively in the second scenario. The red downward triangles, green
diamonds and orange pentagrams represent the performance of optimal measurements
found by AD, PSO, and DE in the third scenario. }
\label{fig:Mopt_input}
\end{figure}
%============================================================================

The optimizations in the second and third scenarios are also demonstrated with the Hamiltonian in
Eq.~(\ref{eq:NV_H}) and dynamics in Eq.~(\ref{eq:NV_ME}). The input measurement is taken as
$\{|ij\rangle\langle ij|\}_{i=0,\pm 1;j=\uparrow,\downarrow}$, which includes 6 operators. In the second
scenario, the number of output POVM operators is set to be 4. As shown in Fig.~\ref{fig:Mopt_input}, the
performance of measurements found by AD (cyan upward triangles), PSO (blue crosses) and DE (yellow circles)
approach to and even reach that of the input measurement (magenta pluses). This fact indicates that in this
case, an optimal 4-operator measurement can reach the performance of the original 6-operator measurement, and
the reduction of operator numbers may benefit the practical precision of the measurements in experiments.
In the third scenario, the performance of optimal measurements found by AD (red downward triangles), PSO
(green diamonds), and DE (orange pentagrams) not only significantly better than that of the input measurement,
but also approach to the ultimate precision limit given by $\mathrm{Tr}(W\mathcal{F}^{-1})$ (black underlines),
indicating that the performance of these optimal measurements are very close to that of the global optimal
measurements, if there exist any. The probe states, the true values of the parameters to be estimated and other
parameters are set to be the same with those in Sec.~\ref{sec:multi}.

\section{Comprehensive optimization}
\label{sec:comprehensive_opt}

%==================================== Table =================================
\begin{table}[tp]
\begin{tabular}{c|c|c|c}
\hline
\hline
~~Algorithms~~ & ~~method=~~ & \multicolumn{2}{c}{~~**kwargs and default values~~}\\
\hline
\multirow{9}{*}{PSO} & \multirow{9}{*}{"PSO"} & "p\_num" & 10 \\
  &   & "psi0"  & [] \\
  &   & "ctrl0"  & [] \\
  &   & "measurement0"  & [] \\
  &   & "max\_episode"  & [1000,100] \\
  &   & "c0"  & 1.0 \\
  &   & "c1"  & 2.0 \\
  &   & "c2"  & 2.0 \\
  &   & "seed"  & 1234 \\
\hline
\multirow{8}{*}{DE} & \multirow{8}{*}{"DE"} & "p\_num" & 10 \\
  &   & "psi0"  & [] \\
  &   & "ctrl0"  & [] \\
  &   & "measurement0"  & [] \\
  &   & "max\_episode"  & 1000 \\
  &   & "c"  & 1.0 \\
  &   & "cr"  & 0.5 \\
  &   & "seed"  & 1234 \\
\hline
\multirow{7}{*}{AD} & \multirow{7}{*}{"AD"} & "Adam"  & False \\
\multirow{7}{*}{(available}  &   & "psi0"  & [] \\
\multirow{7}{*}{{for SC})}  &   &  "ctrl0" & [] \\
  &   & "measurement0"  & [] \\
  &   & "max\_episode"  & 300 \\
  &   & "epsilon" & 0.01 \\
  &   & "beta1" & 0.90 \\
  &   & "beta2" & 0.99 \\
\hline
\hline
\end{tabular}
\caption{Available methods for comprehensive optimization in QuanEstimation and
corresponding default parameter settings. Notice that AD is only available
when {\codefont com.SC()} is called. }
\label{table:CompOpt_paras}
\end{table}
%============================================================================

%================================ Figure ====================================
\begin{figure*}[tp]
\centering\includegraphics[width=16cm]{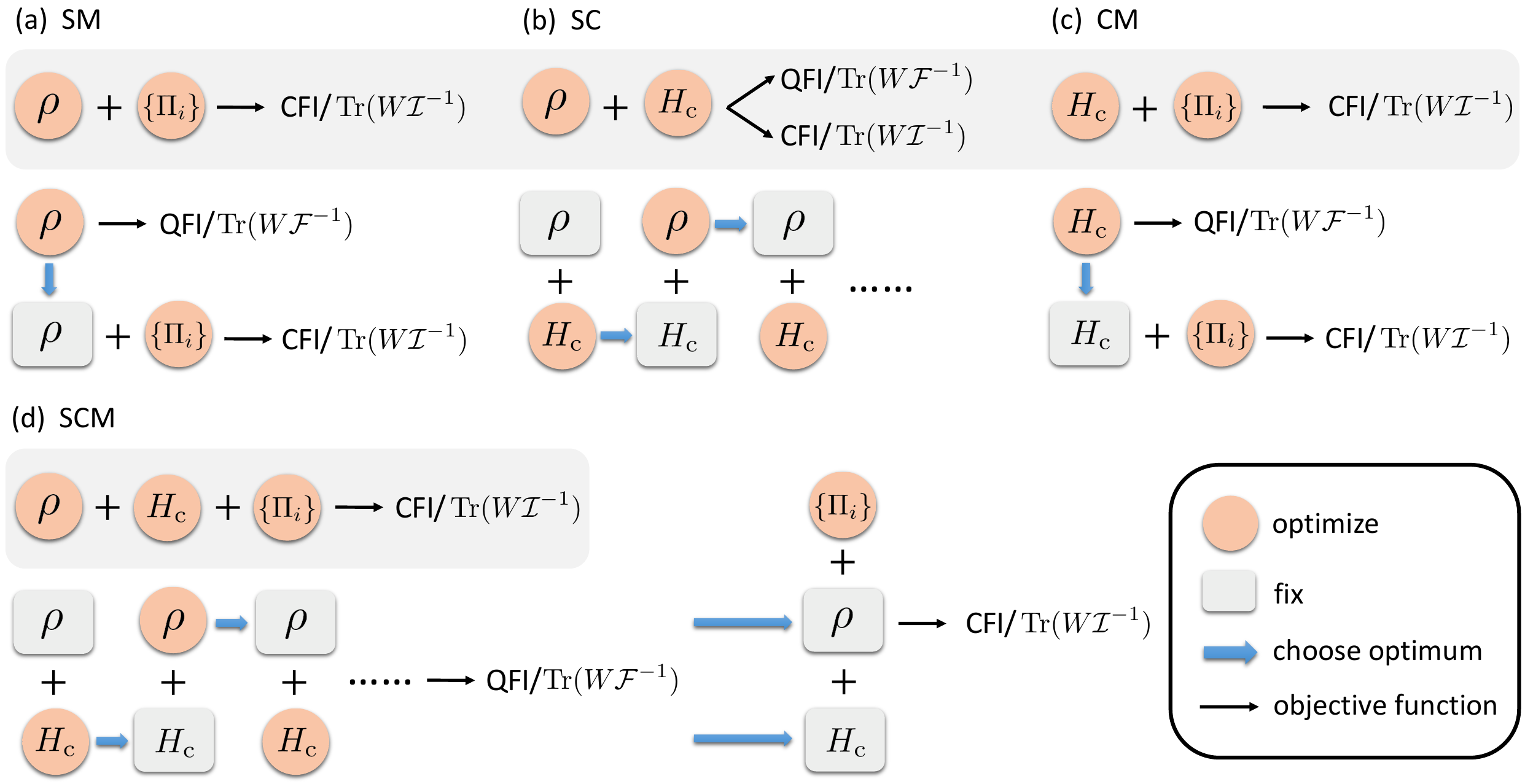}
\caption{Illustration of the comprehensive optimization (first lines with
gray background) and combination of univariate optimizations (second lines)
in four types of multivariate optimizations, including the optimizations of
(a) the probe state and measurement (SM), (b) the probe state and control (SC),
(c) control and measurement (CM), and (d) the probe state, control, and measurement
(SCM).}
\label{fig:compre}
\end{figure*}
%============================================================================

The previous sections focused on the univariate (single variable) optimizations. However, in a
practical scenario the probe state, control (if available) and measurement may all need to be
optimized. More importantly, the optimal results obtained for an univariate optimization may cease
to be optimal when other variables are involved. For example, the optimal probe state and
measurement for the non-controlled case may not be optimal anymore in the controlled case. Hence,
sometimes a comprehensive optimization, i.e., simultaneous multivariate optimization, is in need.

QuanEstimation can deal with four types of multivariate optimizations, including the optimizations of
the probe state and measurement (SM), the probe state and control (SC), control and measurement (CM),
and all three together (SCM). In these scenarios, the key feature of comprehensive optimization is
that all variables are optimized simultaneously. Regarding the objective function, in the cases of SM,
CM, and SCM, namely, when the measurement is involved, it has to be dependent on the measurement.
In current version of the package it is chosen as the CFI or $\mathrm{Tr}(W\mathcal{I}^{-1})$.
In the case of SC, the objective function could be either QFI/$\mathrm{Tr}(W\mathcal{F}^{-1})$ or
CFI/$\mathrm{Tr}(W\mathcal{I}^{-1})$ for a flexible or fixed choice of measurement. The process of
comprehensive optimizations and corresponding objective functions have been illustrated in the first
lines (with gray background) in Figs.~\ref{fig:compre}(a-d). In QuanEstimation, the code for the
execution of comprehensive optimization are:
\begin{lstlisting}[breaklines=true,numbers=none,frame=trBL,mathescape=true]
com = ComprehensiveOpt(savefile=False,
                  method="DE",**kwargs)
com.dynamics(tspan,H0,dH,Hc=[],ctrl=[],
             decay=[],ctrl_bound=[],
             dyn_method="expm")
com.SM(W=[])
com.SC(W=[],M=[],target="QFIM",LDtype="SLD")
com.CM(rho0,W=[])
com.SCM(W=[])
\end{lstlisting}
In the case that the parameterization is described by the Kraus operators, replace {\codefont com.dynamics()}
with the code {\codefont com.Kraus(K,dK)}. All four types of comprehensive optimizations can be called through
{\codefont com.SM()}, {\codefont com.SC()}, {\codefont com.CM()}, and {\codefont com.SCM()}. Notice that if
{\codefont com.Kraus()} is invoked, only {\codefont com.SM()} is available as control is not suitable for the
parameterization process described by the Kraus operators. In {\codefont com.CM()}, the input {\codefont rho0}
is a matrix representing the fixed probe state. In {\codefont com.SC()}, the objective function can be set via
{\codefont target=" "}, including three choices {\codefont target="QFIM"} (default), {\codefont target="CFIM"},
and {\codefont target="HCRB"}. If a set of measurement is input via {\codefont M=[]}, the objective function
will be automatically chosen as the CFIM regardless of the input in {\codefont target=" "}. The type of QFIM
can be adjusted via {\codefont LDtype=" "} ({\codefont "SLD"}, {\codefont "RLD"}, {\codefont "LLD"}). The
available methods for the comprehensive optimization and corresponding default parameter settings are given in
Table~\ref{table:CompOpt_paras}. Notice that AD is only available when {\codefont com.SC()} is called and
the objective function is not the HCRB. At the end of the program, "f.csv" will be generated including the values
of the objective function in all episodes. In the meantime, some or all of the files "controls.csv", "states.csv",
and "measurements.csv" will also be generated according to the type of comprehensive optimization.

Alternatively, the multivariate optimization can also be finished by the combination of univariate
optimizations, as shown in the second lines in Figs.~\ref{fig:compre}(a-d). In the case of SM (or
CM) shown in Fig.~\ref{fig:compre}(a) [Fig.~\ref{fig:compre}(c)], one could first perform the state
(control) optimization with QFI/$\mathrm{Tr}(W\mathcal{F}^{-1})$ the objective function. Next,
take the found optimal state (control) as the fixed input, and further optimize the measurement
with CFI/$\mathrm{Tr}(W\mathcal{I}^{-1})$ the objective function. If the optimized values of the
CFI/$\mathrm{Tr}(W\mathcal{I}^{-1})$ in the second process reaches the optimized values of the
QFI/$\mathrm{Tr}(W\mathcal{F}^{-1})$ in the first process, the entire scheme is then optimal. Things
could be more complex in the multiparameter estimation due to the attainability problem. The existence
of the gap between the optimized $\mathrm{Tr}(W\mathcal{I}^{-1})$ and $\mathrm{Tr}(W\mathcal{F}^{-1})$
does not necessarily mean the scheme is not optimal. Nevertheless, there is no doubt that a smaller gap
always implies a better scheme at least in theory. In the case of SC, the state optimization and
control optimization can be performed in turn with the optimal quantity found in the previous
turn as the fixed input [Fig.~\ref{fig:compre}(b)]. Same with the comprehensive optimization, both
QFI/$\mathrm{Tr}(W\mathcal{F}^{-1})$ and CFI/$\mathrm{Tr}(W\mathcal{I}^{-1})$ can be taken as the
objective function in this case. At last, in the case of SCM the combination strategy in SC
could be performed first with QFI/$\mathrm{Tr}(W\mathcal{F}^{-1})$ the objective function, and
the measurement is further optimized with the found optimal state and control as the fixed input
[Fig.~\ref{fig:compre}(d)]. Same with the scenario of SM, if the optimized CFI/$\mathrm{Tr}(W\mathcal{I}^{-1})$
obtained in the second process reaches the optimized QFI/$\mathrm{Tr}(W\mathcal{F}^{-1})$ in the
first process, the entire scheme is optimal.

\emph{Example.} Now we provide some demonstrations on the comprehensive optimization with QuanEstimation
and compare their performance with the combination strategy. First, consider a non-controlled example
with the single-qubit Hamiltonian $\omega\sigma_3/2$, which is a SM scenario. The dynamics is
governed by Eq.~(\ref{eq:ME_spon}) with decay rates $\gamma_{-}/\omega_{\mathrm{tr}}=0$ and
$\gamma_{+}/\omega_{\mathrm{tr}}=0.1$. The target time $\omega_{\mathrm{tr}}T=20$. In this case, the optimized
values of $\sqrt{\omega_{\mathrm{tr}}T}\delta_{\mathrm{c}}\omega$ in the comprehensive optimization and
combination strategy are both 0.608 (in the units of $\omega_{\mathrm{tr}}$, same below), equivalent to the optimal
$\sqrt{\omega_{\mathrm{tr}}T}\delta_{\mathrm{q}}\omega$ obtained in the solely state optimization, indicating
that the schemes found by both strategies are indeed optimal in theory. Next we invoke the controls described
in Eq.~(\ref{eq:ctrl_demo}). In the case of SC, the optimized $\sqrt{\omega_{\mathrm{tr}}T}\delta_{\mathrm{q}}\omega$
obtained in the combination strategy is 0.441, and that in the comprehensive optimization is 0.440. Furthermore,
in the case of SCM, the optimized $\sqrt{\omega_{\mathrm{tr}}T}\delta_{\mathrm{c}}\omega$ provided by the
combination strategy is 0.441, equivalent to the optimal $\sqrt{\omega_{\mathrm{tr}}T}\delta_{\mathrm{q}}\omega$
obtained in the SC, and that provided by the comprehensive optimization is 0.443. The performance of these two
strategies basically coincide with each other in this example.

\begin{figure}[tp]
\centering\includegraphics[width=8.5cm]{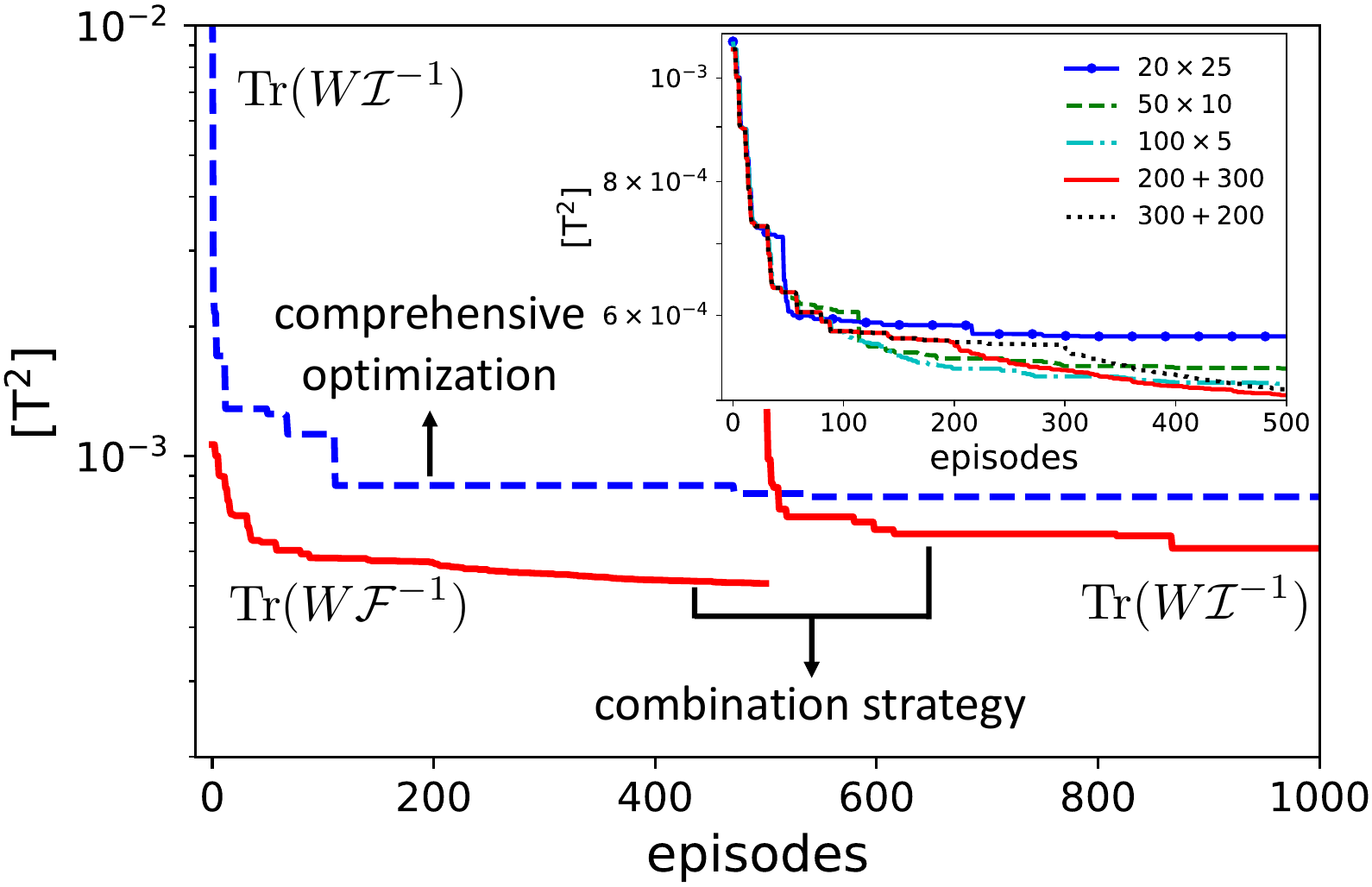}
\caption{Performance comparison between the comprehensive optimization
and combination strategy in the multiparameter estimation in the case of SCM.
The dashed blue line represents the optimization of $\mathrm{Tr}(W\mathcal{I}^{-1})$
in the comprehensive optimization. The solid red lines represent the optimization
of $\mathrm{Tr}(W\mathcal{F}^{-1})$ in the SC (first 500 episodes) and that of
$\mathrm{Tr}(W\mathcal{I}^{-1})$ in the measurement optimization (last 500
episodes) in the combination strategy. The inset shows the performance of
different combination strategies in the SC part due to the episode number
of each optimization. All the optimizations in the figure are finished by DE.}
\label{fig:compre_multi}
\end{figure}

This equivalent performance may due to two facts: the example is simple and the QFI is attainable in
theory. In the multiparameter estimation, these two strategies may show divergent performance as the
QFIM is not always guaranteed to be attainable. For example, in the case of SCM, $\mathrm{Tr}(W\mathcal{F}^{-1})$
are first optimized in the SC. However, it is hard to say whether the optimal probe state and control
for an unattainable $\mathrm{Tr}(W\mathcal{F}^{-1})$ can still provide a good $\mathrm{Tr}(W\mathcal{I}^{-1})$
and benefit the subsequent measurement optimization. To investigate it, we still take the nitrogen-vacancy
center as an example. The free Hamiltonian, control Hamiltonian, and dynamics are described in
Eqs.~(\ref{eq:NV_H}), (\ref{eq:NV_c}), and (\ref{eq:NV_ME}). The performance of comprehensive optimization and
combination strategy in the SCM are shown in Fig.~\ref{fig:compre_multi}. The comprehensive optimization (dashed
blue line), which takes $\mathrm{Tr}(W\mathcal{I}^{-1})$ as the objective function, basically converges at
around $110$ episodes. The combination strategy (solid red line) splits into two parts, the one in the first
500 episodes is the combination optimization of SC, and that in the last 500 episodes is the optimization of the
measurement. The gap between these two lines is actually the gap between the optimal $\mathrm{Tr}(W\mathcal{F}^{-1})$
and the value of $\mathrm{Tr}(W\mathcal{I}^{-1})$ with a random measurement. In the SC part, the alternative
optimizations of the probe state and control can be done in different ways due to the episode number of each
optimization. As shown in the inset of Fig.~\ref{fig:compre_multi}, here we test several selections, including
20 episodes for each optimization (solid circle blue line), 50 episodes for each optimization (dashed green
line), 100 episodes for each optimization (dash-dotted cyan line), 200 episodes for state optimization and 300
episodes for control optimization (solid red line), and 300 episodes for state optimization and 200 episodes
for control optimization (dotted black line). In these selections, the fourth one, 200 episodes for state
optimization and 300 episodes for control optimization, shows the best performance at the end of the 500
episodes, and the corresponding optimal state and control are chosen for the subsequent measurement optimization.
In this example, the final performance of the combination strategy is better than that of the simultaneous
strategy, indicating that the unattainability of $\mathrm{Tr}(W\mathcal{F}^{-1})$ in the SC does not present
negative effects on the final performance. However, this result does not mean the combination strategy is always
better in general. In practice, the comparison of these two strategies might still be needed case by case in
the scheme design.

%================================ Figure ====================================
%============================================================================

\section{Adaptive measurement schemes}
\label{sec:adapt}

Adaptive measurement is another common scenario in quantum parameter estimation. In this scenario, apart from the
unknown parameters $\bold{x}$, the Hamiltonian also includes a set of tunable parameters $\bold{u}$. A typical case
is that the tunable parameters are invoked by the same way with $\bold{x}$, resulting in the total Hamiltonian
$H(\bold{x}+\bold{u})$. In the point estimation approach,  the QFIM and CFIM computed at the true values of $\bold{x}$
may not always provide the practically achievable precision due to the fact that the actual working point may be
slightly away from the true values. Hence, the tunable parameters $\bold{u}$ are invoked to let the Hamiltonian
$H(\bold{x}+\bold{u})$ work at the optimal point $\bold{x}_{\mathrm{opt}}$. An obvious difficulty for the
implementation of this scheme is that one actually does not known the true values in practice, which means $\bold{u}$
has to be given according to the estimated values $\hat{\bold{x}}$, and the entire scheme would only be useful when
it is implemented adaptively. In the meantime, a pre-estimation of $\bold{x}$ is usually needed. The inaccuracy of
$\hat{\bold{x}}$ would result in the inaccuracy of $\bold{u}$, and $\hat{\bold{x}}+\bold{u}$ is then inevitably far
from $\bold{x}_{\mathrm{opt}}$, causing a lousy performance of the scheme. This scheme has been demonstrated by Berni
et al.~\cite{Berni2015} in optical phase estimation with additional real-time feedback controls.

%================================ Figure ====================================
\begin{figure}[tp]
\centering\includegraphics[width=8.5cm]{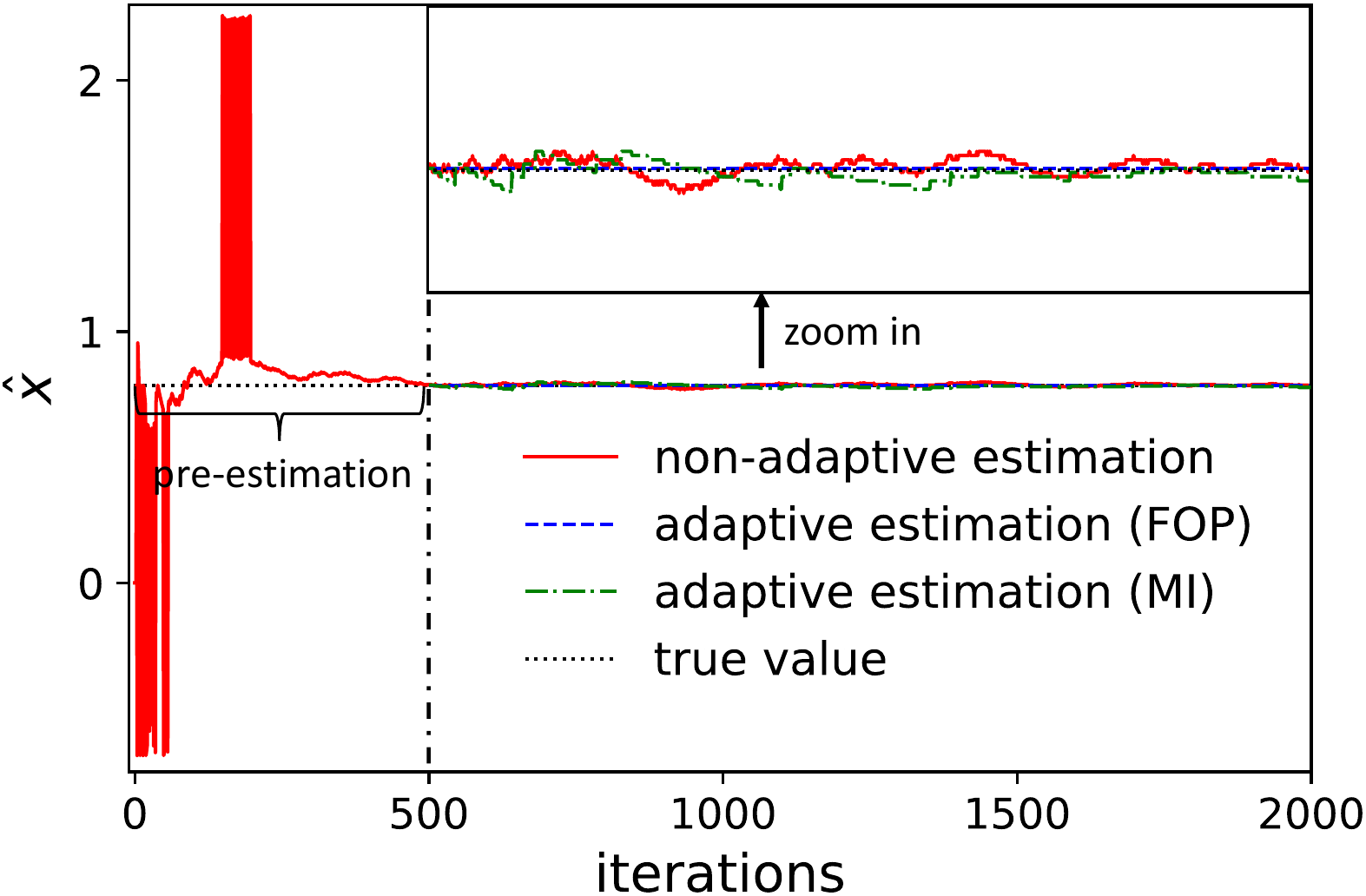}
\caption{Performance comparison between the adaptive schemes realized by FOP
(dashed blue line) and MI (dash-dotted green line), and the non-adaptive schemes
(solid red line). The adaptive measurement starts after 500 rounds of
pre-estimation. The non-adaptive scheme is a full Bayesian estimation. The
dotted black line represents the true value.}
\label{fig:adpt}
\end{figure}
%============================================================================

Now let us introduce in detail all steps required to implement this scheme. Consider the Hamiltonian $H(\bold{x})$
where $\bold{x}$ is restricted in a finite regime with a prior distribution $p(\bold{x})$. The first step is to find
the optimal value $\bold{x}_{\mathrm{opt}}$ in this regime with respect to the minimum $\mathrm{Tr}(W\mathcal{I}^{-1})$
when the measurement is fixed. If the measurement can be altered flexibly in practice, $\bold{x}_{\mathrm{opt}}$,
together with the corresponding optimal measurement, can be obtained with $\mathrm{Tr}(W\mathcal{F}^{-1})$
the objective function. Next, perform the pre-estimation via the Bayesian estimation with the fixed or
optimal measurement and update the prior distribution with the posterior distribution in Eq.~(\ref{eq:Bayes_posterior}).
When $p(\bold{x})$ has been updated to a reasonable narrow distribution, the tunable parameters $\bold{u}$ are then
invoked into the system. In the $n$th round of this step, with the observed result $y^{(n)}$, the posterior distribution
is obtained via the Bayes' rule as
\begin{equation}
p(\bold{x},\bold{u}^{(n)}|y^{(n)})=\frac{p(y^{(n)}|\bold{x},\bold{u}^{(n)})
p(\bold{x})}{\int p(y^{(n)}|\bold{x},\bold{u}^{(n)})p(\bold{x})\mathrm{d}\bold{x}},
\label{eq:adpt_px}
\end{equation}
where $\bold{u}^{(n)}$ is obtained in the $(n-1)$th round. The estimated value $\hat{\bold{x}}^{(n)}$ can be
obtained through the MAP, $\hat{\bold{x}}^{(n)}=\mathrm{argmax}\,p(\bold{x},\bold{u}^{(n)}|y^{(n)})$. The
value of $\bold{u}$ used in the next round is obtained by $\bold{u}^{(n+1)}=\bold{x}_{\mathrm{opt}}-\hat{\bold{x}}^{(n)}$,
and the prior distribution is also replaced by the current posterior distribution. This update method of $\bold{u}$
is referred to as the fixed optimal point method (FOP) in this paper. In QuanEstimation, the pre-estimation
can be finished with the function {\codefont Bayes()} discussed in Sec.~\ref{sec:Bayesian}, and the adaptive process can
be executed with the code:
\begin{lstlisting}[breaklines=true,numbers=none,frame=trBL,mathescape=true]
apt = Adapt(x,p,rho0,method="FOP",
            savefile=False,max_episode=1000,
            eps=1e-8)
apt.dynamics(tspan,H,dH,Hc=[],ctrl=[],
             decay=[],dyn_method="expm")
apt.CFIM(M=[],W=[])
\end{lstlisting}
In the case that the parameterization process is described by the Kraus operators, replace
{\codefont apt.dynamics()} with {\codefont apt.Kraus(K,dK)}. The inputs {\codefont x} and {\codefont p} are the
same with those in {\codefont Bayes()}. The input {\codefont H} is a list of matrices representing the Hamiltonian
with respect to the values in {\codefont x}, and it is multidimensional in the multiparameter case. {\codefont dH}
is a (multidimensional) list with each entry also a list representing $\partial_{\bold{x}}H$ with respect to the
values in {\codefont x}. In the case that specific functions of $H$ and $\partial_{\bold{x}}H$ can be provided,
{\codefont H} and {\codefont dH} can be alternatively generated via the function {\codefont BayesInput()} discussed
in Sec.~\ref{sec:para}. In {\codefont apt.CFIM()}, {\codefont M} is the input measurement and the default one is a
set of SIC-POVM.

During the running of the code, three files "xout.csv", "y.csv", and "pout.csv" will be generated including the
data of $\hat{\bold{x}}$, result $y$ in all rounds of iteration and final obtained $p(\bold{x})$. In the case
that {\codefont savefile=True}, "pout.csv" contains the data of $p(\bold{x})$ in all rounds. If the choice of
measurement is flexible in the experiment, before the invocation of {\codefont apt.CFIM()}, the optimal measurement
with respect to $\bold{x}_{\mathrm{opt}}$ can be first obtained via calling {\codefont M = apt.Mopt(W=[])}.
In the case that the users would like to run the pre-estimation with the optimal measurement, they can just call
{\codefont apt.Mopt()} first and input the optimal measurement to {\codefont Bayes()} for the pre-estimation.

During the running of {\codefont apt.CFIM()}, the users should type the result $y$ obtained in practice on the
screen and receive the values of $\bold{u}$ used for the next round of experiment. In the case that the users
have already done the pre-estimation by themselves, they can directly use {\codefont Adapt()} without calling
{\codefont Bayes()} first.

Apart from the FOP, $\bold{u}$ can also be updated via other strategies. One such choice is utilizing the optimization
of the mutual information (MI), which is defined by
\begin{equation}
I(\bold{u})\!=\!\int\!p(\bold{x}) \sum_{y}p(y|\bold{x},\bold{u})
\mathrm{log}_2\!\left[\frac{p(y|\bold{x},\bold{u})}{\int p(\bold{x})
p(y|\bold{x},\bold{u})\mathrm{d}\bold{x}}\right]\!\mathrm{d}\bold{x}.
\label{eq:MI}
\end{equation}
In the $n$th round, the prior distribution $p(\bold{x})$ is updated via Eq.~(\ref{eq:adpt_px}), and $\bold{u}^{(n+1)}$
is obtained by the equation $\bold{u}^{(n+1)}=\mathrm{argmax}~I(\bold{u})$. In QuanEstimation, this method can be
invoked by setting {\codefont method="MI"} in {\codefont Adapt()}. Notice that in this method a good pre-estimation
should let the prior distribution converges to zero at the boundary of the input {\codefont x}.

Let us still take the Hamiltonian in Eq.~(\ref{eq:Bayes_demo}) as an example. The initial state is $|+\rangle$ and
the target time $\omega_0 T=1$ (Planck units are applied). The prior distribution is uniform in the regime
$(-\pi/4,3\pi/4)$. In this regime, zero is an optimal point and is chosen to be $x_{\mathrm{opt}}$. The measurement
is $\{|+\rangle\langle +|,|-\rangle\langle-|\}$. The results are simulated by generating random values in the regime
$[0,1]$. When it is smaller (larger) than $p(+|x)$, the posterior distribution is calculated with $p(+|x)$ [$p(-|x)$].
As shown in Fig.~\ref{fig:adpt}, after 500 rounds of pre-estimation, the adaptive schemes realized by FOP (dashed blue
line) and MI (dash-dotted greeen line) indeed show better performance, namely, smaller variance, compared to the
non-adaptive scheme (solid red line) which is fully finished by the Bayesian estimation. Notice that this figure is
only a one-time simulation of the experiment. The performance may be different when the results are different.

Another famous adaptive scheme is the online adaptive phase estimation, proposed by Berry et al.~\cite{Berry2000,Berry2001},
in the scenario of Mach-Zehnder interferometer (MZI). In this scheme, after reading the result $y^{(n)}$ in the $n$th
round, the value of the tunable phase $\Phi_{n+1}$ or phase difference $\Delta\Phi_{n+1}$ is generated. The relation
between $\Phi_{n+1}$ and $\Delta\Phi_{n+1}$ can be taken as $\Phi_{n+1}=\Phi_{n}-(-1)^{y^{(n)}}\Delta\Phi_{n+1}$.
Hentschel and Sanders~\cite{Hentschel2010,Hentschel2011} further provided an offline strategy with PSO, and the
optimization methods are further extended to DE~\cite{Lovett2013} and genetic algorithm~\cite{Rambhatla2020} in recent
years. Apart from the original references, details of this scheme can also be found in a recent review~\cite{Liu2022}.
In QuanEstimation, this scheme can be executed by the code:
\begin{lstlisting}[breaklines=true,numbers=none,frame=trBL,mathescape=true]
apt = Adapt_MZI(x,p,rho0)
apt.general()
apt.online(target="sharpness",output="phi")
\end{lstlisting}
The input {\codefont rho0} is a matrix representing the probe state. The output can be tuned between $\Phi$ and
$\Delta\Phi$ by setting {\codefont output="phi"} or {\codefont output="dphi"} in {\codefont apt.online()} in the
demonstrating code. {\codefont target="sharpness"} means the tunable phase is obtained by the maximization of the
sharpness function. The specific formula of the sharpness function can be found in Refs.~\cite{Berry2000,Berry2001,
Hentschel2010,Hentschel2011,Liu2022}. Alternatively, the sharpness function can also be replaced by the mutual
information in Eq.~(\ref{eq:MI}) via setting {\codefont target="MI"}, which has been both theoretically and
experimentally discussed by DiMario and Becerra in 2020~\cite{DiMario2020}.

The offline strategy can also be executed by replacing {\codefont apt.online()} with the code:
\begin{lstlisting}[breaklines=true,numbers=none,frame=trBL,mathescape=true]
apt.offline(target="sharpness",method="DE",
            **kwargs)
\end{lstlisting}
PSO is also available here ({\codefont method="PSO"}). When the entire program is finished, a file named
"xout.csv" including the data of output in all rounds will be generated. In the case of online scheme, an
additional file "y.csv" including the result $y$ in all rounds will also be generated. The design of
{\codefont apt.general()} here is to give us a space for the further inclusion of the adaptive phase estimation
in other optical scenarios such as the SU(1,1) interferometers.

\section{Summary}

In this paper, we present a new open-source toolkit, QuanEstimation, for the design of optimal schemes in the
quantum parameter estimation. The source of the package, as well as the demonstrating code for the calculation
of all examples discussed in this paper, can be download in GitHub~\cite{github1} and the documentation is in
Ref.~\cite{docs2022}. This package is based on both platforms of Python and Julia. The combined structure is
to guarantee the calculation efficiency of Julia is fully utilized, and in the meantime, the people who have
no knowledge of Julia would have no obstacle in using this package. In the meantime, a full Julia version of
the package is also available in GitHub~\cite{github2}, which is suitable for those familiar with Julia.
QuanEstimation includes several well-studied metrological tools in quantum parameter estimation, such as the
various types of Cram\'{e}r-Rao bounds and their quantum correspondences, quantum Ziv-Zakai bound, and Bayesian
estimation. To perform the scheme design, QuanEstimation can execute the optimizations of the probe state,
control, measurement, and the comprehensive optimizations, namely, the simultaneous optimizations among them.
General adaptive measurement schemes as well as the adaptive phase estimation can also be performed with this
toolkit.

QuanEstimation is suitable for many practical quantum systems, especially those with finite-dimensional Hilbert
spaces, such as the trapped ions, nitrogen-vacancy centers, and quantum circuits. Therefore, it is not only
useful for the theorists working in the field of quantum parameter estimation, but could also be particularly
useful for the experimentalists who are not familiar with the theories in this field yet intend to utilize
them to design experimental schemes. More functions and features will be constantly input into the package and
the calculation efficiency for certain specific scenarios will be further improved in the future. Moreover, the
calculations in QuanEstimation are majorly based on the density matrices, which may cause inefficiency when the
dimension of Hilbert space is large. More technologies targeting at the many-body systems, such as the sparse
matrices and matrix product states, will be further involved in the package in the future. We believe that there
is a good chance that this package would become a common toolkit in the field of quantum metrology for the
numerical calculations and scheme designs.

\begin{acknowledgments}
The authors would like to thank Prof.~Libin Fu, Prof.~Re-Bing Wu, Prof.~Lijian Zhang, Prof.~Christiane P. Koch,
Prof.~Syed M. Assad, Prof.~Jun Suzuki, Jinfeng Qin, and Yuqian Xu for helpful discussions. This work was
supported by the National Natural Science Foundation of China (Grants No.\,12175075, No.\,11805073,
No.\,11935012 and No.\,11875231), and the National Key Research and Development Program of China (Grants
No.\,2017YFA0304202 and No.\,2017YFA0205700). H.Y. also acknowledges the support from the Research Grants
Council of Hong Kong (Grant No.\,14307420). R.D.D. was supported by National Science Center (Poland) with
Grant No.~2020/37/B/ST2/02134.
\end{acknowledgments}

\end{document}